\documentclass[journal, twoside, final]{IEEEtran}
\usepackage{amssymb, amsmath, amsthm, amsfonts}
\usepackage{bm, color}
\usepackage{booktabs}
\usepackage{algorithm}
\usepackage{algorithmic}
\usepackage{url}
\usepackage{graphicx}
\usepackage{longtable}
\usepackage{makecell}
\usepackage{cite}
\usepackage{color}

\newtheorem{theorem}{Theorem}
\newtheorem{remark}{Remark}

\newtheorem{prop}{Proposition}
\allowdisplaybreaks[4]
\usepackage{lettrine}
\usepackage{threeparttable}

\begin{document}
\title{Energy-Efficient Non-Orthogonal Multicast and Unicast Transmission of Cell-Free Massive MIMO Systems with SWIPT}

\author{Fangqing Tan, Peiran Wu \IEEEmembership{Member,~IEEE}, Yik-Chung Wu, \IEEEmembership{Senior Member,~IEEE} and Minghua Xia \IEEEmembership{Member,~IEEE}

\thanks{Manuscript received January 30, 2020; revised May 30, 2020; accepted July 16, 2020. This work was supported in part by the China Postdoctoral Science Foundation under Grant 2019M653177, in part the National Natural Science Foundation of China under Grants 61801526 and 61671488, in part by the Major Science and Technology Special Project of Guangdong Province under Grant 2018B010114001, in part by the Guangxi Natural Science Foundation under Grants 2018GXNSFBA138034 and AD18281052, and in part by the Fund of Key Laboratory of Cognitive Radio and Information Processing, Ministry of Education, China, under Grant CRKL180103.
	
Fangqing Tan, Peiran Wu, and Minghua Xia are with the School of Electronics and Information Technology, Sun Yat-sen University, Guangzhou 510006, China. Fangqing Tan is also with the Key Laboratory of Cognitive Radio and Information Processing, Guilin University of Electronic Technology, Guilin 541004, China. Peiran Wu and Minghua Xia are also with the Southern Marine Science and Engineering Guangdong Laboratory, Zhuhai 519082, China (e-mail: \{tanfq, wupr3, xiamingh\}@mail.sysu.edu.cn).}

\thanks{Yik-Chung Wu is with the Department of Electrical and Electronic Engineering, The University of Hong Kong, Hong Kong (e-mail: ycwu@eee.hku.hk).}

\thanks{Color versions of one or more of the figures in this article are available online at https://ieeexplore.ieee.org.

		Digital Object Identifier XXX}
}

\markboth{IEEE Journal on Selected Areas in Communications} {Tan \MakeLowercase{\textit{et al.}}: Energy-Efficient Non-Orthogonal Multicast and Unicast Transmission of Cell-Free Massive MIMO Systems with SWIPT}

\maketitle

\IEEEpubid{\begin{minipage}{\textwidth} \ \\[12pt] \centering 0733-8716 \copyright\ 2020 IEEE. Personal use is permitted, but republication/redistribution requires IEEE permission. \\
See \url{https://www.ieee.org/publications/rights/index.html} for more information.\end{minipage}}

\begin{abstract}
	This work investigates the energy-efficient resource allocation for layered-division multiplexing (LDM) based non-orthogonal multicast and unicast transmission in cell-free massive multiple-input multiple-output (MIMO) systems, where each user equipment (UE) performs wireless information and power transfer simultaneously. To begin with, the achievable data rates for multicast and unicast services are derived in closed form, as well as the received radio frequency (RF) power at each UE. Based on the analytical results, a nonsmooth and nonconvex optimization problem for energy efficiency (EE) maximization is formulated, which is however a challenging fractional programming problem with complex constraints. To suit the massive access setting, a first-order algorithm is developed to find both initial feasible point and the nearly optimal solution. Moreover, an accelerated algorithm is designed to improve the convergence speed. Numerical results demonstrate that the proposed first-order algorithms can achieve almost the same EE as that of second-order approaches yet with much lower computational complexity, which provides insight into the superiority of the proposed algorithms for massive access in cell-free massive MIMO systems.
\end{abstract}

\IEEEpubidadjcol

\begin{IEEEkeywords}
	Cell-free massive multiple-input multiple-output, energy efficiency, first-order algorithm, layered-division multiplexing, non-orthogonal multicast and unicast transmission.
\end{IEEEkeywords}

\section{Introduction}
\lettrine[lines=2]{W}{ith} the phenomenal growth of Internet-of-Things (IoT) applications for pervasive interconnectivity of the global physical environment, there will be more than $75.4$ billion IoT devices by 2025 \cite{IHS}. To enable massive connectivity among wireless networks over limited radio resources, massive multiple-input multiple-output (MIMO) systems use few hundred antennas to simultaneously serve large number of user terminals \cite{Marzetta2016}.  For the implementation of massive MIMO systems in real-world applications, a large number of antennas can be deployed in central or distributed manners. In the former case, all antennas are located in a compact space, without  backhaul requirement. In the latter case, however, different antennas are geometrically separated yet connected to a central processing unit (CPU) via a backhaul network. As user equipments (UEs) are closer to antennas, the distributed MIMO (also known as cell-free massive MIMO) is capable of extending coverage area and/or reducing outage probability, compared with the central MIMO \cite{7386643, CF_MIMO_1}.  On the other hand, simultaneous wireless information and power transfer (SWIPT) technique was widely accepted as a promising technique for 5G wireless networks, especially for low-power consumption applications such as IoT networks \cite{SWIPT_1, 7070727}. When SWIPT is integrated into cell-free massive MIMO systems, higher degrees of macro-diversity and lower path loss rendered by cell-free massive MIMO can be exploited to boost the performance of SWIPT \cite{SWIPT_7, SWIPT_9}. Therefore, the integration of cell-free massive MIMO and SWIPT is appealing for future energy-efficient wireless networks. 

Unlike traditional orthogonal transmission without inter-user interference, non-orthogonal transmission enables more UEs to be simultaneously served, where mild interference among UEs are tolerated \cite{6151774}. In recent years, the power-domain non-orthogonal multiple access (NOMA) technology has emerged for the massive connectivity in wireless networks. For instance, the energy efficiency (EE) of SWIPT-enabled NOMA networks was investigated in \cite{8350092, 8854318, 8891923}, and the fairness issue among UEs was studied in \cite{8247249}. To further enhance the capability of multiple access in 5G networks, more and more multicast services such as live video sharing emerge, apart from unicast services like private communications. Conventionally, multicast and unicast transmissions use distinct time or frequency resources. These traditional orthogonal or NOMA schemes are easy to implement in practice but yield low spectral efficiency (SE). To improve SE, layered division multiplexing (LDM) scheme, which was originally introduced for joint multicast and unicast transmission in the new generation digital television broadcasting \cite{7378924, 7378930}, finds many applications in cellular networks \cite{NOUM_1, NOUM_2, NOUM_3, BF_1}.

\IEEEpubidadjcol

Unlike the power-domain NOMA where multiple messages share a same beamformer but with different powers, LDM applies a layered transmission structure to transmit multiple signals with different beamformers and robustness against different services and reception environment. The work \cite{NOUM_1} demonstrated that LDM outperforms orthogonal multiplexing schemes in terms of SE by jointly optimizing multicast and unicast beamforming. The work \cite{NOUM_2} investigated the energy-efficient precoding for the LDM-based non-orthogonal multicast and unicast transmission, and \cite{NOUM_3} studied the energy-efficient hybrid precoding for integrated multicast-unicast millimeter wave system with SWIPT. The authors of \cite{BF_1} designed the non-orthogonal multicast and unicast transmission for massive MIMO systems, which provides higher achievable data rate than the counterpart orthogonal schemes. However, to the authors' best knowledge, there is few research concerning the energy-efficient LDM-based non-orthogonal multicast and unicast transmission in the open literature.

This paper designs an energy-efficient LDM-based non-orthogonal multicast and unicast transmission for cell-free massive MIMO systems, where a large number of multi-antenna access points (APs) cooperating via a capacity-limited backhaul network jointly provide multicast and unicast services for massive UEs with simultaneous information and power transfer. The main contributions of this paper are summarized as follows:

\begin{itemize}
	\item The achievable data rates of multicast and unicast services as well as the received radio frequency (RF)  power at each UE are derived in closed form. The resultant expressions enable to precisely quantify the effects of the transmit power of APs and the power splitting factors of UEs on the system EE, thus facilitating the energy-efficient resource allocation strategies. 
	
	\item By using a practical power consumption model, an EE maximization problem is formulated for the joint design of transmit power and power splitting factors under the constraints of multicast and unicast service rates, per-AP power and backhaul capacity. However, the problem is neither smooth nor convex. For ease of mathematical tractability, the original formulation is transformed into a sequence of quasi-convex problems by using the smooth approximation and the successive convex approximation (SCA) techniques. Then, to suit the massive access setting, a first-order algorithm is developed to solve the resultant optimization problem, which achieves almost the same EE as that of traditional second-order algorithms yet with much lower computational complexity.

	\item To improve the convergence speed, an accelerated first-order algorithm is further designed by exploiting the momentum technique. As each iterative update in the accelerated algorithm is more aggressive than the conventional gradient step, it converges more than twice as fast as the first-order algorithm, without loss of EE.
	
	\item To identify a feasible initial point, the original feasibility problem is equivalently transformed into a nonconvex optimization problem. Then, another first-order algorithm is developed to efficiently solve the transformed optimization problem. Finally, simulation results demonstrate the effectiveness of the proposed algorithms.
\end{itemize}
As the proposed first-order algorithms are Hessian-free and attain almost the same EE as the optimal second-order algorithm but with much lower computational complexity and remarkably faster convergence speed, they are eminently suitable for massive access in large-scale wireless networks. 

To detail the aforementioned contributions, the rest of this paper is organized as follows. Section~\ref{Sec_sys_m} describes the system model. Section~\ref{Sec_pro_f} formulates the EE maximization problem. Section~\ref{Sec_EE_RA} designs a first-order algorithm and its accelerated version to solve the problem. Section~\ref{Sec_ffp} develops another first-order algorithm to find a feasible initial point. Simulation results are discussed in Section~\ref{Sec_sim} and, finally, Section~\ref{Sec_con} concludes the paper.

\section{System Model}
\label{Sec_sys_m}
\begin{figure}[t]
	\centering
	\includegraphics[width=3.5in]{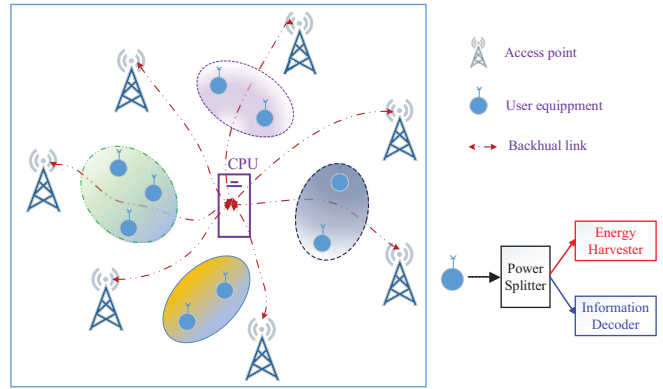}
	\caption{Non-orthogonal multicast and unicast downlink transmission in a cell-free massive MIMO system with SWIPT.}
	\label{Fig_1}
\end{figure}

As shown in Fig.~\ref{Fig_1}, we consider the downlink transmission of a backhaul-constrained cell-free massive MIMO system with SWIPT, which consists of $N$ APs each equipped with $M$ transmit antennas, and $K$ single-antenna UEs. To enable massive access, the system provides both group-specific multicast service and user-specific unicast service while each UE has a dedicated unicast request and subscribes a group-specific multicast service. For the multicast service, different UEs inquiring the same content are served in the same group. Mathematically, the set of UEs in the $g^{\rm th}$ group is denoted as $\mathcal{K}_{g}$, $g \in \mathcal{G}$, where $\mathcal{G}=\{1,\cdots, G\}$ with $G$ being the total number of multicast groups. Also, it is assumed that each UE joins only one multicast group, i.e., $\mathcal{K}_{i} \cap \mathcal{K}_{g} = \emptyset, \ \forall i \neq g, \ i, g \in \mathcal{G}$ and $\sum_{i\in{\cal G}}|\mathcal{K}_{g}|=K$.  For the unicast service, on the other hand, different UEs are served individually.  

Assume that the system operates in time-division duplex  mode and the channel coherence interval $\tau_{\rm c}$ is divided into two phases, namely, uplink channel estimation phase and downlink data transmission and power transfer phase. In practice, the channel from the $k^{\rm th}$ UE to the $n^{\rm th}$ AP can be modeled as
\begin{equation}
	\bm{g}_{n, k} = \beta_{n, k}^{\frac{1}{2}}\bm{h}_{n, k},
\end{equation}
where $\beta_{n, k}$ denotes large-scale fading while $\bm{h}_{n,k} \in \mathbb{C}^{M\times1} \sim \mathcal{CN}\left( \bm{0},\bm{I}_{M} \right)$ refers to small-scale Rayleigh fading.

In the downlink phase, the APs provide both multicast and unicast services by exploiting a two-layer LDM structure, where the first layer is used for the multicast service while the second layer for the unicast service and the two layers are superposed with different beamforming vectors at each AP. Let $s_{{\rm m},g}$ be the multicast symbol requested by the $g^{\rm th}$ group with normalized energy, i.e., $\mathbb{E}\left[|s_{{\rm m},g}|^{2}\right]=1$, and $s_{{\rm u},k}$ be the unicast symbol requested by the $k^{\rm th}$ UE with $\mathbb{E}\left[|s_{{\rm u},k}|^{2}\right]=1$. Then, the signal transmitted by the $n^{\rm th}$ AP can be shown as
\begin{align} \label{TS}
	\bm{x}_{n} = \sum_{g \in \mathcal{G}} \sqrt{q_{n,g}}\bm{w}_{n,g}s_{{\rm m},g} + \sum_{k \in \mathcal{K}}\sqrt{p_{n,k}}\bm{v}_{n,k}s_{{\rm u},k},
\end{align}
where $q_{n,g}$ and $\bm{w}_{n,g}$ denote the transmit power and beamforming vector for the multicast service to the $g^{\rm th}$ group, respectively, while $p_{n,k}$ and $\bm{v}_{n,k}$ stand for the transmit power and beamforming vector for the unicast service to the $k^{\rm th}$ UE with $k \in \mathcal{K} \triangleq \{1, 2, \cdots, K\}$. In this paper, it is assumed that the APs use conjugate beamforming, that is,
\begin{equation} \label{PM_1}
	\bm{w}_{n, g} = \frac{\hat{\bm{f}}_{n, g}}{\sqrt{\mathbb{E} [\|\hat{\bm{f}}_{n, g}\|^{2}]}}, \quad
	\bm{v}_{n, k} = \frac{\hat{\bm{g}}_{n, k}}{\sqrt{\mathbb{E} [\|\hat{\bm{g}}_{n, k}\|^2]}},
\end{equation}
where $\hat{\bm{f}}_{n,g}$ is an estimate of the linear combination of all UE channels within the $g^{\rm th}$ group, denoted $\bm{f}_{n,g} =\sum_{k\in\mathcal{K}_{g}}\bm{g}_{n,k}$, while $\hat{\bm{g}}_{n,k}$ is an estimate of ${\bm{g}}_{n,k}$. For more details on how to obtain the channel estimates $\hat{\bm{f}}_{n,g}$ and $\hat{\bm{g}}_{n,k}$ in real-world applications, please refer to Appendix~\ref{Appendix-A}.

In light of \eqref{TS}, the received signal at the $k^{\rm th}$ UE can be expressed as
\begin{equation} \label{yrk}
	y_{k} = \sum_{n \in \mathcal{N}}\bm{g}_{n,k}^{\rm H}\bm{x}_{n}+n_{k},
\end{equation}
where the superscript $(\cdot)^{\rm H}$ denotes Hermitian transpose of a vector or matrix, $\mathcal{N} \triangleq \{1, 2, \cdots, N\}$, and $n_{k} \sim \mathcal{CN} \left(0, \delta_{k}^{2}\right)$ denotes the additive white Gaussian noise (AWGN) at the $k^{\rm th}$ UE. Next, the received signal $y_{k}$ is split into two parts: one for information decoder and the other for energy harvester, as each discussed below.

\subsection{The Achievable Data Rates}
At the information decoder, the signal is given by $y_{k}^{\rm I} =\sqrt{\rho_{k}}y_{k}+z_{k}$,
where $\rho_{k}\in[0,1]$ is the power splitting factor, and $z_{k} \sim \mathcal{CN}\left(0, \sigma_{k}^{2}\right)$ is the thermal noise introduced by the power splitter. In this paper, it is assumed that each UE has only channel statistics rather than instantaneous channel state information (CSI). Then, substituting \eqref{TS} into \eqref{yrk}, the received signal at the $k^{\rm th}$ UE for multicast message $s_{{\rm m},g_{k}}$ can be rewritten as  
\begin{align}
y_{{\rm m},k}^{\rm I}
	&= \underbrace{\sqrt{\rho_{k}}\mathbb{E}\left[\sum_{n \in \mathcal{N}} \sqrt{q_{n,g_{k}}} \bm{g}_{n,k}^{\rm H}\bm{w}_{n,g_{k}}\right] s_{{\rm m},g_{k}}}_{\text{desired signal}}+\underbrace{\sqrt{\rho_{k}} n_{k}+z_{k}}_{\text{total noise}}\nonumber\\
	&\quad {}+\underbrace{\sqrt{\rho_{k}}\sum_{n \in \mathcal{N}}\left(\bm{g}_{n,k}^{\rm H}\bm{w}_{n,g_{k}}-\mathbb{E}\left[\bm{g}_{n,k}^{\rm H}\bm{w}_{n,g_{k}}\right]\right)s_{{\rm m},g_{k}}}_{\text{ beamforming uncertainty}}\nonumber  \\
	&\quad {}+\underbrace{\sqrt{\rho_{k}} \sum_{n \in \mathcal{N}} \bm{g}_{n,k}^{\rm H}\sum_{g \in \mathcal{G} \setminus g_{k}} \sqrt{q_{n,g}} \bm{w}_{n,g}s_{{\rm m},g}}_{\text{interference from other multicast groups}} \nonumber\\
	&\quad {}+\underbrace{\sqrt{\rho_{k}}\sum_{n\in \mathcal{N}} \bm{g}_{n,k}^{\rm H} \sum_{j \in \mathcal{K}} \sqrt{p_{n,j}} \bm{v}_{n,j} s_{{\rm u},j}}_{\text{interference from unicast services}}, \label{Mkk}
\end{align}
where $g_{k} \in \mathcal{G}$ denotes the index of the group to which the $k^{\rm th}$ UE belongs. The beamforming uncertainty term in \eqref{Mkk} indicates the gap between the instantaneous channel gain and the statistical mean. When the number of APs (i.e., $N$) turns large, the instantaneous channel gain fluctuates only slightly around its mean due to the channel hardening effect; thus, the value of the beamforming uncertainty is very small in real-world massive MIMO systems. As a result, signal detection at each UE can use only the statistical CSI rather than the instantaneous one \cite{Marzetta2016}.

By using the capacity bound in \cite{5898372},  the achievable data rate in the unit of nat/s  for the multicast message of the $g_{k}^{\rm th}$ group is given by
\begin{subequations}
	\begin{equation}\label{Rate_g}
		R_{{\rm m},g_{k}} =\min_{k\in\mathcal{K}_{g_{k}}}\left\{R_{k}\right\},
	\end{equation}
	where
	\begin{equation}\label{rate_m}
		R_{k}  =\ln\left(1+\frac{|{\rm D}_{{\rm m},g_{k}}|^{2}}{\mathbb{E}\left[|{\rm I}_{{\rm u}}|^{2}+|{\rm I}_{{\rm m},\tilde{g}_{k}}|^{2}+|{\rm V}_{{\rm m},g_{k}}|^{2}\right]+\delta_{k}^{2}+\frac{\sigma_{k}^{2}}{\rho_{k}}}\right),
	\end{equation}
	with
	\begin{align}
		{\rm D}_{{\rm m},g_{k}}&\triangleq \mathbb{E}\left[\sum_{n\in \mathcal{N}}\sqrt{q_{n,g_{k}}} \bm{g}_{n,k}^{\rm H} \bm{w}_{n,g_{k}}\right],\label{DSg}\\
		{\rm V}_{{\rm m},g_{k}}&\triangleq \sum_{n \in \mathcal{N}} \sqrt{q_{n,g_{k}}}\left(\bm{g}_{n,k}^{\rm H} \bm{w}_{n,g_{k}}-\mathbb{E} \left[\bm{g}_{n,k}^{\rm H} \bm{w}_{n,g_{k}}\right]\right),\label{Vgk}\\
		{\rm I}_{{\rm m},\tilde{g}_{k}} &\triangleq \sum_{g \in \mathcal{G}\setminus g_{k}}\sum_{n \in \mathcal{N}}\sqrt{q_{n,g}}\bm{g}_{n,k}^{\rm H}\bm{w}_{n,g},\label{Igk}\\
		{\rm I}_{\rm u} &\triangleq\sum_{j \in \mathcal{K}}\sum_{n \in \mathcal{N}}\sqrt{p_{n,j}} \bm{g}_{n,k}^{\rm H} \bm{v}_{n,j}.\label{Iu}
	\end{align}
\end{subequations}

For unicast messages, on the other hand, by recalling the principle of LDM \cite{7378924}, each UE first decodes its multicast message by treating the unicast signal as noise, and then encodes its unicast message after canceling the multicast signal. Consequently, the received signal for unicast message of  the $k^{\rm th}$ UE, can be written as
\begin{equation}\label{ykk}
y_{{\rm u},k}^{\rm I}=y_{k}^{\rm I}-\mathbb{E}\left[\sum_{n \in \mathcal{N}}\sqrt{q_{n,g_{k}}} \bm{g}_{n,k}^{\rm H} \bm{w}_{n,g_{k}}\right]s_{{\rm m},g_{k}}.
\end{equation}
Like \eqref{rate_m}, the achievable data rate for the unicast message of the $k^{\rm th}$ UE is given by
\begin{subequations}
	\begin{equation}\label{rate_u}
		R_{{\rm u},k}  =\ln\left(1+\frac{|{\rm D}_{{\rm u},k}|^2}{\mathbb{E}\left[{\rm VI}_{{\rm u},k}\right]+\delta_{k}^{2}+\frac{\sigma_{k}^{2}}{\rho_{k}}}\right),
	\end{equation}
	where ${\rm VI}_{{\rm u},k}=|{\rm V}_{{\rm u},k}|^2+|{\rm I}_{{\rm u},\tilde{k}}|^{2}+|{\rm I}_{{\rm m},\tilde{g}_{k}}|^{2}+|{\rm V}_{{\rm m},g_{k}}|^{2}$, with ${\rm V}_{{\rm m},g_{k}}$  and ${\rm I}_{{\rm m},\tilde{g}_{k}}$  are given by the preceding \eqref{Vgk} and \eqref{Igk}, respectively, and
	\begin{align}
		{\rm D}_{{\rm u},k} &\triangleq \mathbb{E}\left[\sum_{n\in\mathcal{N}} \sqrt{p_{n,k}}
		\bm{g}_{n,k}^{\rm H}\bm{v}_{n,k}\right],\label{DSk}\\
		{\rm V}_{{\rm u},k} &\triangleq \sum_{n \in \mathcal{N}} \sqrt{p_{n,k}} \left(\bm{g}_{n,k}^{\rm H}
		\bm{v}_{n,k}-\mathbb{E}\left[\bm{g}_{n,k}^{\rm H}\bm{v}_{n,k}\right]\right),\label{Vk}\\
		{\rm I}_{{\rm u},\tilde{k}} &\triangleq \sum_{j \in \mathcal{K} \setminus k}
		\sum_{n \in \mathcal{N}} \sqrt{p_{n,j}} \bm{g}_{n,k}^{\rm H}\bm{v}_{n,j}. \label{Iuk}
	\end{align}
\end{subequations}

\subsection{The Harvested Energy}
At the energy harvester of  the $k^{\rm th}$ UE, the received RF power can be expressed as
\begin{align}
	p_{{\rm in},k}
	&= \left(1-\rho_{k}\right) \mathbb{E}\left[|y_{k}|^2\right]\nonumber\\
	&= \left(1-\rho_{k}\right)\Bigg(\sum_{j \in \mathcal{K}}\mathbb{E} \left[\left|{\sum_{n \in \mathcal{N}}}\sqrt{p_{n, j}}\bm{g}_{n, k}^{\rm H}\bm{v}_{n, j}\right|^{2}\right]\nonumber\\
	&\quad{}+\sum_{g \in \mathcal{G}} \mathbb{E}\left[\left|{\sum_{n \in \mathcal{N}}}\sqrt{q_{n, g}}\bm{g}_{n, k}^{\rm H}\bm{w}_{n, g}\right|^{2}\right]+ \delta_{k}^{2}\Bigg), \label{EH_k}
\end{align}
where \eqref{yrk} is used to derive \eqref{EH_k}. Accordingly, the harvested energy at the $k^{\rm th}$ UE is denoted by $\mathcal{E}_{k} = \mathcal{F}\left(p_{{\rm in}, k}\right)$, where $\mathcal{F}$ is a function representing the energy conversion process. In this paper, we adopt a practical non-linear energy-harvesting model \cite{8322450}: 
\begin{equation}
	\!\mathcal{F}\left(p_{{\rm in},k}\right) = \left[\frac{P_{\max}}{\eth_{1}}\left(\frac{1+\eth_{1}}{1+\exp(-\iota_{1} p_{{\rm in},k}+\iota_{2})}-1\right)\right]^{+}\!,
\end{equation}	
where $[x]^{+} \triangleq \max\{0, x\}$, $\eth_{1} = \exp\left(-\iota_{1} P_{0}+\iota_{2}\right)$, and the parameter $P_{0}$ denotes the harvester's sensitivity threshold while $P_{\max}$ refers to the maximum harvested power when the energy harvesting circuit is saturated. Also, the parameters $\iota_{1}$ and $\iota_{2}$ are used to capture the nonlinear dynamics of energy harvesting circuits.

\section{Problem Formulation}
\label{Sec_pro_f}
In this section, analytical expressions for the achievable data rates of either unicast or multicast service and the received RF power at each UE are first derived. Then, the total power consumption of the system is quantified. Finally, an EE maximization problem is formulated.

\subsection{Achievable Data Rates and Received RF Power}
Based on \eqref{rate_m} and \eqref{rate_u}, the analytical expressions of the achievable data rates $R_{{\rm m},g}$ and $R_{{\rm u},k}$ can be derived, as formalized in the following theorem.

\begin{theorem}
	\label{Theorem-1}
	With conjugate beamforming, the achievable data rate for unicast service of the $k^{\rm th}$ UE can be explicitly expressed as
	\begin{subequations}
		\begin{equation} \label{RA_U}
			R_{{\rm u}, k}(\mathcal{V})
			= \ln\left(1+\frac{M|\hat{\bm{\xi}}_{k}^{\rm H}\bar{\bm{p}}_{k}|^{2}}{\varphi_{{\rm u},k}(\mathcal{V})}\right),
		\end{equation}
		where $\hat{\bm{\xi}}_{k} \triangleq \left[\hat{\beta}_{1,k}^{\frac{1}{2}},\cdots,\hat{\beta}_{N,k}^{\frac{1}{2}}\right]^{\rm H}$, and
		\begin{align}
			\varphi_{{\rm u},k}(\mathcal{V})
				&\triangleq \sum_{j\in \mathcal{K}} \|\bm{\Xi}_{k}\bar{\bm{p}}_{j}\|^{2}+M\sum_{j\in \mathcal{K}_{g_{k}}\setminus k}		
					\|\hat{\bm{\Xi}}_{k}\bar{\bm{p}}_{j}\|^{2}\nonumber\\
				&\quad{}+\sum_{g\in \mathcal{G}} \|\bm{\Xi}_{k}\bar{\bm{q}}_{g}\|^{2}+\delta_{k}^{2}+\frac{\sigma_{k}^{2}}{\rho_{k}},
		\end{align}
		with
		$\bm{\Xi}_{k} \triangleq {\rm diag}\left\{ \beta_{1, k}^{\frac{1}{2}},\cdots, \beta_{N, k}^{\frac{1}{2}}\right\}$,
		$\hat{\bm{\Xi}}_{k} \triangleq {\rm diag}\left\{\hat{\beta}_{1,k}^{\frac{1}{2}}, \cdots,\hat{\beta}_{N,k}^{\frac{1}{2}}\right\}$,
		$\bar{\bm{p}}_{j} \triangleq \left[p_{1,j}^{\frac{1}{2}},\cdots,p_{N,j}^{\frac{1}{2}}\right]^{\rm H}$, and
		$\bar{\bm{q}}_{g} \triangleq \left[q_{1,g}^{\frac{1}{2}},\cdots,q_{N,g}^{\frac{1}{2}}\right]^{\rm H}$.
		
		On the other hand, the achievable data rate for multicast service of the $g^{\rm th}$ group is given by
		\begin{equation} \label{RA_M}
			R_{{\rm m}, {g}}(\mathcal{V})= \min_{k\in\mathcal{K}_{g}} \left\{ R_{k}(\mathcal{V}) \right\},
		\end{equation}
		where 
		\begin{equation}\label{RA_MM}
			R_{k}(\mathcal{V})=\ln \left( 1+\frac{M|\hat{\bm{\xi}}_{k}^{\rm H}\bar{\bm{q}}_{g_{k}}|^{2}}{\varphi_{{\rm m},g_{k}}(\mathcal{V})} \right),
		\end{equation}
		and $\mathcal{V} \triangleq \left\{ \left\{ \bar{\bm{q}}_{g}\right\}_{g\in\mathcal{G}}, \left\{\bar{\bm{p}}_{k},\ \rho_{k}\right\} _{k\in\mathcal{K}} \right\}$ denotes the set of all transmit power and power splitting factors, and
		\begin{align}
			\varphi_{{\rm m}, g_{k}}(\mathcal{V})&\triangleq 
			\sum_{g\in \mathcal{G}} \|\bm{\Xi}_{k}\bar{\bm{q}}_{g}\|^{2} +\sum_{j\in\mathcal{K}}\|\bm{\Xi}_{k}\bar{\bm{p}}_{j}\|^{2}\nonumber\\
			&\quad{}+M\sum_{j\in\mathcal{K}_{g_{k}}}\|\hat{\bm{\Xi}}_{k}\bar{\bm{p}}_{j}\|^{2}+\delta_{k}^{2}+\frac{\sigma_{k}^{2}}{\rho_{k}}.
		\end{align}
	\end{subequations}
\end{theorem}

\begin{IEEEproof}
	See Appendix~\ref{Appendix-B}.
\end{IEEEproof}

Likewise, in the following theorem we formalize an analytical expression for the received RF power $p_{{\rm in}, k}$ defined by~\eqref{EH_k}.
\begin{subequations}
	\begin{theorem}
		With conjugate beamforming, the received RF power at the $k^{\rm th}$ UE is given by
		\begin{equation} \label{EH_uuk}
			p_{{\rm in},k}(\mathcal{V}) = \left(1-\rho_{k}\right)\varphi_{{\rm e},k}(\mathcal{V}),
		\end{equation}
		where
		\begin{align}
			\varphi_{{\rm e},k}(\mathcal{V}) &\triangleq \sum_{g\in \mathcal{G}} \|\bm{\Xi}_{k}\bar{\bm{q}}_{g}\|^{2} +M\|\hat{\bm{\Xi}}_{k}\bar{\bm{q}}_{g_{k}}\|^{2}+\sum_{j\in \mathcal{K}} \|\bm{\Xi}_{k}\bar{\bm{p}}_{j}\|^{2}\nonumber\\
			&\quad{}+M\sum_{j\in \mathcal{K}_{g_{k}}} \|\hat{\bm{\Xi}}_{k}\bar{\bm{p}}_{j}\|^{2}+\delta_{k}^{2}.
		\end{align}
	\end{theorem}
\end{subequations}

\begin{IEEEproof}
	The proof follows a similar procedure as that of Theorem~\ref{Theorem-1} and thus is omitted.
\end{IEEEproof}

\subsection{System Power Consumption}
A practical power consumption model for a cell-free massive MIMO system consists of two parts: power consumption of APs and power consumption of backhaul links \cite{PC_mod, EE_C-RAN}. In particular, the power consumption of the $n^{\rm th}$ AP can be modeled as 
\begin{subequations}
\begin{align} \label{PC_AP}
		P_{{\rm AP},n}(\mathcal{V})
		&\!=\!\begin{cases}
		\begin{array}{rl}
			\frac{1}{\xi_{n}}p_{{\rm tr}, n}+p_{n}^{\rm ac}, & \text{ if } p_{{\rm tr}, n} > 0 \\
			p_{n}^{\rm sl}, & \text{ if } p_{{\rm tr}, n} = 0
		\end{array},
		\end{cases}
	\end{align}
	where $p_{{\rm tr},n} = \mathbb{E}\left[\|\bm{x}_{n}\|^{2}\right] = \sum_{k\in\mathcal{K}}p_{n,k} + \sum_{g\in\mathcal{G}}q_{n,g}$, which can be rewritten as
	\begin{equation}
		p_{{\rm tr}, n}(\mathcal{V})= \sum_{k\in\mathcal{K}}\bar{\bm{p}}_{k}^{\rm H}\bm{E}_{n}\bar{\bm{p}}_{k}
		+ \sum_{g\in\mathcal{G}}\bar{\bm{q}}_{g}^{\rm H}\bm{E}_{n}\bar{\bm{q}}_{g},
	\end{equation}
\end{subequations}
with $\bm{E}_{n} \in \mathbb{R}^{N \times N}$ having zero entries except $\left[\bm{E}_{n}\right]_{n, n}=1$. Moreover, in \eqref{PC_AP}, $\xi_n$ indicates the efficiency of power amplifier at the $n^{\rm th}$ AP, and $p_{n}^{\rm ac}$ and $p_{n}^{\rm sl}$ stand for the circuit power consumption when the $n^{\rm th}$ AP is in active and sleep modes, respectively. In general, since $p_{n}^{\rm ac}$ is much larger than $p_{n}^{\rm sl}$, it motivates us strategically to switch off the APs to save power in case of very light traffic.

On the other hand, since the backhaul links are used to transfer data between the APs and the CPU (cf. Fig.~\ref{Fig_1}), their total power consumption is proportional to the aggregate data rate transmitted over each backhaul link. To be specific, for the backhaul link connecting the $n^{\rm th}$ AP and the CPU, its power consumption can be calculated as \cite{CF_MIMO_4}
\begin{subequations}
	\begin{equation}\label{PC_BH}
		P_{{\rm BH}, n}(\mathcal{V}) = p_{{\rm bh}, n}C_{{\rm bh}, n}(\mathcal{V}) + p_{0, n},
	\end{equation}
	where $p_{{\rm bh}, n}$ is the traffic-dependent power consumption while $p_{0, n}$ is a fixed power consumption of each backhaul link (traffic-independent power), and $C_{{\rm bh}, n}(\mathcal{V})$ is the aggregate data rate transmitted over the backhaul link between the $n^{\rm th}$ AP and the CPU, given by
\begin{align} \label{C_fh}
		C_{{\rm bh}, n}(\mathcal{V})&= \sum_{g \in \mathcal{G}}\mathbb{I}\left(\bar{\bm{q}}_{g}^{\rm H}
		\bm{E}_{n}\bar{\bm{q}}_{g}\right)R_{{\rm m}, g}(\mathcal{V})\nonumber\\
		&\quad{}+ \sum_{k \in \mathcal{K}} \mathbb{I}\left(\bar{\bm{p}}_{k}^{\rm H}
		\bm{E}_{n}\bar{\bm{p}}_{k}\right)R_{{\rm u}, k}(\mathcal{V}),
	\end{align}
	where $\mathbb{I}(\cdot)$ is the indicator function.
\end{subequations}

As a result, combining \eqref{PC_AP} and \eqref{PC_BH} gives the total power consumption of the system:
\begin{align}
	P_{\rm tot}(\mathcal{V}) &= \sum_{n \in \mathcal{N}} \left(p_{n}^{{\rm sl}}+\xi_{n}^{-1}p_{{\rm tr},n}(\mathcal{V}) 
	+ \mathbb{I}\left(p_{{\rm tr}, n}(\mathcal{V})\right)\triangle p_{n}\right) \nonumber \\
	&\quad {}+ \sum_{n \in \mathcal{N}} \Bigg[p_{{\rm bh}, n}\bigg(\sum_{g \in \mathcal{G}}
	\mathbb{I}\left(\bar{\bm{q}}_{g}^{\rm H} \bm{E}_{n}\bar{\bm{q}}_{g}\right) R_{{\rm m},g}(\mathcal{V})\nonumber \\
	&\quad{}+ \sum_{k \in \mathcal{K}} \mathbb{I}\left(\bar{\bm{p}}_{k}^{\rm H} \bm{E}_{n}\bar{\bm{p}}_{k}\right)
	R_{{\rm u}, k}(\mathcal{V})\bigg)+ p_{0, n}\Bigg], \label{P_tt1}
\end{align}
in which $\triangle p_{n} \triangleq  p_{n}^{\rm ac} - p_{n}^{\rm sl}$ is a constant.

\subsection{Problem Formulation}
Now, we are in a position to maximize the EE of the system by dynamically allocating transmit power and power splitting factors. By definition of EE and in light of \eqref{RA_U}, \eqref{RA_M} and \eqref{P_tt1}, the  problem can be formulated as
\begin{subequations} \label{P0}
	\begin{align}
	\max_{\mathcal{V}} \ & \frac{\left(1-\frac{\tau_{\rm p}}{\tau_{\rm c}}\right)\left(\sum\limits_{g \in \mathcal{G}}
		R_{{\rm m},g}(\mathcal{V})+\sum\limits_{k\in\mathcal{K}}R_{{\rm u},k}(\mathcal{V})\right)} {P_{\rm tot}(\mathcal{V})}, \label{Obj} \\
	{\rm s.t.} \  & \ R_{{\rm m},g}(\mathcal{V})\geq \bar{r}_{{\rm m},g}, \ \forall g \in \mathcal{G}, \label{C1} \\
	& \ R_{{\rm u},k}(\mathcal{V})\geq \bar{r}_{{\rm u},k}, \ \forall k \in \mathcal{K}, \label{C2} \\
	& \ \mathcal{E}_{k}(\mathcal{V})\geq \bar{e}_{k}, \ \forall k \in \mathcal{K}, \label{C3} \\
	& \ C_{{\rm bh},n}(\mathcal{V})\leq \bar{c}_{n,\max}, \ \forall n \in \mathcal{N}, \label{C4} \\
	& \ p_{{\rm tr},n}(\mathcal{V})\leq \bar{p}_{n,\max}, \ \forall n \in \mathcal{N}, \label{C5}
	\end{align}
\end{subequations}
where the factor $\left(1-\tau_{\rm p}/\tau_{\rm c}\right)$ in \eqref{Obj} comes from the fact that during each coherence interval of $\tau_{\rm c}$, $\tau_{\rm p}$ symbol intervals are spent on the uplink channel estimation; $\bar{r}_{\rm{m}, g}$ in \eqref{C1} and $\bar{r}_{{\rm u}, k}$ in \eqref{C2} are the minimal data rates required for multicast service of the $g^{\rm th}$ group and unicast service of the $k^{\rm th}$ UE, respectively; the symbol $\bar{e}_{k}$ in \eqref{C3} is the minimal energy required for the $k^{\rm th}$ UE; $\bar{c}_{n, \max}$ in \eqref{C4} is the maximal capacity of the backhaul link between the CPU and the $n^{\rm th}$ AP, and $\bar{p}_{n, \max}$ in \eqref{C5} is the allowable maximal transmit power at the $n^{\rm th}$ AP.

It is not hard to see that the optimization problem given by \eqref{P0} is neither smooth nor convex, and even finding a feasible point satisfying the nonconvex constraints \eqref{C1}-\eqref{C4} is NP-hard. Moreover, for a massive access setting, the total number of APs and UEs is very large and thus the joint optimization of transmit power and power splitting factors would induce prohibitively high complexity. To tackle these challenges, in the next section a low-complexity first-order algorithm and its accelerated algorithm are developed, provided that a feasible initial point is available. Afterwards, another first-order algorithm is designed to find a feasible initial point.

\section{Energy-efficient Resource Allocation}
\label{Sec_EE_RA}
In this section, the original problem \eqref{P0} is first transformed into a smooth one by introducing auxiliary variables and approximating the indicator functions. Then, the resulting smooth and nonconvex problem is approximated by a sequence of quasi-convex problems by using the SCA framework. Finally, a first-order algorithm and its accelerated algorithm are developed to solve each SCA subproblem, instead of traditional second-order algorithms with high complexity.

\subsection{Resolving Nonsmoothness of Objective \eqref{Obj} and Constraints \eqref{C1} and \eqref{C4}}
We begin with settling the nonsmoothness of $R_{\rm{m}, g}(\mathcal{V})$ in \eqref{Obj}-\eqref{C1}. By introducing a set of auxiliary variables $\mathcal{R}\triangleq \left\{ \mathcal{R}_{{\rm m}, g} \right\} _{g \in \mathcal{G}}$, such that
\begin{equation} \label{R_gk}
	\mathcal{R}_{{\rm m}, g}\leq R_{k}(\mathcal{V}),\ \forall k \in \mathcal{K}_{g},
\end{equation}
the nonsmooth multicast rate function $R_{\rm{m},g}(\mathcal{V})$ can be replaced by $\mathcal{R}_{{\rm m},g}$.

Then, we deal with the nonsmoothness of the indicator function $\mathbb{I}\left(x\right)$ in \eqref{Obj} and \eqref{C4}. Similar to \cite{8579566}, the indicator can be approximated by the smooth function $f_{\theta}\left(x\right) = x/(x+\theta)$, where $\theta$ is a small constant that controls the smoothness of approximation.\footnote{In this paper, we set $\theta=10^{-5}$, which can achieve an attractive trade-off between smoothness and approximation accuracy \cite{8579566}. The effect of the smooth approximation on the data rates of multicast and unicast serves can be negligible. This is mainly because, in the practical communication systems, there is a circuit sensitivity threshold and the transceiver cannot be activated when the transmit power is very small.} By approximating $\mathbb{I}\left(x\right)$ with $f_{\theta}\left(x\right)$, the constraint shown in \eqref{C4} is approximated as
\begin{align} \label{C_fh22}
	C_{{\rm bh}, n}(\mathcal{V})
	&\approx \sum_{g \in \mathcal{G}}f_{\theta}\left(\bar{\bm{q}}_{g}^{\rm H}\bm{E}_{n}\bar{\bm{q}}_{g}\right)\mathcal{R}_{{\rm m}, g}\nonumber\\
	&\quad{}+ \sum_{k \in \mathcal{K}}f_{\theta}\left(\bar{\bm{p}}_{k}^{\rm H}\bm{E}_{n}\bar{\bm{p}}_{k}\right)R_{{\rm u}, k}\left(\mathcal{V}\right)
	\leq \bar{c}_{n, \max},
\end{align}
Accordingly, $P_{\rm tot}(\mathcal{V})$ in \eqref{P_tt1} can be approximated as
\begin{align}
	P_{{\rm tot}}(\mathcal{V})
	& \approx \sum_{n \in \mathcal{N}}\left(p_{n}^{\rm sl}+\xi_{n}^{-1}p_{{\rm tr},n}(\mathcal{V}) + f_{\theta}\left(p_{{\rm tr}, n}(\mathcal{V})\right)\triangle{p_{n}}\right) \nonumber \\
	& \quad{}+ \sum_{n \in \mathcal{N}}\Bigg[p_{{\rm bh},n} \bigg(\sum_{g \in \mathcal{G}}
	f_{\theta}\left(\bar{\bm{q}}_{g}^{\rm H}\bm{E}_{n}\bar{\bm{q}}_{g}\right)\mathcal{R}_{{\rm m}, g}\nonumber\\
	&\quad{}+ \sum_{k\in \mathcal{K}}f_{\theta}\left(\bar{\bm{p}}_{k}^{\rm H}\bm{E}_{n}\bar{\bm{p}}_{k}\right)R_{{\rm u},k}\left(\mathcal{V}\right)\bigg) + p_{0,n}\Bigg]. \label{P_tt2}
\end{align}

With \eqref{R_gk}-\eqref{P_tt2}, the optimization problem shown in \eqref{P0} is transformed into a smooth one. However, the constraint \eqref{C_fh22} is nonconvex because there are couples between the rate $\mathcal{R}_{{\rm m}, g}$ (resp., $R_{{\rm u}, k}(\mathcal{V})$) and $f_{\theta}\left(\bar{\bm{q}}_{g}^{\rm H}\bm{E}_{n}\bar{\bm{q}}_{g}\right)$ (resp., $f_{\theta}\left(\bar{\bm{p}}_{ k}^{\rm H}\bm{E}_{n}\bar{\bm{p}}_{k}\right)$). Moreover, since the functions $R_{{\rm u}, k}(\mathcal{V})$, $R_{k}(\mathcal{V})$  and $\mathcal{E}_{k}(\mathcal{V})$ are nonconcave, the constraints \eqref{C2}, \eqref{C3} and \eqref{R_gk} are also nonconvex, making the problem \eqref{P0} extremely hard to solve. To address the nonconvexity, in the next subsection we use the SCA technique and construct a sequence of convex constraints to approximate \eqref{C2}, \eqref{C3}, \eqref{R_gk} and \eqref{C_fh22}.

\subsection{Tackling Nonconvex Constraints \eqref{C2}, \eqref{C3}, \eqref{R_gk} and \eqref{C_fh22} }
We start with the constraint \eqref{C_fh22}. In particular, we first replace $\mathcal{R}_{{\rm m},g}$ (resp., $R_{{\rm u},k}(\mathcal{V})$) with $\mathcal{R}_{{\rm m}, g}^{(t)}$ (resp., $R_{{\rm u},k}(\mathcal{V}^{(t)})$) obtained from the previous iteration \cite{l0_l1, 7942111}. Accordingly, \eqref{C_fh22} can be simplified as
\begin{align} \label{C5_12}
	C_{{\rm bh}, n}^{(t)}(\mathcal{V})
	&= \sum_{g \in \mathcal{G}}f_{\theta}\left(\bar{\bm{q}}_{g}^{\rm H}\bm{E}_{n}\bar{\bm{q}}_{g}\right)\mathcal{R}_{{\rm m}, g}^{(t)}\nonumber\\
	&\quad{}+ \sum_{k \in \mathcal{K}}f_{\theta}\left(\bar{\bm{p}}_{k}^{\rm H}\bm{E}_{n}\bar{\bm{p}}_{k}\right)R_{{\rm u}, k}(\mathcal{V}^{(t)})
	\leq \bar{c}_{n, \max}.
\end{align}
Then, by exploiting the concavity of $f_{\theta}\left(x\right)$, we have
\begin{subequations}
	\begin{align}
		\hspace{-1em} f_{\theta}(\bar{\bm{q}}_{g}^{\rm H}\bm{E}_{n}\bar{\bm{q}}_{g})
			& \leq f_{\theta}(\bar{\bm{q}}_{g}^{(t){\rm H}}\bm{E}_{n}\bar{\bm{q}}_{g}^{(t)}) + \bm{\vartheta}_{{\rm m}, n, g}^{(t)}(\bar{\bm{q}}_{g} - \bar{\bm{q}}_{g}^{(t)}) , \label{sca_1a} \\
		\hspace{-1em} f_{\theta}\left(\bar{\bm{p}}_{k}^{\rm H}\bm{E}_{n}\bar{\bm{p}}_{k}\right)
			& \leq f_{\theta}(\bar{\bm{p}}_{k}^{(t){\rm H}}\bm{E}_{n}\bar{\bm{p}}_{k}^{(t)}) + \bm{\vartheta}_{{\rm u}, n, k}^{(t)}(\bar{\bm{p}}_{k}
	-\bar{\bm{p}}_{k}^{(t)}), \label{sca_1b}
	\end{align}
\end{subequations}
where $\bm{\vartheta}_{{\rm m}, n, g}^{(t)} \triangleq {\partial f_{\theta}\left(\bar{\bm{q}}_{g}^{\rm H}\bm{E}_{n}\bar{\bm{q}}_{g}\right)}/{\partial\bar{\bm{q}}_{g}}\left|_{\bar{\bm{q}}_{g}^{(t)}}\right.$ 
and
$\bm{\vartheta}_{{\rm u},n,k}^{(t)} \triangleq {\partial f_{\theta}\left(\bar{\bm{p}}_{k}^{\rm H}\bm{E}_{n}\bar{\bm{p}}_{k}\right)}/{\partial\bar{\bm{p}}_{k}}\left|_{\bar{\bm{p}}_{k}^{(t)}}\right.$.
Substituting \eqref{sca_1a} and \eqref{sca_1b} into \eqref{C5_12} yields
\begin{align} \label{fh_tt}
	C_{{\rm bh}, n}^{(t)}(\mathcal{V})
	&\!\leq\! \sum_{g \in \mathcal{G}}\left(\bm{\vartheta}_{{\rm m},n,g}^{(t)}\bar{\bm{q}}_{g} + \mathcal{H}_{{\rm m},n,g}^{(t)}\right)\mathcal{R}_{{\rm m},g}^{(t)}+\sum_{k \in \mathcal{K}}\!\!\big(\bm{\vartheta}_{{\rm u}, n, k}^{(t)}\bar{\bm{p}}_{k}\nonumber\\
	&\quad{}+\mathcal{H}_{{\rm u}, n, k}^{(t)}\big)R_{{\rm u}, j}(\mathcal{V}^{(t)})\triangleq \bar{C}_{{\rm bh},b}^{(t)}\leq \bar{c}_{n, \max},
\end{align}
where $\mathcal{H}_{{\rm m},n,g}^{(t)}\triangleq f_{\theta}\left(\bar{\bm{q}}_{g}^{(t){\rm H}}\bm{E}_{n}\bar{\bm{q}}_{g}^{(t)}\right)-\bm{\vartheta}_{{\rm m},n,g}^{(t)}\bar{\bm{q}}_{g}^{(t)}$ and $\mathcal{H}_{{\rm u},n,k}^{(t)}\triangleq f_{\theta}\left(\bar{\bm{p}}_{k}^{(t){\rm H}}
\bm{E}_{n}\bar{\bm{p}}_{k}^{(t)}\right)-\bm{\vartheta}_{{\rm u},n,k}^{(t)}\bar{\bm{p}}_{k}^{(t)}$.
Since $\bar{C}_{{\rm bh}, n}^{(t)}(\mathcal{V})$ is linear over $\mathcal{V}$, the constraint shown in \eqref{fh_tt} is convex.

By using a similar approach as above, $P_{\rm tot}(\mathcal{V})$ in \eqref{P_tt2} can be approximated as
\begin{align}
	\! P_{\rm tot}(\mathcal{V})
		\!&\leq \sum_{n \in \mathcal{N}}\bigg[p_{n}^{\rm sl}+\xi_{n}^{-1}\left(\sum_{k \in \mathcal{K}}
			\bar{\bm{p}}_{k}^{\rm H}\bm{E}_{n}\bar{\bm{p}}_{k}+\sum_{g \in \mathcal{G}}\bar{\bm{q}}_{g}
			^{\rm H}\bm{E}_{n}\bar{\bm{q}}_{g}\right)\nonumber\\
		&\quad{}+\bigg(\sum_{k\in\mathcal{K}}\bm{\varkappa}_{{\rm u},n,k}^{(t)}
			\bar{\bm{p}}_{k}+\sum_{g \in \mathcal{G}}\bm{\varkappa}_{{\rm m},n,g}^{(t)}\bar{\bm{q}}_{g}+\mathcal{H}_{{\rm mu},n}^{(t)}\bigg)\triangle p_{n}\bigg] \nonumber \\
		&\quad {} + \sum_{n\in\mathcal{N}}\!\!\bigg[p_{{\rm 0},n}+p_{{\rm bh},n}\Big(\sum_{g \in \mathcal{G}}(\bm{\vartheta}_{{\rm m},n,g}^{(t)}
			\bar{\bm{q}}_{g}+\mathcal{H}_{{\rm m},n,g}^{(t)})\mathcal{R}_{{\rm m},g}^{(t)} \nonumber \\
		&\quad {}+\sum_{k \in \mathcal{K}}(\bm{\vartheta}_{{\rm u},n,k}^{(t)}\bar{\bm{p}}_{k}+\mathcal{H}_{{\rm u},n,k}^{(t)})R_{{\rm u},k}(\mathcal{V}^{(t)})\Big)\bigg] \nonumber\\
		& \triangleq \bar{P}_{\rm tot}^{(t)}(\mathcal{V}), \label{PP_t}
\end{align}
where $\mathcal{H}_{{\rm mu},n}^{(t)}\triangleq f_{\theta}\left(p_{{\rm tr},n}\left(\mathcal{V}^{(t)}\right)\right)-\sum_{k \in \mathcal{K}}\bm{\varkappa}_{{\rm u},n,k}^{(t)}\bar{\bm{p}}_{k}^{(t)}-\sum_{g \in \mathcal{G}}\bm{\varkappa}_{{\rm m},n,g}^{(t)}\bar{\bm{q}}_{g}^{(t)}$ with
$\bm{\varkappa}_{{\rm m},n,g}^{(t)} = {\partial f_{\theta}\left(p_{{\rm tr},n}(\mathcal{V})\right)}/{\partial\bar{\bm{q}}_{g}}\left|_{\bar{\bm{q}}_{g}^{(t)}}\right.$ and
$\bm{\varkappa}_{{\rm u},n,k}^{(t)} = {\partial f_{\theta}\left(p_{{\rm tr},n}(\mathcal{V})\right)}/{\partial\bar{\bm{p}}_{k}}\left|_{\bar{\bm{p}}_{k}^{(t)}}\right.$. It is not hard to see that $\bar{P}_{\rm tot}^{(t)}(\mathcal{V})$ given by \eqref{PP_t} is concave quadratic over $\mathcal{V}$.

Next, we deal with the nonconvex constraints \eqref{C2} and \eqref{R_gk}. By using the path-following algorithms \cite{path_f_1}, we obtain the following lower bounding concave approximations of $R_{k}(\mathcal{V})$ and $R_{{\rm u}, k}(\mathcal{V})$ at the $t^{\rm th}$ iteration, say, $\mathcal{V}^{(t)}$.

\begin{prop} \label{Prop_1}
	Given a fixed point $\mathcal{V}^{(t)}$, the lower bounding concave approximations for $R_{k}(\mathcal{V})$ and $R_{{\rm u}, k}(\mathcal{V})$ are given by
	\begin{align} \label{rate_mm}
	R_{k}(\mathcal{V})
		&\geq \zeta_{{\rm m}, g_{k}}^{(t)} + 2\bm{\psi}_{{\rm m}, g_{k}}^{(t)}\bar{\bm{q}}_{g_{k}}  \nonumber \\
		&\quad {}- \phi_{{\rm m}, g_{k}}^{(t)}\left(\varphi_{{\rm m}, g_{k}}(\mathcal{V}) + M\left|\hat{\bm{\xi}}_{k}^{\rm H}\bar{\bm{q}}_{g_{k}}\right|^{2}\right) \nonumber\\
		&\triangleq \bar{R}_{k}^{(t)}(\mathcal{V}),
	\end{align}
	with
	$\zeta_{{\rm m}, g_{k}}^{(t)}\!\triangleq\!R_{k}(\mathcal{V}^{(t)})- M|\hat{\bm{\xi}}_{k}^{\rm H}\bar{\bm{q}}_{g_{k}}^{(t)}|^{2}/\varphi_{{\rm m}, g_{k}}(\mathcal{V}^{(t)})$,
	$\phi_{{\rm m}, g_{k}}^{(t)} \triangleq M|\hat{\bm{\xi}}_{k}^{\rm H}\bar{\bm{q}}_{g_{k}}^{(t)}|^{2}/\varphi_{{\rm m}, g_{k}}(\mathcal{V}^{(t)})\left(\varphi_{{\rm m}, g_{k}}(\mathcal{V}^{(t)}) + M|\hat{\bm{\xi}}_{k}^{\rm H}\bar{\bm{q}}_{g_{k}}^{(t)}|^{2}\right)$, and
	$\bm{\psi}_{{\rm m}, g_{k}}^{(t)} \triangleq {M\bar{\bm{q}}_{g_{k}}^{(t){\rm H}}\hat{\bm{\xi}}_{k}\hat{\bm{\xi}}_{k}^{\rm H}}/{\varphi_{{\rm m}, g_{k}}(\mathcal{V}^{(t)})}$, and
	\begin{align} \label{rate_uu}
		R_{{\rm u}, k}(\mathcal{V})
			&\geq \zeta_{{\rm u}, k}^{(t)} + 2\bm{\psi}_{{\rm u}, k}^{(t)}\bar{\bm{p}}_{k} - \phi_{{\rm u}, k}^{(t)}\left(\varphi_{{\rm u}, k}(\mathcal{V})
				+ M|\hat{\bm{\xi}}_{k}^{\rm H}\bar{\bm{p}}_{k}|^{2}\right)\nonumber\\
			&\triangleq \bar{R}_{{\rm u}, k}^{(t)}(\mathcal{V}),
	\end{align}
	with
	$\zeta_{{\rm u}, k}^{(t)} \triangleq R_{{\rm u}, k}(\mathcal{V}^{(t)}) - M|\hat{\bm{\xi}}_{k}^{\rm H}\bar{\bm{p}}_{k}^{(t)}|^{2}/\varphi_{{\rm u}, k}({\mathcal{V}}^{(t)})$,
	$\phi_{{\rm u},k}^{(t)} \triangleq M|\hat{\bm{\xi}}_{k}^{\rm H}\bar{\bm{p}}_{k}^{(t)}|^{2}/\varphi_{{\rm u}, k}(\mathcal{V}^{(t)})\left(\varphi_{{\rm u}, k}(\mathcal{V}^{(t)}) + M|\hat{\bm{\xi}}_{k}^{\rm H}\bar{\bm{p}}_{k}^{(t)}|^{2}\right)$, and
	$\bm{\psi}_{{\rm u}, k}^{(t)} \triangleq M\bar{\bm{p}}_{k}^{(t){\rm H}}\hat{\bm{\xi}}_{k}\hat{\bm{\xi}}_{k}^{\rm H}/\varphi_{{\rm u}, k}(\mathcal{V}^{(t)})$.
\end{prop}

\begin{IEEEproof}
	The proof follows a similar approach as in \cite{path_f_1}, which is omitted here for brevity.
\end{IEEEproof}

With Proposition~\ref{Prop_1}, the nonconvex constraints \eqref{C2} and \eqref{R_gk} can be approximated as
\begin{align}
	\bar{r}_{{\rm u}, k}-\bar{R}_{{\rm u}, k}^{(t)}(\mathcal{V}) & \leq 0, \ \forall k \in \mathcal{K}, \label{C22} \\
	\mathcal{R}_{{\rm m}, g_{k}}-\bar{R}_{k}^{(t)}(\mathcal{V}) & \leq 0, \ \forall k \in \mathcal{K}. \label{C55}
\end{align}
Since $\bar{R}_{k}^{(t)}(\mathcal{V})$ and $\bar{R}_{{\rm u},k}^{(t)}(\mathcal{V})$ given by \eqref{rate_mm} and \eqref{rate_uu}, respectively, are concave quadratic over $\cal V$, the constraints shown in \eqref{C22}-\eqref{C55} are convex.

Finally, we address the nonconvex \eqref{C3}, which can be rewritten as
\begin{align} \label{EHC}
	\frac{\mathcal{F}^{-1}\left(\bar{e}_{k}\right)}{1-\rho_{k}}-\varphi_{{\rm e}, k}(\mathcal{V}) \leq 0, \ \forall n \in \mathcal{N},
\end{align}
where
\begin{align} \label{neh}
	\mathcal{F}^{-1}\left(x\right) &= \begin{cases}
	\begin{array}{ll}
		+\infty, & \text{if } x \geq P_{\max} \\
		\frac{\iota_{2}}{\iota_{1}}-\frac{1}{\iota_{1}}\ln\left(\frac{1+\eth_{1}}{1+\eth_{2}x}-1\right), & \text{if } 0<x<P_{\max} \\
		0, & \text{if } x \leq 0
	\end{array} \end{cases}
\end{align}
denotes the pseudo-inverse of $\mathcal{F}\left(x\right)$, with  $\eth_{2}=P_{\max}^{-1}\exp\left(-\iota_{1} P_{0}+\iota_{2}\right)$  \cite{8322450}. Since $\|\bm{\Xi}_{k}\bar{\bm{q}}_{g}\|^{2}$ is a convex function of $\bar{\bm{q}}_{g}$, by using the first-order Taylor expansion at a given point $\bar{\bm{q}}_{g}^{(t)}$, it can be approximated by \cite{SCA_1}
\begin{equation}\label{fot}
	\|\bm{\Xi}_{k}\bar{\bm{q}}_{g}\|^{2} \geq -\|\bm{\Xi}_{k}\bar{\bm{q}}_{g}^{(t)}\|^{2} + 2\bar{\bm{q}}_{g}^{(t)
	{\rm H}}\bm{\Xi}_{k}^{2}\bar{\bm{q}}_{g}.
\end{equation}
Like \eqref{fot}, $\|\hat{\bm{\Xi}}_{k}\bar{\bm{q}}_{g_{k}}^{(t)}\|^{2}$, $\|\bm{\Xi}_{k}\bar{\bm{p}}_{j}^{(t)}\|^{2}$ and $\|\hat{\bm{\Xi}}_{k}\bar{\bm{p}}_{j}^{(t)}\|^{2}$ can be approximated by their corresponding first-order Taylor expansions. Thus, an approximation of $\varphi_{{\rm e}, k}(\mathcal{V})$ can be given by
\begin{equation} \label{phi_ee1}
	\varphi_{{\rm e}, k}(\mathcal{V}) \geq \bar{\varphi}_{{\rm e}, k}^{(t)}(\mathcal{V}), \ \forall \mathcal{V}
\end{equation}
where
\begin{align} \label{phi_ee}
	\bar{\varphi}_{{\rm e}, k}^{(t)}(\mathcal{V})
	&\triangleq \zeta_{{\rm e}, k}^{(t)} + 2\sum_{g\in\mathcal{G}}\bar{\bm{q}}_{g}^{(t){\rm H}}\bm{\Xi}_{k}^{2}\bar{\bm{q}}_{g}
		+ 2\sum_{j \in \mathcal{K}}\bar{\bm{p}}_{j}^{(t){\rm H}}\bm{\Xi}_{k}^{2}\bar{\bm{p}}_{j}\nonumber\\
	&\quad {}+ 2M\sum_{j \in \mathcal{K}_{g_{k}}}\bar{\bm{p}}_{j}^{(t){\rm H}}\hat{\bm{\Xi}}_{k}^{2}\bar{\bm{p}}_{j}
		+ 2M\bar{\bm{q}}_{g_{k}}^{(t){\rm H}}\hat{\bm{\Xi}}_{k}^{2}\bar{\bm{q}}_{g_{k}},
\end{align}
with $\zeta_{{\rm e}, k}^{(t)} \triangleq -\sum_{g \in \mathcal{G}} \|\bm{\Xi}_{k}\bar{\bm{q}}_{g}^{(t)}\|^{2} -M\|\hat{\bm{\Xi}}_{k}\bar{\bm{q}}_{g_{k}}^{(t)}\|^{2}-\sum_{j \in \mathcal{K}} \|\bm{\Xi}_{k}\bar{\bm{p}}_{j}^{(t)}\|^{2}
-M\sum_{j\in\mathcal{K}_{g_{k}}}\|\hat{\bm{\Xi}}_{k}\bar{\bm{p}}_{j}^{(t)}\|^{2}+\delta_{k}^{2}$. Therefore, with a fixed point $\mathcal{V}^{(t)}$, the nonconvex constraint \eqref{C3} can be approximated by
\begin{equation} \label{C44}
	\frac{\mathcal{F}^{-1}\left(\bar{e}_{k}\right)}{1-\rho_{k}} - \bar{\varphi}_{{\rm e}, k}^{(t)}(\mathcal{V}) \leq 0.
\end{equation}
Since ${\mathcal{F}^{-1}\left(\bar{e}_{k}\right)}/{\left(1-\rho_{k}\right)}$ and $\bar{\varphi}_{{\rm e}, k}^{(t)}(\mathcal{V})$ in \eqref{C44} are  convex  and linear over $\mathcal{V}$, respectively, the constraint \eqref{C44} is convex.

Now, with the convex constraints attained in \eqref{fh_tt}, \eqref{C22}-\eqref{C55} and \eqref{C44}, the problem \eqref{P0} can be iteratively solved in the SCA framework, with the $t^{\rm th}$ SCA subproblem explicitly given by
\begin{subequations} \label{P_t1}
	\begin{align}
	\left\{\mathcal{V}^{(t+1)}, \mathcal{R}^{(t+1)}\right\}
	 = \min_{\mathcal{V}, \mathcal{R}}\ \frac{\bar{P}_{{\rm tot}}^{(t)}(\mathcal{V})}{\left(1-\frac{\tau_{\rm p}}{\tau_{\rm c}}\right)R_{\rm sum}(\cal{V,R})}, \label{objj}
	\end{align}
	\vspace{-1.25em}
	\begin{align}
	\hspace{-2em} {\rm s.t.} \ &\  \bar{r}_{{\rm m},g}-{\mathcal{R}}_{{\rm m},g} \leq 0, \ \forall g\in\mathcal{G}, \label{rgg}\\ 
	& \ \mathcal{R}_{{\rm m},g_{k}}-\bar{R}_{k}^{(t)}(\mathcal{V}) \leq 0, \ \forall k\in\mathcal{K}, \label{rgg1} \\
	& \ \bar{r}_{{\rm u},k}-\bar{R}_{{\rm u},k}^{(t)}(\mathcal{V}) \leq 0, \ \forall k\in\mathcal{K}, \label{ruu} \\
	& \ \frac{\mathcal{F}^{-1}(\bar{e}_{k})}{1-\rho_{k}}-\bar{\varphi}_{{\rm e}, k}^{(t)}(\mathcal{V}) \leq 0, \ \forall k \in \mathcal{K}, \label{egg} \\
	& \ \bar{C}_{{\rm bh}, n}^{(t)}(\mathcal{V})-\bar{c}_{n, \max} \leq 0, \ \forall n \in \mathcal{N}, \label{cgg} \\
	& \ p_{{\rm tr},n}(\mathcal{V})-\bar{p}_{n,\max} \leq 0, \ \forall n\in\mathcal{N}. \label{tgg}
	\end{align}
\end{subequations}
where $R_{\rm sum}({\cal V,R})\triangleq\sum_{g \in \mathcal{G}}\mathcal{R}_{{\rm m}, g} + \sum_{k \in \mathcal{K}}\bar{R}_{{\rm u}, k}^{(t)}(\mathcal{V})$. Clearly, the subproblem \eqref{P_t1} is quasi-convex since the objective function \eqref{objj} is quasi-convex and all the constraints \eqref{rgg}-\eqref{tgg} are convex. To solve this quasi-convex problem, the classic method is to transform it into convex forms via Charnes-Cooper transform or Dinkelbach's transform \cite{zappone2015energy}. Afterwards, the interior-point method can be applied to solve the transformed convex problem. However, the interior-point method results in cubic-order complexity, which is impractical in massive access scenario. To suit massive access applications, a low-complexity first-order algorithm is developed in the next subsection.

\subsection{The Proposed First-Order Algorithm for Solving \eqref{P_t1}}
In general, second-order optimization algorithms depend upon the gradient and curvature information of an objective function and thus have fast convergence but are prone to computation and memory resources. Instead, first-order algorithms involve only gradient information and hence benefit lower computational complexity (albeit slow convergence), making them more suitable for large-scale optimization problems \cite{Beck2017}.

In principle, first-order optimization algorithms for constrained problems consist of two steps: gradient step and gradient projection step. However, since \eqref{P_t1} is imposed by coupling constraints \eqref{rgg}-\eqref{cgg}, the gradient projection onto the constraints \eqref{rgg}-\eqref{cgg} would be highly complicated. Fortunately, by applying the Dinkelbach's transform, the cost function will be strongly convex over $\mathcal{V}$ and linear over $\mathcal{R}$. In particular, applying the Dinkelbach's transform to \eqref{P_t1}, we obtain a sequence of convex problems, with the $c^{\rm th}$ problem expressed as
\begin{subequations}\label{P_tt22}
	\begin{align}
	\min_{\mathcal{V},\mathcal{R}}\  & \bar{P}_{\rm tot}^{(t)}(\mathcal{V})-\eta^{(c)}\left(1-\frac{\tau_{\rm p}}{\tau_{\rm c}}\right) R_{\rm sum}({\cal V,R}),\label{obj_2}\\
	{\rm s.t.\ } & \eqref{rgg}-\eqref{tgg},
	\end{align}
\end{subequations}
where $\eta^{(c)}$ is obtained by substituting the optimal $\left(\mathcal{V},\mathcal{R}\right)$ of the $\left(c-1\right)^{\rm th}$
problem into \eqref{objj}. As $c$ increases, the solution sequence of \eqref{P_tt22} is guaranteed to converge to the global optimal solution of \eqref{P_t1}.

Substituting \eqref{PP_t} and \eqref{rate_uu} into \eqref{P_tt22}, it is not hard to see that \eqref{obj_2} is strongly convex over $\mathcal{V}$ and linear over $\mathcal{R}$. Dropping the constants independent of $\left(\mathcal{V}, \mathcal{R}\right)$, the problem \eqref{P_tt22} can be reformulated as
\begin{subequations}\label{P_ptt22}
	\begin{align}
	\ \min_{\mathcal{V}, \mathcal{R}} \ & \Gamma(\mathcal{V})-\eta^{(c)}\left(1-\frac{\tau_{\rm p}}{\tau_{\rm c}}\right)\sum_{g \in \mathcal{G}} \mathcal{R}_{{\rm m}, g}, \\
	{\rm s.t.} \ & \eqref{rgg}-\eqref{tgg},
	\end{align}
\end{subequations}
where
\begin{align}
	\Gamma(\mathcal{V})
	& \triangleq \sum_{n \in \mathcal{N}}\Bigg[\sum_{k \in \mathcal{K}}\left(\xi_{n}^{-1}\bar{\bm{p}}_{k}^{{\rm H}}\bm{E}_{n}\bar{\bm{p}}_{k}
	+ \Delta p_{n}\bm{\varkappa}_{{\rm u}, n, k}^{(t)}\bar{\bm{p}}_{k}\right) \nonumber\\
	&\quad{}+\sum_{g \in \mathcal{G}}\left(\xi_{n}^{-1}\bar{\bm{q}}_{g}^{{\rm H}}\bm{E}_{n}\bar{\bm{q}}_{g} 
	+ \Delta p_{n}\bm{\varkappa}_{{\rm m}, n, g}^{(t)}\bar{\bm{q}}_{g}\right)+p_{{\rm bh}, n}\nonumber\\
	&\quad{}\times\Big(\sum_{g \in \mathcal{G}}\bm{\vartheta}_{{\rm m}, n, g}^{(t)}\bar{\bm{q}}_{g}\mathcal{R}_{{\rm m}, g}^{(t)}+\sum_{k \in \mathcal{K}}\bm{\vartheta}_{{\rm u}, n, k}^{(t)}\bar{\bm{p}}_{k}R_{{\rm u}, k}(\mathcal{V}^{(t)})\Big)\Bigg]\nonumber\\
	&\quad {}-\eta^{(c)}\left(1-\frac{\tau_{{\rm p}}}{\tau_{{\rm c}}}\right)
	\sum_{k \in\mathcal{K}}\Big[2\bm{\psi}_{{\rm u}, k}^{(t)}\bar{\bm{p}}_{k}-\phi_{{\rm u}, k}^{(t)}\big(\varphi_{{\rm u}, k}(\mathcal{V})\nonumber\\ 
	&\quad{}+ M|\hat{\bm{\xi}}_{k}^{\rm H}\bar{\bm{p}}_{k}|^{2}\big)\Big]. \label{PP_D}
\end{align}
Since $\Gamma(\mathcal{V})$ is of a positive definite quadratic form, the objective function of the problem \eqref{P_ptt22} is strongly convex over $\mathcal{V}$ and linear over $\mathcal{R}$. With the strong convexity, we can derive the dual problem of \eqref{P_ptt22}. Specifically, let $\mathcal{L} \triangleq \left\{ \left\{ \lambda_{{\rm m}, g}\right\} _{g \in \mathcal{G}}, \left\{ \bar{\lambda}_{{\rm m}, k}, \lambda_{{\rm u}, k}, \lambda_{{\rm e}, k}\right\} _{k \in \mathcal{K}}, \left\{ \lambda_{{\rm c}, n}, \lambda_{{\rm p}, n}\right\} _{n \in \mathcal{N}}\right\} $ be the dual variables corresponding to the constraints \eqref{rgg}-\eqref{tgg}, the dual problem of \eqref{P_ptt22} is formalized in the following proposition.
\begin{prop}
	\label{Prop_2}
	\begin{subequations}
		The dual problem of \eqref{P_ptt22} is given by
		\begin{align}
			\max_{\mathcal{L} \in \mathcal{P}} \ \mathcal{D}(\mathcal{L})
			& \triangleq \Gamma(\mathcal{V}^{\circ}) + \sum_{g \in \mathcal{G}}\lambda_{{\rm m}, g}\bar{r}_{{\rm m}, g}
				- \sum_{k \in \mathcal{K}}\Bigg[\bar{\lambda}_{{\rm m}, k}\bar{R}_{k}^{(t)}(\mathcal{V}^{\circ})\nonumber\\
			&{}+ \lambda_{{\rm u}, k}\left(\bar{r}_{{\rm u}, k} - \bar{R}_{{\rm u}, k}^{(t)}(\mathcal{V}^{\circ})\right)+ \lambda_{{\rm e}, k}\bigg(\frac{\mathcal{F}^{-1}\left(\bar{e}_{k}\right)}{1-\rho_{k}^{\circ}} \nonumber \\
			&{}- \bar{\varphi}_{{\rm e},k}^{(t)}(\mathcal{V}^{\circ})\bigg)\Bigg] + \sum_{n \in \mathcal{N}}\!\!\lambda_{{\rm c}, n}\left(\bar{C}_{{\rm bh}, n}^{(t)}(\mathcal{V}^{\circ}) - \bar{c}_{n, \max}\right)\nonumber  \\
			&{}+ \sum_{n \in \mathcal{N}}\lambda_{{\rm p}, n}\left(p_{{\rm tr}, n}(\mathcal{V}^{\circ})-\bar{p}_{n,\max}\right),\label{dual} 
		\end{align}
		where $\mathcal{V}^{\circ} \triangleq \left\{ \left\{ \bar{\bm{q}}_{g}^{\circ}\right\}_{g \in \mathcal{G}}, \left\{ \bar{\bm{p}}_{k}^{\circ}, \rho_{k}^{\circ}\right\}_{k \in \mathcal{K}}\right\}$ is uniquely given by
		\begin{align}
			\bar{\bm{q}}_{g}^{\circ} &= \left(\mathcal{A}_{{\rm m},g}^{(t)}\right)^{-1}\bm{a}_{{\rm m},g}^{(t)}, \label{dp_m1} \\
			\bar{\bm{p}}_{k}^{\circ} &= \left(\mathcal{B}_{{\rm u},k}^{(t)}\right)^{-1}\bm{b}_{{\rm u},k}^{(t)}, \label{dp_m2} \\
			\rho_{k}^{\circ} &= \frac{\sqrt{\chi_{{\rm u}, k}^{(t)} + \chi_{{\rm m}, k}^{(t)}}\sigma_{k}}{\sqrt{\chi_{{\rm u}, k}^{(t)} 
							+ \chi_{{\rm m}, k}^{(t)}}\sigma_{k} + \sqrt{\lambda_{{\rm e}, k}\mathcal{F}^{-1}\left(\bar{e}_{k}\right)}}, \label{dp_m3}
		\end{align}
		with
		\begin{align}
			\mathcal{A}_{{\rm m},g}^{(t)}
			& \triangleq \sum_{n\in\mathcal{N}}2\left(\xi_{n}^{-1}+\lambda_{{\rm p},n}\right)\bm{E}_{n}
				+2M\sum_{j \in \mathcal{K}_{g}}\chi_{{\rm m},j}^{(t)}\hat{\bm{\xi}}_{j}\hat{\bm{\xi}}_{j}^{\rm H} \nonumber\\
			&\quad {}+2\sum_{j \in \mathcal{K}}\left(\chi_{{\rm u},j}^{(t)}+\chi_{{\rm m},j}^{(t)}\right)\bm{\Xi}_{j}^{2},\\
			\bm{a}_{{\rm m},g}^{(t)}
			& \triangleq 2\hspace{-0.25em}\sum_{j \in \mathcal{K}_{g}}\!\!\left(M\lambda_{{\rm e},j}\hat{\bm{\Xi}}_{j}^{2}\bar{\bm{q}}_{g}^{(t)}+\bar{\lambda}_{{\rm m},j}\bm{\psi}_{{\rm m},g_{j}}^{(t){\rm H}}\right) + 2\!\sum_{j \in \mathcal{K}}\lambda_{{\rm e},j}\bm{\Xi}_{j}^{2}\bar{\bm{q}}_{g}^{(t)}\nonumber \\
			&\quad {}- \sum_{n \in \mathcal{N}}\!\!\left[\Delta p_{n}\bm{\varkappa}_{{\rm m},n,g}^{(t){\rm H}}-(\lambda_{{\rm c},n}+p_{{\rm bh},n})\mathcal{R}_{{\rm m},g}^{(t)}\bm{\vartheta}_{{\rm m},n,g}^{(t){\rm H}}\right],  \\
			\mathcal{B}_{{\rm u},k}^{(t)}
			& \triangleq 2\hspace{-0.25em}\sum_{n \in \mathcal{N}}\!\!(\xi_{n}^{-1}+\lambda_{{\rm p},n})\bm{E}_{n}
				+2\sum_{j \in \mathcal{K}}(\chi_{{\rm u},j}^{(t)}+\chi_{{\rm m},j}^{(t)})\bm{\Xi}_{j}^{2}+2M \nonumber \\
			&\quad {}\times\hspace{-0.5em}\sum_{j \in \mathcal{K}_{g_{k}}}\!\!(\chi_{{\rm u},j}^{(t)}+\chi_{{\rm m},j}^{(t)})\hat{\bm{\Xi}}_{j}^{2}
				+ 2M\chi_{{\rm u},k}^{(t)}(\hat{\bm{\xi}}_{k}\hat{\bm{\xi}}_{k}^{\rm H}-\hat{\bm{\Xi}}_{k}^{2}), \\
			\bm{b}_{{\rm u},k}^{(t)}
			&\triangleq 2M\hspace{-0.5em}\sum_{j \in \mathcal{K}_{g_{k}}}\!\!\lambda_{{\rm e},j}\hat{\bm{\Xi}}_{j}^{2}\bar{\bm{p}}_{k}^{(t)}
				\!+\!2\left(\eta^{(c)}-\frac{\tau_{\rm p}\eta^{(c)}}{\tau_{\rm c}}+\lambda_{{\rm u},k}\right)\bm{\psi}_{{\rm u},k}^{(t){\rm H}} \nonumber \\
			&\quad {}- \hspace{-0.25em}\sum_{n \in \mathcal{N}} \hspace{-0.25em} \left[\Delta p_{n}\bm{\varkappa}_{{\rm u},n,k}^{(t){\rm H}} \!+\!\left(\lambda_{{\rm c},n}+p_{{\rm bh},n}\right)R_{{\rm u},k}(\mathcal{V}^{(t)})\bm{\vartheta}_{{\rm u},n,k}^{(t){\rm H}}\right] \nonumber \\
			&\quad {}+ 2\sum_{j \in \mathcal{K}}\lambda_{{\rm e},j}\bm{\Xi}_{j}^{2}\bar{\bm{p}}_{k}^{(t)}, \\
			\chi_{{\rm u},k}^{(t)}
			& \triangleq \left(\eta^{(c)}-\frac{\tau_{\rm p} \eta^{(c)}}{\tau_{\rm c}} + \lambda_{{\rm u},k}\right)\phi_{{\rm u},k}^{(t)},  \\
			\chi_{{\rm m},k}^{(t)}
			& \triangleq \bar{\lambda}_{{\rm m},k}\phi_{{\rm m},g_{k}}. \label{jj}		
		\end{align}
		Moreover, the domain of the dual function $\mathcal{D}\left(\mathcal{L}\right)$ is
		\begin{align} \label{dom}
			\mathcal{P} &\!=\! \begin{cases}
			\begin{array}{ll}
			\bar{\lambda}_{{\rm m}, k},\ \lambda_{{\rm u}, k},\ \lambda_{{\rm e}, k} \geq 0, & \forall k \in \mathcal{K}, \\
			\lambda_{{\rm c}, n},\ \lambda_{{\rm p}, n} \geq 0, & \forall n \in \mathcal{N}, \\
			\lambda_{{\rm m}, g} \geq 0, & \forall g \in \mathcal{G}, \\
			\sum\limits_{k \in \mathcal{K}_{g}}\bar{\lambda}_{{\rm m}, k} - \eta^{(c)}-\frac{\tau_{\rm p}\eta^{(c)}}{\tau_{\rm c}} - \lambda_{{\rm m}, g}=0, & \forall g \in \mathcal{G}.
			\end{array}
			\end{cases}
		\end{align}	
	\end{subequations}
\end{prop}
\begin{IEEEproof}
	See Appendix~\ref{Appendix-C}.
\end{IEEEproof}

In light of Proposition~\ref{Prop_2}, for any fixed $\mathcal{L}$, the value of $\mathcal{V}^{\circ}$ is uniquely determined by \eqref{dp_m1}-\eqref{dp_m3}. Next, we can obtain
the partial derivatives as
${\partial \mathcal{D}(\mathcal{L})}/{\partial \lambda_{{\rm m},g}} = \bar{r}_{{\rm m},g}$, $\forall g \in \mathcal{G}$;
${\partial \mathcal{D}(\mathcal{L})}/{\partial \bar{\lambda}_{{\rm m},k}} = - \bar{R}_{k}^{(t)}(\mathcal{V}^{\circ})$,
${\partial \mathcal{D}(\mathcal{L})}/{\partial \lambda_{{\rm u},k}} = \bar{r}_{{\rm u},k}-\bar{R}_{{\rm u},k}^{(t)}(\mathcal{V}^{\circ})$,
${\partial \mathcal{D}(\mathcal{L})}/{\partial \lambda_{{\rm e},k}} = \frac{\mathcal{F}^{-1}\left(\bar{e}_{k}\right)}{1-\rho_{k}^{\circ}}
-\bar{\varphi}_{{\rm e}, k}^{(t)}(\mathcal{V}^{\circ})$, $\forall k \in \mathcal{K}$; and
${\partial \mathcal{D}(\mathcal{L})}/{\partial \lambda_{{\rm c},n}} = \bar{C}_{{\rm bh},n}^{(t)}(\mathcal{V}^{\circ})-\bar{c}_{n, \max}$,
${\partial \mathcal{D}(\mathcal{L})}/{\partial \lambda_{{\rm p},n}} = p_{{\rm tr},n}(\mathcal{V}^{\circ})-\bar{p}_{n,\max}$, $\forall n \in \mathcal{N}$.
Thus, the gradient projection step increasing
$\mathcal{D}\left(\left\{\left\{\mu_{{\rm m}, g}\right\} _{g\in\mathcal{G}}, \left\{\bar{\mu}_{{\rm m}, k}, \mu_{{\rm u}, k}, \mu_{{\rm e}, k}\right\}_{k \in \mathcal{K}}, \left\{\mu_{{\rm c},n}, \mu_{{\rm p}, n}\right\}_{n \in \mathcal{N}}\right\}\right)$ can be expressed as
\begin{align} \label{La_fac}
	\begin{cases}
	\begin{array}{ll}
	\mu_{{\rm m}, g} \leftarrow \lambda_{{\rm m}, g} + \nu_{s}\bar{r}_{{\rm m}, g}, & \forall g \in \mathcal{G}, \\
	\bar{\mu}_{{\rm m}, k} \leftarrow \bar{\lambda}_{{\rm m}, k} - \nu_{s}\bar{R}_{k}^{(t)}(\mathcal{V}^{\circ}), & \forall k \in \mathcal{K},\\
	\mu_{{\rm u}, k} \leftarrow \lambda_{{\rm u}, k} + \nu_{s}\left(\bar{r}_{{\rm u}, k} - \bar{R}_{{\rm u}, k}^{(t)}(\mathcal{V}^{\circ})\right), & \forall k \in \mathcal{K},\\
	\mu_{{\rm e}, k} \leftarrow \lambda_{{\rm e}, k} + \nu_{s}\left(\frac{\mathcal{F}^{-1}\left(\bar{e}_{k}\right)}{1-\rho_{k}^{\circ}} - \bar{\varphi}_{{\rm e}, k}^{(t)}(\mathcal{V}^{\circ})\right), & \forall k \in \mathcal{K}, \\
	\mu_{{\rm c}, n} \leftarrow \lambda_{{\rm c}, n} + \nu_{s}\left(\bar{C}_{{\rm bh}, n}^{(t)}(\mathcal{V}^{\circ}) - \bar{c}_{n, \max}\right), & \forall n \in \mathcal{N}, \\
	\mu_{{\rm p}, n} \leftarrow \lambda_{{\rm p}, n} + \nu_{s}\left(p_{{\rm tr}, n}(\mathcal{V}^{\circ}) - \bar{p}_{n, \max}\right), & \forall n \in \mathcal{N},
	\end{array}
	\end{cases}
\end{align}
where $\nu_{s}$ is the step size at the $s^{\rm th}$ iteration to guarantee the convergence. To satisfy the constraints in \eqref{dom}, we need to further project
$\left\{\left\{\mu_{{\rm m}, g}\right\}_{g \in \mathcal{G}}, \left\{\bar{\mu}_{{\rm m}, k}, \mu_{{\rm u}, k}, \mu_{{\rm e}, k}\right\}_{k \in \mathcal{K}}, \left\{\mu_{{\rm c}, n}, \mu_{{\rm p}, n}\right\}_{n \in \mathcal{N}}\right\}$
onto $\mathcal{P}$ to find its nearest feasible point, which is equivalent to
\begin{align}
	\min_{\mathcal{L} \in \mathcal{P}}
	& \ \sum_{g \in \mathcal{G}}\left(\lambda_{{\rm m}, g}-\mu_{{\rm m}, g}\right)^{2} + \sum_{k \in \mathcal{K}}
	\Big[\left(\bar{\lambda}_{{\rm m}, k}-\bar{\mu}_{{\rm m}, k}\right)^{2}\nonumber\\ 
	&\quad {}+ \left(\lambda_{{\rm u}, k}-\mu_{{\rm u}, k}\right)^{2} + \left(\lambda_{{\rm e}, k}-\mu_{{\rm e}, k}\right)^{2}\Big] \nonumber \\
	&\quad {}+ \sum_{n \in \mathcal{N}}\left[\left(\lambda_{{\rm c}, n}-\mu_{{\rm c}, n}\right)^{2} + \left(\lambda_{{\rm p}, n}-\mu_{{\rm p}, n}\right)^{2}\right]. \label{pro_1}
\end{align}
Since $\mathcal{P}$ in \eqref{dom} consists of separable linear constraints, we can efficiently find the projection result of \eqref{pro_1} as (for more details, please refer to  Appendix~\ref{Appendix-D}):
\begin{align}
	\lambda_{{\rm m},g}
	&= \left(\mu_{{\rm m},g}+\frac{\varpi_{g}}{2}\right)^{+}, \quad \forall g\in{\cal G}; \nonumber\\ 
	\bar{\lambda}_{{\rm m},k} &= \left(\bar{\mu}_{{\rm m},k} - \frac{\varpi_{g}}{2}\right)^{+}, \quad \forall k\in{\cal K}_{g},\label{pro_22} \\
	\lambda_{{\rm u},k}
	&= \mu_{{\rm u},k}^{+}, \quad
	\lambda_{{\rm e},k} = \mu_{{\rm e}, k}^{+}, \quad \forall k\in{\cal K}; \nonumber\\
	\lambda_{{\rm c},n} &=\mu_{{\rm c},n}^{+}, \quad
	\lambda_{{\rm p},n} =\mu_{{\rm p},n}^{+}, \quad \forall n\in{\cal N}\label{pro_2}
\end{align}
where $\left(\cdot\right)^{+}$ means the non-negative projection on \eqref{dom}, and $\varpi_{g}$ is a parameter satisfying
\begin{equation}
	\sum_{k \in \mathcal{K}_{g}}\left(\bar{\mu}_{{\rm m}, k}-\frac{\varpi_{g}}{2}\right)^{+} - \left(\mu_{{\rm m}, g}
	+\frac{\varpi_{g}}{2}\right)^{+} = \eta^{(c)}-\frac{\tau_{\rm p} \eta^{(c)}}{\tau_{\rm c}},
\end{equation}
whose value can be readily determined by using the bisection method.

By iteratively updating $\mathcal{L}$ as per \eqref{La_fac} and \eqref{pro_22}-\eqref{pro_2}, we can get the optimal solution of $\mathcal{L}$  to the dual problem \eqref{dual}. Then, the optimal solution of $\mathcal{V}$ to the primal problem \eqref{P_tt22} is obtained by substituting the optimal $\mathcal{L}$ into \eqref{dp_m1}-\eqref{dp_m3}. In light of \eqref{rgg1}, the optimal solution of $\left\{\mathcal{R}_{{\rm m},g}\right\}_{g\in \mathcal{G}}$ to \eqref{P_tt22} satisfies its equality, i.e., 
\begin{equation} \label{R_ggg}
	\mathcal{R}_{{\rm m}, g} = \min_{k \in \mathcal{K}_{g}}\bar{R}_{k}^{(t)}(\mathcal{V}), \quad \forall g \in \mathcal{G}.
\end{equation}

To sum up, the procedure for solving the problem \eqref{P0} is formalized in Algorithm~\ref{Algorithm1}. In particular, \eqref{P0} is iteratively solved with iterations over $t$ in the SCA framework, and each SCA subproblem \eqref{P_t1} is solved with iterations over $c$ and~$s$. Since the domain of $\mathcal{D}(\mathcal{L})$ in \eqref{dom} is closed and convex, the iteration with respect to $s$ is guaranteed to converge to the global optimum of \eqref{dual} at a rate of $\mathcal{O}(1/s)$, if the step size $\nu_{s}$ is smaller than the inverse of the Lipschitz constant of $\Delta \mathcal{D}(\mathcal{L})$ \cite{beck2009gradient}. Moreover, since the primal problem \eqref{P_ptt22} is convex, the convergent optimum of \eqref{dual} is also the global optimum of \eqref{P_ptt22}, provided that \eqref{P_ptt22} is strictly feasible \cite{boyd_convex_2004}. In addition, since \eqref{P_ptt22} is the Dinkelbach's transform of the quasi-convex problem \eqref{P_t1}, the iteration with respective to $c$ is guaranteed to converge to the global optimal solution of \eqref{P_t1} \cite{dinkelbach1967nonlinear}. Moreover, according to \cite{l0_l1, SCA_1}, the iteration over $t$ is guaranteed to converge to a stationary point of the problem \eqref{P0}. Last but not the least, since Algorithm~\ref{Algorithm1} is based on the SCA framework, it requires to be initialized from a feasible point, which is addressed in the next section.

\begin{remark}[On the coordination overhead] 
		As channel estimation and conjugate beamforming are performed locally at each AP, there is no overhead for exchanging the instantaneous CSI among the APs. On the other hand, each iteration of Algorithm~\ref{Algorithm1} can be executed in parallel, which benefits lower coordination overhead. In particular, according to \eqref{dp_m1}-\eqref{jj}, the all primal variables $\mathcal{V}^{\circ} \triangleq \left\{ \left\{ \bar{\bm{q}}_{g}^{\circ}\right\}_{g \in \mathcal{G}}, \left\{ \bar{\bm{p}}_{k}^{\circ}, \rho_{k}^{\circ}\right\}_{k \in \mathcal{K}}\right\}$ can be updated in parallel. Since the update of each primal variable only depends on a few dual variables, the message passing overhead is small. Likewise, according to \eqref{La_fac} and \eqref{pro_22}-\eqref{pro_2}, the dual variables can also be updated in parallel and each depends on only a few primal variables. Thanks to the parallel structure, Algorithm~\ref{Algorithm1} has the potential of leveraging the modern multicore multi-thread processor architecture for speeding up the computation, with extremely low coordination overhead.
\end{remark}

\subsection{Accelerated First-Order Algorithm}
Although Algorithm~\ref{Algorithm1} only involves  the gradient information, it may require a large number of iterations to converge. To improve the convergence speed, we further exploit the momentum technique \cite{beck2009fast} to accelerate Algorithm~\ref{Algorithm1}. In particular, the projected gradient step in \eqref{La_fac} is modified by
\begin{subequations}
	\begin{align} \label{La_fac_AA_1}
		\begin{cases}
		\begin{array}{ll}
		\tilde{\mu}_{{\rm m}, g}^{(s)} \leftarrow \lambda_{{\rm m}, g} + \nu_{s}\bar{r}_{{\rm m}, g}, & \forall g \in \mathcal{G}, \\
		\tilde{\bar{\mu}}_{{\rm m}, k}^{(s)} \leftarrow \bar{\lambda}_{{\rm m}, k} - \nu_{s}\bar{R}_{k}^{(t)}(\mathcal{V}^{\circ}), & \forall k \in \mathcal{K},\\
		\tilde{\mu}_{{\rm u}, k}^{(s)} \leftarrow \lambda_{{\rm u}, k} + \nu_{s}\left(\bar{r}_{{\rm u}, k} - \bar{R}_{{\rm u}, k}^{(t)}(\mathcal{V}^{\circ})\right), & \forall k \in \mathcal{K},\\
		\tilde{\mu}_{{\rm e}, k}^{(s))} \leftarrow \lambda_{{\rm e}, k} + \nu_{s}\left(\frac{\mathcal{F}^{-1}\left(\bar{e}_{k}\right)}{1-\rho_{k}^{\circ}} - \bar{\varphi}_{{\rm e}, k}^{(t)}(\mathcal{V}^{\circ})\right), & \forall k \in \mathcal{K}, \\
		\tilde{\mu}_{{\rm c}, n}^{(s)} \leftarrow \lambda_{{\rm c}, n} + \nu_{s}\left(\bar{C}_{{\rm bh}, n}^{(t)}(\mathcal{V}^{\circ}) - \bar{c}_{n, \max}\right), & \forall n \in \mathcal{N}, \\
		\tilde{\mu}_{{\rm p}, n}^{(s)} \leftarrow \lambda_{{\rm p}, n} + \nu_{s}\left(p_{{\rm tr}, n}(\mathcal{V}^{\circ}) - \bar{p}_{n, \max}\right), & \forall n \in \mathcal{N},
		\end{array}
		\end{cases}
	\end{align}
	and
	\begin{align} \label{La_fac_AA_2}
		\begin{cases}
		\begin{array}{ll}
		\mu_{{\rm m},g}\leftarrow\tilde{\mu}_{{\rm m}, g}^{(s)}+\frac{\pi^{(s-1)}-1}{\pi^{(s)}}
		\left(\tilde{\mu}_{{\rm m}, g}^{(s)}- \tilde{\mu}_{{\rm m}, g}^{(s-1)}\right), & \forall g \in \mathcal{G}, \\
		\bar{\mu}_{{\rm m},k}\leftarrow\tilde{\bar{\mu}}_{{\rm m}, k}^{(s)}+\frac{\pi^{(s-1)}-1}{\pi^{(s)}}
		\left(\tilde{\bar{\mu}}_{{\rm m}, k}^{(s)}-\tilde{\bar\mu}_{{\rm m}, k}^{(s-1)}\right), & \forall k \in \mathcal{K}, \\
		\mu_{{\rm u},k}\leftarrow\tilde{\mu}_{{\rm u}, k}^{(s)}+\frac{\pi^{(s-1)}-1}{\pi^{(s)}}
		\left(\tilde{\mu}_{{\rm u}, k}^{(s)}-\tilde{\mu}_{{\rm u}, k}^{(s-1)}\right), & \forall k \in \mathcal{K}, \\
		\mu_{{\rm e},k}\leftarrow\tilde{\mu}_{{\rm e}, k}^{(s)}+\frac{\pi^{(s-1)}-1}{\pi^{(s)}}
		\left(\tilde{\mu}_{{\rm e}, k}^{(s)}-\tilde{\mu}_{{\rm e}, k}^{(s-1)}\right), & \forall k \in \mathcal{K}, \\
		\mu_{{\rm c},n}\leftarrow\tilde{\mu}_{{\rm c}, n}^{(s)}+\frac{\pi^{(s-1)}-1}{\pi^{(s)}}
		\left(\tilde{\mu}_{{\rm c}, n}^{(s)}-\tilde{\mu}_{{\rm c}, n}^{(s-1)}\right), & \forall n \in \mathcal{N}, \\
		\mu_{{\rm p},n}\leftarrow\tilde{\mu}_{{\rm p}, n}^{(s)}+\frac{\pi^{(s-1)}-1}{\pi^{(s)}}
		\left(\tilde{\mu}_{{\rm p}, n}^{(s)}-\tilde{\mu}_{{\rm p}, n}^{(s-1)}\right), & \forall n \in \mathcal{N}, 
		\end{array}
		\end{cases}
	\end{align}
\end{subequations}
where $\pi^{(s)}$ is the weighting parameter to dynamically control the momentums 
$\tilde{\mu}_{{\rm m}, g}^{(s)}-\tilde{\mu}_{{\rm m}, g}^{(s-1)}$, 
$\tilde{\bar\mu}_{{\rm m}, k}^{(s)}-\tilde{\bar\mu}_{{\rm m}, k}^{(s-1)}$, 
$\tilde{\mu}_{{\rm u}, k}^{(s)}-\tilde{\mu}_{{\rm u}, k}^{(s-1)}$,
$\tilde{\mu}_{{\rm e}, k}^{(s)}-\tilde{\mu}_{{\rm e}, k}^{(s-1)}$, 
$\tilde{\mu}_{{\rm c}, n}^{(s)}-\tilde{\mu}_{{\rm c}, n}^{(s-1)}$ and 
$\tilde{\mu}_{{\rm p}, n}^{(s)}-\tilde{\mu}_{{\rm p}, n}^{(s-1)}$. To achieve fast convergence,   $\pi^{\left(s\right)}$ is updated by \cite{beck2009fast}
\begin{align}\label{La_fac_AA_3}
	\pi^{(0)}=1,\quad \pi^{(s)}=\frac{1}{2}\left(1+\sqrt{1+4\left(\pi^{(s-1)}\right)^{2}}\right).
\end{align}

The great insight of the acceleration lies in the momentums $\tilde{\mu}_{{\rm m}, g}^{(s)}-\tilde{\mu}_{{\rm m}, g}^{(s-1)}$, $\tilde{\bar\mu}_{{\rm m}, k}^{(s)}-\tilde{\bar\mu}_{{\rm m}, k}^{(s-1)}$, $\tilde{\mu}_{{\rm u}, k}^{(s)}-\tilde{\mu}_{{\rm u}, k}^{(s-1)}$, $\tilde{\mu}_{{\rm e}, k}^{(s)}-\tilde{\mu}_{{\rm e}, k}^{(s-1)}$, $\tilde{\mu}_{{\rm c}, n}^{(s)}-\tilde{\mu}_{{\rm c}, n}^{(s-1)}$ and $\tilde{\mu}_{{\rm p}, n}^{(s)}-\tilde{\mu}_{{\rm p}, n}^{(s-1)}$ in \eqref{La_fac_AA_2} (without these momentums, the accelerated algorithm would reduce to Algorithm~\ref{Algorithm1}). These momentums utilize previous updates to generate an overshoot, so that the updates using \eqref{La_fac_AA_1} and \eqref{La_fac_AA_2} in the accelerated algorithm is more aggressive than the conventional gradient step \eqref{La_fac} in Algorithm~\ref{Algorithm1}. On the other hand, to ensure these overshoots to be well behaved, the momentums are controlled by a sequence of weighting parameters $\left\{\pi^{(s)}\right\}$.  With $\left\{\pi^{(s)}\right\}$ updated according to \eqref{La_fac_AA_3}, the accelerated algorithm is guaranteed to converge to the global optimum of \eqref{P_t1} at a rate of $\mathcal{O}(1/s^{2})$ \cite{beck2009fast}. As the other steps of the accelerated algorithm are identical to Algorithm~\ref{Algorithm1} except the projected gradient step, the detailed algorithm procedure is omitted for brevity.

\begin{algorithm}[!t]
	\caption{First-Order Algorithm for Solving Problem \eqref{P0}}
	\label{Algorithm1}
	\begin{algorithmic}[1]
		\STATE \textbf{Initialization:} Generate $\mathcal{V}^{(0)}$ via Algorithm~\ref{Algorithm2} (to be detailed in Section V).
		\STATE {$t = 0$;}
		\REPEAT
		\STATE {Compute $\zeta_{{\rm m}, g_{k}}^{(t)}$,
			$\phi_{{\rm m}, g_{k}}^{(t)}$,
			$\bm{\psi}_{{\rm m}, g_{k}}^{(t)}$,
			$\zeta_{{\rm u}, k}^{(t)}$,
			$\phi_{{\rm u}, k}^{(t)}$,
			$\bm{\psi}_{{\rm u}, k}^{(t)}$,
			$\zeta_{{\rm e}, k}^{(t)}$,
			$R_{{\rm u}, k}(\mathcal{V}^{(t)})$, $\forall k \in \mathcal{K}$, and $\mathcal{R}_{{\rm m},g}^{(t)}$, 
			$\bm{\vartheta}_{{\rm m}, n, g}^{(t)}$,
			$\bm{\varkappa}_{{\rm m}, n, g}^{(t)}$,
			$\mathcal{R}_{{\rm m}, g}^{(t)}$,			
			$\mathcal{H}_{{\rm m}, n, g}^{(t)}$,
			$\bm{\vartheta}_{{\rm u}, n, k}^{(t)}$,
			$\bm{\varkappa}_{{\rm u}, n, k}^{(t)}$,
			$\mathcal{H}_{{\rm u}, n, k}^{(t)}$,
			$\mathcal{H}_{{\rm mu}, n}^{(t)}$, $\forall n \in \mathcal{N}, g \in \mathcal{G}, k \in \mathcal{K}$ according to Proposition~\ref{Prop_1};}
		\STATE {$\mathcal{V} = \mathcal{V}^{(t)}$};
		\REPEAT
		\STATE {Update $\mathcal{R}_{{\rm m}, g}$ as per \eqref{R_ggg};}
		\STATE {Update $\eta^{(c)}$ with the cost function value of \eqref{objj};}
		\STATE {Set $\lambda_{{\rm m}, g} = 0$,
			$\bar{\lambda}_{{\rm m}, k} = \left(\eta^{(c)} - {\tau_{\rm p} \eta^{(c)}}/{\tau_{\rm c}}\right)/{|\mathcal{K}_{g_{k}}|}$,
			$\lambda_{{\rm u}, k} = 0$,
			$\lambda_{{\rm e}, k} = 0$,
			$\lambda_{{\rm p}, n} = 0$,
			$\lambda_{{\rm c}, n} = 0$, $\forall n \in \mathcal{N}, k \in \mathcal{K}, g \in \mathcal{G}$;}
		\REPEAT
		\STATE{Update $\mathcal{V}^{\circ}$ as per \eqref{dp_m1}-\eqref{dp_m3};}
		\STATE{Update $\mathcal{L}$ according to \eqref{La_fac} and \eqref{pro_22}-\eqref{pro_2};}
		\UNTIL {convergence}
		\STATE {Output $\mathcal{V} = \mathcal{V}^{\circ}$;}
		\UNTIL {convergence}
		\STATE {Output $\mathcal{V}^{(t+1)} = \mathcal{V}$;}
		\STATE {$t \leftarrow t+1$;}
		\UNTIL {convergence}
	\end{algorithmic}
\end{algorithm}

\section{Finding Feasible Initial Point via First-Order Algorithm}
\label{Sec_ffp}
In the last section, we have developed Algorithm~\ref{Algorithm1} and  its accelerated algorithm for solving \eqref{P0} provided that an initial feasible point is available. In practice, however, it is challenging to find a feasible point for the nonconvex constrained problem \eqref{P0}. To address this issue, the feasibility problem is transformed into an equivalent nonconvex optimization problem with only a simple set of constraints. Then, by designing a first-order algorithm, we solve the transformed nonconvex problem in the SCA framework.

\subsection{Problem Transformation}
By applying \eqref{R_gk} and \eqref{C_fh22}, the feasibility problem of \eqref{P0} can be written as
\begin{subequations} \label{fea}
	\begin{align}
	\text{find} & \ \mathcal{V} \\
	{\rm s.t.}  & \ R_{k}(\mathcal{V})\geq  \bar{r}_{{\rm m}, g_{k}}, \ \forall  k \in \mathcal{K}, \\
	& \ \eqref{C2},\ \eqref{C3},\ \eqref{C5},\ \eqref{C_fh22}.
	\end{align}
\end{subequations}
To establish its (in)feasibility, \eqref{fea} can be equivalently transformed into the exact penalty formulation with only a simple set of constraints \cite{8003316,8474356}:
\begin{subequations} \label{fea_11}
	\begin{align}
		\min_{\mathcal{V}}\ h(\mathcal{V})
		&\triangleq \sum_{j \in \mathcal{K}}\bigg[ \left(\bar{r}_{{\rm m}, g_{j}}-R_{j}(\mathcal{V})\right)^{+} + \left(\bar{r}_{{\rm u},j}-R_{{\rm u},j}(\mathcal{V})\right)^{+}\nonumber\\
		&\quad{}+ \left(\frac{\mathcal{F}^{-1}\left(\bar{e}_{j}\right)}{1-\rho_{j}}-\varphi_{{\rm e}, j}(\mathcal{V})\right)^{+}\bigg] \nonumber \\
		&\quad {}+ \sum_{n \in \mathcal{N}}\left(C_{{\rm bh}, n}(\mathcal{V})-\bar{c}_{n,\max}\right)^{+}, \label{fea_1} \\
		{\rm s.t.} \ 
		&\sum_{k \in \mathcal{K}}\bar{\bm{p}}_{k}^{\rm H}\bm{E}_{n}\bar{\bm{p}}_{k} + \sum_{g \in \mathcal{G}}\bar{\bm{q}}_{g}^{\rm H}\bm{E}_{n}\bar{\bm{q}}_{g} \leq \bar{p}_{n, \max}, \ \forall n \in \mathcal{N}. \label{fea_c}
	\end{align}
\end{subequations}
In particular, there are $3K+N$ separable components in the objective function $h(\mathcal{V})$, with each measuring the degrees of violating the constraints. As $h(\mathcal{V})$ given by \eqref{fea_1} is nonconvex, it is very hard to directly solve the problem \eqref{fea_11}, if not impossible. To proceed, in the next subsection we apply SCA to approximate $h(\mathcal{V})$ via a sequence of convex upper bounds. This would at least guarantee the convergence to a stationary point of the problem \eqref{fea_11}. Despite without theoretical guarantee for global optimality, SCA based algorithms for finding a feasible point have been empirically demonstrated to be highly successful in converging to the global minimum (i.e., a feasible point) with a few number of iterations \cite{8003316,6954488}.

\subsection{First-Order Algorithm for Solving \eqref{fea_11}}
To tackle the nonconvexity of $h(\mathcal{V})$ shown in \eqref{fea_1}, by using \eqref{fh_tt}, \eqref{rate_mm}, \eqref{rate_uu} and \eqref{phi_ee1}, we construct a sequence of convex upper bounds:
\begin{align}
	h(\mathcal{V})&\leq \sum_{j \in \mathcal{K}}\Bigg[\underbrace{\left(bar{r}_{{\rm m},g_{j}}
	-\bar{R}_{j}^{(t)}(\mathcal{V})\right)^{+}}_{\hbar_{{\rm m}, j}}+\underbrace{\left(\bar{r}_{{\rm u}, j}-\bar{R}_{{\rm u}, j}^{(t)}(\mathcal{V})\right)^{+}}_{\hbar_{{\rm u}, j}}\nonumber\\
	&\quad{}+ \underbrace{\left(\frac{\mathcal{F}^{-1}\left(\bar{e}_{j}\right)}{1-\rho_{j}}-\bar{\varphi}_{{\rm e}, j}^{(t)}(\mathcal{V})\right)^{+}}_{\hbar_{{\rm e}, j}}\Bigg] \nonumber \\
	&\quad {}+ \sum_{n \in \mathcal{N}}\underbrace{\left[\bar{C}_{{\rm bh}, n}^{(t)}(\mathcal{V})-\bar{c}_{n\max}\right]^{+}}_{\hbar_{{\rm c}, n}}
	\triangleq \bar{h}^{(t)}(\mathcal{V}). \label{fea_2}
\end{align}
With \eqref{fea_2}, the $t^{\rm th}$ SCA subproblem of \eqref{fea_11} can be rewritten as
\begin{equation} \label{fea_3}
	\mathcal{V}^{(t+1)} = \arg\min_{\mathcal{V}}\ \bar{h}^{(t)}(\mathcal{V}),  \text{ s.t.} \ \eqref{fea_c}.
\end{equation}
Thanks to the simplicity of constraint \eqref{fea_c}, a first-order algorithm can be designed to solve each SCA subproblem \eqref{fea_3}. Specifically, by performing a subgradient step, $\left\{ \left\{ \bar{\bm{q}}_{g}\right\} _{g\in\mathcal{G}}, \left\{ \bar{\bm{p}}_{k},\rho_{k}\right\}_{k \in \mathcal{K}}\right\}$ is updated to $\left\{\left\{ {\bm{z}}_{{\rm m}, g}\right\}_{g \in \mathcal{G}}, \left\{ {\bm{z}}_{{\rm u}, k}, \gamma_{k}\right\}_{k \in \mathcal{K}}\right\}$ according to
\begin{subequations} \label{fea_33}
	\begin{align}
		\bm{z}_{{\rm m}, g} 
		&= \bar{\bm{q}}_{g}-\bar{\nu}_{s}\frac{\partial \bar{h}^{(t)}(\mathcal{V})}{\partial\bar{\bm{q}}_{g}}, \quad \forall g\in\mathcal{G}, \\
		\bm{z}_{u,k} 
		&= \bar{\bm{p}}_{k}-\bar{\nu}_{s}\frac{\partial \bar{h}^{(t)}(\mathcal{V})}{\partial\bar{\bm{p}}_{k}}, \quad \forall k \in \mathcal{K}, \\
		\gamma_{k} 
		&= \rho_{k}-\bar{\nu}_{s}\frac{\partial \bar{h}^{(t)}(\mathcal{V})}{\partial\rho_{k}}, \quad \forall k \in \mathcal{K},
	\end{align}
\end{subequations}
where $\bar{\nu}_{s}$ is the step size at the $s^{\rm th}$ iteration,  ${\partial \bar{h}^{(t)}(\mathcal{V})}/{\partial\bar{\bm{q}}_{g}}$, ${\partial \bar{h}^{(t)}(\mathcal{V})}/{\partial\bar{\bm{p}}_{k}}$ and 
${\partial \bar{h}^{(t)}(\mathcal{V})}/{\partial\rho_{k}}$ 
are the subgradients of $\bar{h}^{(t)}(\mathcal{V})$ with respect to $\bar{\bm q}_{g}$, $\bar{\bm p}_{k}$, and  $\rho_{k}$, respectively (for more details, please refer to Appendix~\ref{Appendix-E}). Then, to satisfy the constraint \eqref{fea_c},  $\left\{ \left\{ {\bm{z}}_{{\rm m},g}\right\} _{g \in \mathcal{G}},\left\{ {\bm{z}}_{{\rm u},k},\gamma_{k}\right\}_{k \in \mathcal{K}}\right\}$ is further projected onto the constraint set of \eqref{fea_c}, which is equivalent to
\begin{subequations}\label{fea_44}
	\begin{align}
		\min_{\mathcal{V}}  
		& \sum_{g \in \mathcal{G}}\|\bar{\bm{q}}_{g}-\bm{z}_{{\rm m}, g}\|^{2}
		+ \sum_{k \in \mathcal{K}}\left(\|\bar{\bm{p}}_{k}-\bm{z}_{{\rm u}, k}\|^{2} + |\rho_{k}-\gamma_{k}|^{2}\right), \label{fea_4} \\
		{\rm s.t.} \ 
		& \eqref{fea_c}.
	\end{align}
\end{subequations}
Since \eqref{fea_44} is a quadratic programming with only a simple set of constraints, by resorting to the Karush-Kuhn-Tucker (KKT) conditions detailed in Appendix~\ref{Appendix-F}, its solution can be expressed as 
\begin{subequations} \label{fea_5}
	\begin{align}
		\bar{\bm{q}}_{g} &= 
		\begin{cases}
		\begin{array}{ll}
		\bm{z}_{{\rm m}, g}, & \text{if} \ \Lambda_{n} \leq \bar{p}_{n, \max} \\
		\sum\limits_{n \in \mathcal{N}}\sqrt{\frac{\Lambda_{n}}{\bar{p}_{n, \max}}}\bm{E}_{n}\bm{z}_{{\rm m}, g}, & \text{otherwise }
		\end{array}, & \forall g \in \mathcal{G},
		\end{cases} \\
		\bar{\bm{p}}_{k} &= 
		\begin{cases}
		\begin{array}{ll}
		\bm{z}_{{\rm u},k}, & \text{if}\ \Lambda_{n} \leq \bar{p}_{n, \max} \\
		\sum\limits_{n \in \mathcal{N}}\sqrt{\frac{\Lambda_{n}}{\bar{p}_{n, \max}}}\bm{E}_{n}\bm{z}_{{\rm u}, k}, & \text{otherwise }
		\end{array}, & \forall k \in \mathcal{K},
		\end{cases} \\
		\rho_{k} &= \gamma_{k}, \quad \forall k \in \mathcal{K}
	\end{align}
\end{subequations}
where $\Lambda_{n} \triangleq \sum_{k \in \mathcal{K}}\bm{z}_{{\rm u}, k}^{\rm H}\bm{E}_{n}\bm{z}_{{\rm u}, k} + \sum_{g \in \mathcal{G}}\bm{z}_{{\rm m}, g}^{\rm H}\bm{E}_{n}\bm{z}_{{\rm m}, g}$.

In summary, the first-order algorithm developed for finding a feasible point of the problem \eqref{P0} is formalized in Algorithm~\ref{Algorithm2}. In particular, the problem \eqref{fea_11} is iteratively solved with iteration over $t$ in the SCA framework while each SCA subproblem \eqref{fea_3} is solved with iterations over $s$. Since the constraints set of \eqref{fea_c} is closed and convex, with the step size chosen as $\bar{\nu}_{s}=\mathcal{O}\left(1/{\sqrt{s}}\right)$, the iteration with respect to $s$ is guaranteed to converge to the global minimum of \eqref{fea_3} \cite{7080879}. Moreover, like Algorithm~\ref{Algorithm1}, the iteration with respective to $t$ is guaranteed to converge to a stationary point of \eqref{fea_11}. It is remarkable that, since Algorithm~\ref{Algorithm2} is also based on the SCA framework, it needs to be initialized from a feasible point of the problem \eqref{fea_11}. For simplicity, as shown in Step~1 of Algorithm~\ref{Algorithm2}, a feasible initial point is readily obtained by equally allocating the total transmit power of an AP to each service, and thus each power initial splitting factor is obtained by satisfying the minimum harvested energy requirement.

It is noteworthy that to guarantee the global optimal solution to \eqref{fea_11} is challenging because it is an NP-hard problem. If the cost function $h(\mathcal{V})$ given by \eqref{fea_1} converges to zero, we can claim that the stationary point obtained by Algorithm~\ref{Algorithm2} must also be a global optimal solution and, thus, it can serve as a feasible point of the problem \eqref{P0}. Otherwise, we need to rerun Algorithm~\ref{Algorithm2} with different initial points. Similarly, the feasibility problem \eqref{fea} is also NP-hard due to the nonconvex constraints. Therefore, the existing approaches based on the interior-point method cannot guarantee the convergence to a feasible point but only to a stationary point \cite{6954488}.

\begin{algorithm}[t]
	\caption{First-Order Algorithm for Finding a Initial Feasible Point of  \eqref{P0}}
	\label{Algorithm2}
	\begin{algorithmic}[1]
		\STATE \textbf{Initialization:} Generate an initial point denoted
		$\bar{\bm{q}}_{g}^{(0)} = \sum_{n\in\mathcal{N}}\sqrt{\frac{\bar{p}_{n, \max}}{G+K}}\bm{E}_{n}\bm{1}_{N}, \forall g \in \mathcal{G}$,
		$\bar{\bm{p}}_{k}^{(0)} = \sum_{n \in \mathcal{N}}\sqrt{\frac{\bar{p}_{n, \max}}{G+K}}\bm{E}_{n}\bm{1}_{N}, \forall k \in \mathcal{K}$ and
		$\rho_{k}^{(0)} = 1-{\mathcal{F}^{-1}\left(\bar{e}_{k}\right)}/{\varphi_{{\rm e}, k}\left(\left\{\bar{\bm{q}}_{g}^{(0)}\right\}_{g \in \mathcal{G}}, \left\{\bar{\bm{p}}_{k}^{(0)}\right\}_{k \in \mathcal{K}}\right)}, \forall k \in \mathcal{K}$, where $\bm{1}_{N}$ is a $N \times1$ vector with each element being unity;
		\STATE {$t = 0$;}
		\REPEAT
		\STATE {Compute $\zeta_{{\rm m}, g_{k}}^{(t)}$,
			$\phi_{{\rm m}, g_{k}}^{(t)}$,
			$\bm{\psi}_{{\rm m}, g_{k}}^{(t)}$,
			$\zeta_{{\rm u}, k}^{(t)}$,
			$\phi_{{\rm u}, k}^{(t)}$,
			$\bm{\psi}_{{\rm u}, k}^{(t)}$,
			$\zeta_{{\rm e}, k}^{(t)}$,
			$R_{{\rm u}, k}\left(\mathcal{V}^{(t)}\right)$, $\forall k \in \mathcal{K}$, and
			$\mathcal{R}_{{\rm m}, g}^{(t)}$,
			$\bm{\vartheta}_{{\rm m}, n, g}^{(t)}$,
			$\mathcal{H}_{{\rm m}, n, g}^{(t)}$,
			$\bm{\vartheta}_{{\rm u}, n, k}^{(t)}$,
			$\mathcal{H}_{{\rm u}, n, k}^{(t)}$,
			$\forall n\in\mathcal{N}, g \in \mathcal{G}, k \in \mathcal{K}$;}
		\STATE {$\mathcal{V} = \mathcal{V}^{(t)}$;}
		\REPEAT
		\STATE {Compute $\big\{{\partial \bar{h}^{(t)}(\mathcal{V})}/{\partial\bar{\bm{p}}_{k}}, {\partial \bar{h}^{(t)}(\mathcal{V})}/{\partial\rho_{k}}\big\}_{k \in \mathcal{K}}$ and $\big\{{\partial \bar{h}^{(t)}(\mathcal{V})}/{\partial\bar{\bm{q}}_{g}}\big\}_{g \in \mathcal{G}}$ as per  \eqref{D.4a}-\eqref{D.4c};}
		\STATE {Update $\left\{ \bm{z}_{{\rm m}, g}\right\}_{g \in \mathcal{G}}$ and $\left\{ \bm{z}_{{\rm u}, k}, \gamma_{k}\right\}_{k \in \mathcal{K}}$ as per \eqref{fea_33};}
		\STATE {Update $\left\{\bar{\bm{q}}_{g}\right\}_{g \in \mathcal{G}}$ and $\left\{\bar{\bm{p}}_{k}, \rho_{k}\right\}_{k \in \mathcal{K}}$ according to \eqref{fea_5};}
		\UNTIL {convergence}
		\STATE {Output $\mathcal{V}^{(t+1)} = \mathcal{V}$;}
		\STATE {$t \leftarrow t+1$;}
		\UNTIL {convergence}
	\end{algorithmic}
\end{algorithm}

\subsection{Complexity Analysis}
To solve \eqref{fea_11}, we first need to solve the convex subproblem \eqref{fea_3} at each SCA iteration. According to Algorithm~\ref{Algorithm2}, the computational complexity is dominated by computing subgradients  $\left\{{\partial \bar{h}^{(t)}(\mathcal{V})}/{\partial\bar{\bm{q}}_{g}}\right\}_{g \in \mathcal{G}}$ and $\left\{{\partial \bar{h}^{(t)}(\mathcal{V})}/{\partial\bar{\bm{p}}_{k}}, {\partial \bar{h}^{(t)}(\mathcal{V})}/{\partial\rho_{k}}\right\}_{k \in \mathcal{K}}$, which requires $\mathcal{O}(N(K+G)+K)$ floating-point operations for each iterations \cite{8322450}. Then, we perform $\mathcal{O}({(2K+G)}/\epsilon_{2})$ scalar operations to update the resource coefficients, where $\epsilon_{2}$ is the target convergence accuracy of inner iterations in Algorithm~\ref{Algorithm2}. Therefore, the overall complexity of solving \eqref{fea_11} via Algorithm~\ref{Algorithm2} is $\mathcal{O}(T_{\max2}(N(K+G)+K+(2K+G)/\epsilon_2))$, where $T_{\max2}$ is the number of outer iterations for Algorithm~\ref{Algorithm2} to converge. 

By using a similar approach as above, the computational complexity of solving  the optimization problem \eqref{P0} via Algorithm~\ref{Algorithm1} and its accelerated version can be computed and given by $\mathcal{O}(T_{\max1}C_{\max}(N(K+G)+K+(2K+G)/{\epsilon_1}))$ and $\mathcal{O}(T_{\max1}C_{\max}(N(K+G)+K+(2K+G)/\sqrt{\epsilon_1}))$, respectively, where $\epsilon_{1}$ is the target convergence accuracy of inner iterations in Algorithm~\ref{Algorithm1}, $T_{\max1}$ is the number of outer iterations for Algorithm~\ref{Algorithm1} to converge, and $C_{\max}$ is the number of Dinkelbach's method iteration to reach convergence. 

For comparison purposes, the following traditional second-order algorithms are accounted for:
\begin{itemize}
	\item \textbf{FP-IPM}: In this algorithm, each convex subproblem \eqref{fea_3} is first approximated by a sequence of second order cone programmings (SOCPs) and, then, the resultant SOCPs are solved via an interior-point method solver, e.g., MOSEK, yielding computational complexity $\mathcal{O}(T_{\max2}(N(K+G)+K)^{3.5})$ \cite{NOUM_1}.
	\item \textbf{IPM}: In this algorithm, each SCA subproblem \eqref{P_tt22} is modeled as a conic form and then solved directly by an interior-point method solver, e.g., MOSEK, yielding computational complexity $\mathcal{O}(T_{\max1}C_{\max}(N(K+G)+K)^{3.5})$ \cite{7120191}. 
	\item \textbf{SCS}: In this algorithm, each SCA subproblem \eqref{P_tt22} is transformed into the standard cone programming, which can be solved by the ADMM based solver, e.g., the splitting conic solver, with computational complexity $\mathcal{O}(T_{\max1}C_{\max}(N^{3.5}(K^{3.5}+G^{3.5})+K^{3.5}))$ \cite{7120191}.
	\item \textbf{Optimal branch-and-reduce-and-bound  (BRB) algorithm}: This algorithm is capable of finding the global optimal solution of \eqref{P0} by using monotonic optimization, yet with extremely high computational complexity $\mathcal{O}(T_{\max3}C_{\max}(N(K+G)+K)^{3.5})$, where $T_{\max3}$ is the number of iterations for the ``BRB'' algorithm to converge and it is very large if a predetermined desired accuracy $\epsilon$ is small \cite[Eq. (30)]{NOUM_1}. 
\end{itemize}

To sum up, Table~\ref{Table-I} shown at the top of the next page compares the computational complexity of the proposed algorithms and traditional ones. It is clear from Table~\ref{Table-I} that when the number of APs (say, $N$) and the number of UEs (say, $K$) are very large, the computational complexity of Algorithm~\ref{Algorithm2} is much less than that of the ``FP-IPM'' method. On the other hand, it is not hard to see that the computational complexity of Algorithm~\ref{Algorithm1} and its accelerated version is much lower than those of the traditional methods including ``IPM'', ``SCS'' and ``BRB''. As $N$ and $K$ increase further, the superiority of Algorithm~\ref{Algorithm1} and its accelerated algorithm becomes more evident, which demonstrate that the proposed algorithms are eminently suitable for massive access scenario.

\begin{table*}[!t]
	\caption{Comparison of Computational Complexity}
	\label{Table-I}
	\centering
		\begin{tabular}{lll}
			\Xhline{1.2pt}
			\textbf{Algorithm} & \textbf{Computational Complexity}  \\
			\Xhline{1.2pt}
			Algorithm~\ref{Algorithm1}  & $\mathcal{O}\left(T_{\max1}C_{\max}\left(N(K+G)+K+\frac{2K+G}{\epsilon_1}\right)\right)$  \\
			Accelerated Algorithm~\ref{Algorithm1}  & $\mathcal{O}\left(T_{\max1}C_{\max}\left(N(K+G)+K+\frac{2K+G}{\sqrt{\epsilon_1}}\right)\right)$  \\
			Algorithm~\ref{Algorithm2}  & $\mathcal{O}\left(T_{\max2}\left(N(K+G)+K+\frac{2K+G}{\epsilon_2}\right)\right)$  \\
			FP-IPM & $\mathcal{O}\left(T_{\max2}(N(K+G)+K)^{3.5}\right)$ \\
			IPM & $\mathcal{O}\left(T_{\max1}C_{\max}(N(K+G)+K)^{3.5}\right)$ \\
			SCS  & $\mathcal{O}\left(T_{\max1}C_{\max}(N^{3.5}(K^{3.5}+G^{3.5})+K^{3.5})\right)$ \\
			BRB & $\mathcal{O}\left(T_{\max3}C_{\max}(N(K+G)+K)^{3.5}\right)$ \\
			\Xhline{1.2pt}
	\end{tabular}	
\end{table*}

\begin{table*}[!t]
	\caption{Simulation Parameter Setting}
	\label{Table-II}
	\centering
	\begin{tabular}{c l l !{\vrule width1.2pt} c l l}
		\Xhline{1.2pt}
		\bf{Symbol}	& \bf{Parameter} & \bf{Value} & \bf{Symbol} &  \bf{Parameter} & \bf{Value}  \\
		\Xhline{1.2pt}
		$D$	& Length of coverage area	& 300 	& $d_{0}$	& Reference distance & 5 m \\
		$\varsigma$	& Path-loss exponent	& 3.76	& $\sigma$	& Shadow fading's standard deviation & 8 dB \\
		$B$	& Transmission bandwidth		& 20 MHz & $K$ & Number of UEs & 12 \\
		$G$	& Number of multicast groups	& 4 & $M$		& Number of antennas at each AP  & 2 \\
		$\rho_{\rm p}$  & Pilot transmit power	& 100 mW		& $\tau_{\rm c}$ & Channel coherent interval  & 200 \\
		$\tau_{\rm p}$  & Length of pilot sequence	& 4 mW	& $P_{0}$ & The sensitivity of energy harvester  & 0.08 mW \\
		$P_{\max}$	& The maximum harvested power	& 37.5 mW	& $\iota_{1}$	& Energy conversion parameter & 116 \\
		$\iota_{2}$	& Energy conversion parameter	& 2.3		& $p_{n}^{\rm ac}$ & Active power at each AP & 10.65 W \\
		$p_{n}^{\rm sl}$	& Sleep power at each AP	& 5.05 W	& $\xi_{n}$ & Power amplifier efficiency & 0.25 \\
		$p_{{\rm bh}, n}$	& Traffic-dependent power of backhaul	& 0.25 W/Gbits & $p_{0, n}$	& Fixed power of backhaul & 0.825 W \\
		$\delta_{k}^2$	& Antenna noise power & -70 dBm	& $\sigma_{k}^2$ & Thermal noise power & -104 dBm \\
		\Xhline{1.2pt}
	\end{tabular}
\end{table*}

\section{Simulation Results and Discussions}
\label{Sec_sim}
In this section, Monte-Carlo simulations are carried out to evaluate the proposed first-order algorithms for energy-efficient resource allocation in LDM-based cell-free massive MIMO systems. All experiments are performed on MATLAB R2019b running on a Windows x64 machine with 3.7 GHz CPU and 32 GB RAM.

In the simulation setup, the APs and UEs are randomly distributed over a coverage area of $D \times D \text{ m}^2$. The large-scale fading coefficient $\beta_{n,k}$ is modeled as the product of path-loss and shadowing, i.e., $\beta_{n, k} = \left(d_{0}/d_{n, k}\right)^{\varsigma}\times 10^{\epsilon_{n, k}/10}$, where $d_{n, k}$ is the distance between the $n^{\rm th}$ AP and the $k^{\rm th}$ UE, $d_{0}$ is the reference distance, and $\varsigma$ is the path-loss exponent and $10^{\epsilon_{n, k}/10}$ captures log-normal shadowing with $\epsilon_{n, k}\sim \mathcal{N}\left(0, \sigma^2\right)$. In agreement with the simulation parameter setting in \cite{CF_MIMO_4, EE_C-RAN, 8322450}, the main system parameters used in our simulation experiments are summarized in Table~\ref{Table-II}. For simplicity, we set $\bar{r}_{{\rm m}, g} = r_{{\rm m}}, \forall g\in\mathcal{G}$; $\bar{r}_{{\rm u}, k} = r_{{\rm u}}$, $\bar{e}_{k} = e_{\min}, \forall k \in \mathcal{K}$; and $\bar{p}_{n, \max} = p_{\max}$ and $\bar{c}_{n, \max} = c_{\max}, \forall n \in \mathcal{N}$.

The step size $\nu_{s}$ used in Algorithm~\ref{Algorithm1} is fixed to $1$ and the step size $\bar{\nu}_s$ in Algorithm~\ref{Algorithm2} is set to $2/(M\sqrt{s})$. The iteration of either algorithm terminates when the relative change of the corresponding objective function between two consecutive iterations is less than $10^{-4}$. All the simulation results are obtained by averaging over $10^{4}$ simulation trials. 

\subsection{Convergence Behavior of the Proposed Algorithms}
\begin{figure}[!t]
	\centering
	\includegraphics[width=3.5in]{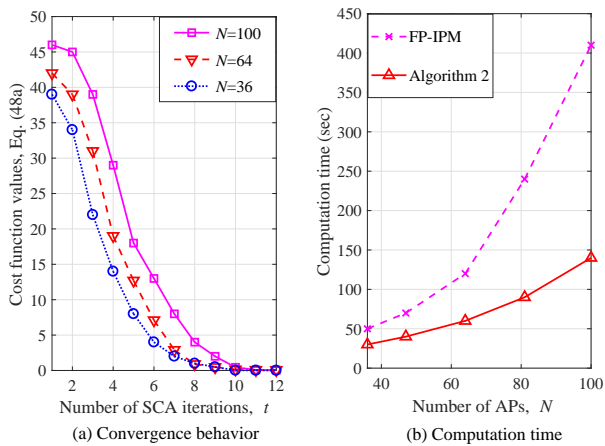}
	\vspace{-15pt}
	\caption{The convergence behavior and computation time of Algorithm~\ref{Algorithm2} ($p_{\max} = 30 \text{ dBm}$, $c_{\max} = 10 \text{ bps/Hz}$, $e_{\min} = 30 \text{ mW}$, and $r_{\rm m} = r_{\rm u} = 0.5 \text{ bps/Hz}$). }
	\label{Fig_2}
\end{figure}

To verify the convergence of Algorithm~\ref{Algorithm2} for finding a feasible point of the problem \eqref{P0}, Fig.~\ref{Fig_2}(a) shows the value of the cost function given by $h(\mathcal{V})$ in \eqref{fea_1} versus the number of SCA iterations. It is seen that the values of \eqref{fea_1} decrease to the global minimum $0$ through up to $12$ SCA iterations under different number of APs (i.e., $N$), which demonstrates that Algorithm~\ref{Algorithm2} can quickly find a feasible initial point of \eqref{P0}. On the other hand, Fig. \ref{Fig_2}(b) shows the computation time versus $N$. It is clear that Algorithm~\ref{Algorithm2} greatly reduces the computation time, compared with the ``FP-IPM'' algorithm. Moreover, the reduction of computation time becomes more evident as $N$ increases. This demonstrates that the proposed Algorithm~\ref{Algorithm2} is more suitable for cell-free massive MIMO systems.

\begin{figure}[t]
	\centering
	\includegraphics[width=3.75in]{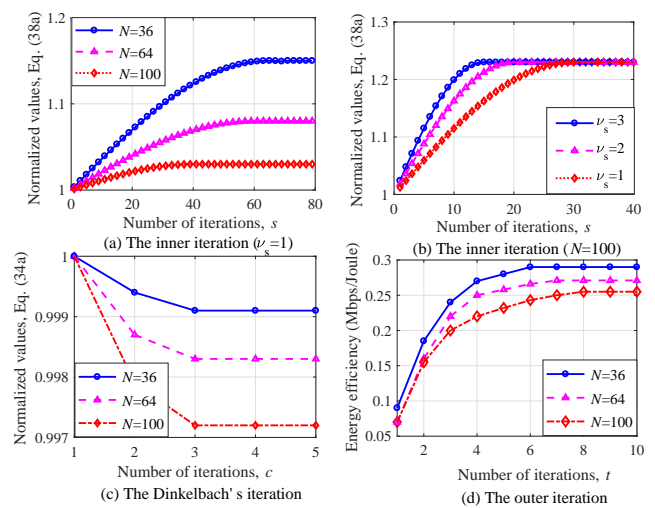}
	\vspace{-15pt}
	\caption{The convergence behavior of the three layer iterations in Algorithm~\ref{Algorithm1} ($p_{\max} = 30 \text{ dBm}$, $c_{\max} = 10 \text{ bps/Hz}$, $e_{\min} = 30 \text{ mW}$, and $r_{\rm m} = r_{\rm u} = 0.5 \text{ bps/Hz}$). }
	\label{Fig_3}
	\vspace{-10pt}
\end{figure}

Figure~\ref{Fig_3} shows the convergence behavior of three layer iterations inherent in Algorithm~\ref{Algorithm1}. In particular, Fig.~\ref{Fig_3}(a) shows the convergence of the inner iteration. It is seen that the inner iterations converge after $60$ iterations in the case of $N=36$, and the convergence speed becomes faster as $N$ increases. Figure~\ref{Fig_3}(b) shows the convergence of inner iteration ($N=100$) with varying step size $\nu_{s}$. It is clear that the convergence becomes faster as $\nu_{s}$ increases. Figure~\ref{Fig_3}(c) illustrates that the Dinkelbach's algorithm achieves convergence with only $2$ iteration, irrespective of the value of $N$. This fast convergence is due to the super-linear convergence rate of the Dinkelbach's algorithm \cite{dinkelbach1967nonlinear} and a warm start provided by the last SCA loop. Finally, Fig. \ref{Fig_3}(d) shows that the outer iteration converges within $10$ iterations even if $N=100$.

\begin{figure}[t]
	\centering
	\includegraphics[width=3.125in]{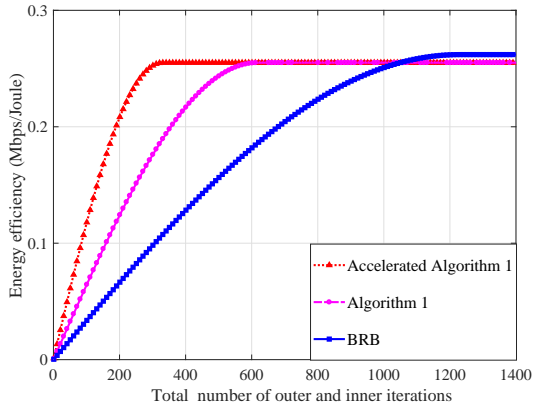}
	\caption{The convergence behavior of Algorithm~\ref{Algorithm1} and its accelerated algorithm ($N=100$, $p_{\max}=30 \text{ dBm}$, $c_{\max}=10 \text{ bps/Hz}$, $e_{\min}=30 \text{ mW}$, and $r_{\rm m}=r_{\rm u}=0.5 \text{ bps/Hz}$).}
	\label{Fig_4}
\end{figure}

Figure~\ref{Fig_4} compares the convergence behavior of the proposed algorithms (including Algorithm~\ref{Algorithm1} and its accelerated version) and the optimal ``BRB'' algorithm. It is seen that the optimal BRB algorithm gets the energy efficiency about $0.262\ {\rm Mbps/Joule}$ while the proposed algorithms get that about $0.255\ {\rm Mbps/Joule}$. Clearly, the loss of energy efficiency caused by the smooth approximations \eqref{C_fh22}-\eqref{P_tt2} is only $2.67\%$ but the proposed algorithms converge more than twice as fast as the BRB algorithm.

\subsection{Computational Complexity}
\begin{figure}[t]
	\centering
	\includegraphics[width=3.5in]{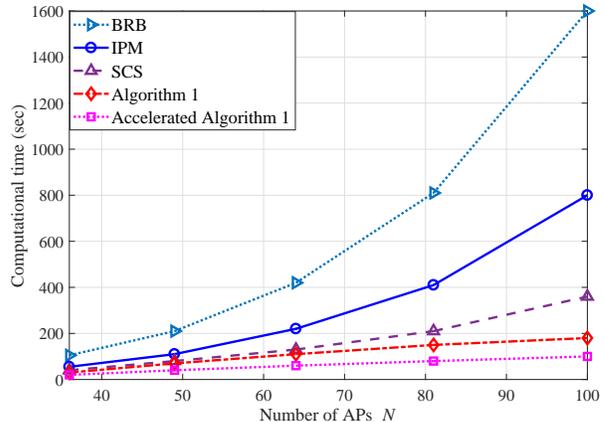}
	\caption{The computation time versus $N$ ($p_{\max}=30 \text{ dBm}$,  and $c_{\max}=10 \text{ bps/Hz}$, $e_{\min}=30 \text{ mW}$, $r_{\rm m}=r_{\rm u}=0.5 \text{ bps/Hz}$). }
	\label{Fig_5}
\end{figure}

To verify the complexity of the proposed algorithms, besides the computational complexity summarized in Table~\ref{Table-I}, Fig.~\ref{Fig_5} illustrates the computation time of the proposed algorithms, in comparison with that of ``IPM'', ``SCS'' and ``BRB'' algorithms. It is clear that, for each algorithm the computation time increases with the number of APs (i.e., $N$), as expected. However, the proposed Algorithm~\ref{Algorithm1} has much shorter computation time than the traditional second-order algorithms (i.e., ``IPM'', ``SCS'' and BRB'' algorithms), and the accelerated algorithm spends the shortest computation time, which corroborates with the complexity analysis in Section V-C. The computation time of the optimal ``BRB'' algorithm is the longest as it is in essence exhaustive searching. In view of the results of Table~\ref{Table-I} and Fig.~\ref{Fig_5}, the proposed algorithms have the lowest computational complexity and the shortest computation time.

\subsection{Energy Efficiency}

\begin{figure}[t]
	\centering
	\includegraphics[width=3.5in]{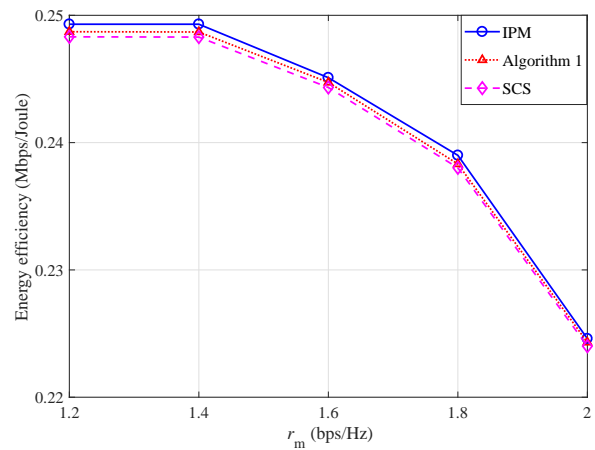}
	\caption{The energy efficiency versus $r_{\rm m}$ ($N=100$, $p_{\max}=30 \text{ dBm}$, $r_{\rm u}=0.5 \text{ bps/Hz}$, $c_{\max}=10 \text{ bps/Hz}$, and $e_{\min}=30 \text{ mW}$).  }
	\label{Fig_6}
\end{figure}

Figure~\ref{Fig_6} illustrates the EE versus the minimum data rate requirement for multicast service (say, $r_{\rm m}$). As Algorithm~\ref{Algorithm1} obtains the same  EE as its accelerated algorithm (cf. Fig.~\ref{Fig_4}), and the optimal ``BRB'' algorithm achieves almost the same EE as the proposed first-order algorithms yet with extremely high complexity (cf. Fig.~\ref{Fig_4} and Table~\ref{Table-I}), both the accelerated algorithm and the optimal ``BRB'' algorithm are not included in the figure for brevity. It is seen from Fig.~\ref{Fig_6} that, the ``IPM'' algorithm obtains a slightly higher EE than those of Algorithm~\ref{Algorithm1} and the ``SCS'' algorithm. This is because Algorithm~\ref{Algorithm1} is an iterative algorithm that solves each SCA subproblem \eqref{P_tt22}. Hence, the accuracy of the obtained optimal solution is affected by the predefined convergence tolerance of each iteration layer. Moreover, the ``SCS'' algorithm requires a large number of auxiliary variables for standard form transformation, and hence the resulting optimal solution is affected. On the other hand, it is observed from Fig.~\ref{Fig_6}  that, for each scheme, the EE remains constant if $r_{\rm m}$ is small whereas it decreases as $r_{\rm m}$ turns large. The reason behind this observation is that, when $r_{\rm m}$ is small, the resource allocation that maximizes the EE can also easily satisfy the data rate requirement of multicast service; however, when $r_{\rm m}$ is large, more APs need to be active and more power has to be transmitted so as to satisfy the data rate requirement, which incurs more power consumption and hence decreases the EE. This also implies that almost all the APs need to be switched on when the data rate requirements are extremely high. Finally, it is remarkable that, as the minimum data rate requirement for unicast service (say, $r_{\rm u}$) has a similar impact on the EE as that of $r_{\rm m}$, it is not plotted for brevity. 

\begin{figure}[t]
	\includegraphics[width=3.5in]{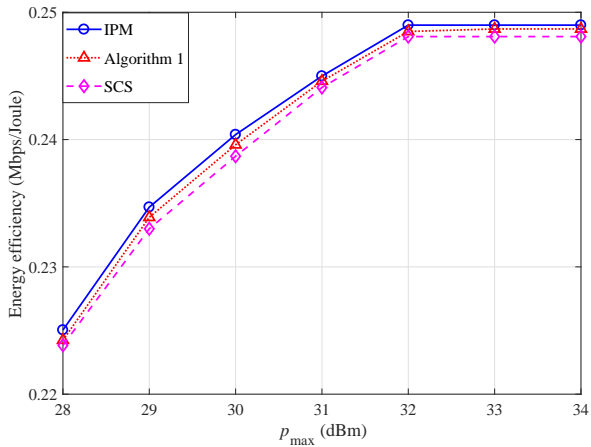}
	\centering{}
	\caption{The energy efficiency versus $p_{\max}$ ($N = 100$, $c_{\max} = 10 \text{ bps/Hz}$, $e_{\min} = 30 \text{ mW}$, and $r_{\rm m} = r_{\rm u}=0.5 \text{ bps/Hz}$). }
	\label{Fig_7}
\end{figure}

Figure~\ref{Fig_7} shows the EE versus the allowable maximal transmit power at APs (i.e., $p_{\max}$). As seen, the proposed Algorithm~\ref{Algorithm1} obtains almost the same EE as the second-order ``IPM'' and ``SCS'' algorithms. Their EEs increase as $p_{\max}$ but when $p_{\max} > 32\ \text{dBm}$, further increasing $p_{\max}$ does not benefit higher EE. The reason behind this observation is that the sum data rate first increases with transmit power consumption, yielding an increasing EE. However, when the power budget becomes sufficiently large, the gain of sum data rate cannot compensate for the sharp increase of the power consumption. Consequently, the EE becomes saturated when the transmit power is large enough.

\begin{figure}[t]
	\includegraphics[width=3.5in]{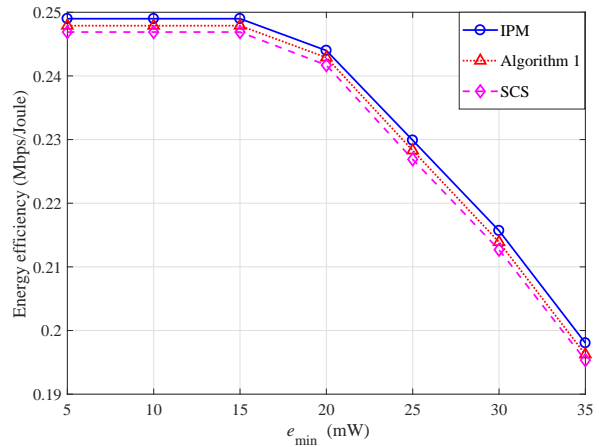}
	\centering{}
	\caption{The energy efficiency versus $e_{\min}$ ($N = 100$, $p_{\max} = 30 \text{ dBm}$, $c_{\max} = 10 \text{ bps/Hz}$, and $r_{\rm m} = r_{\rm u} = 0.5 \text{ bps/Hz}$.}
	\label{Fig_8}
\end{figure}

Figure~\ref{Fig_8} illustrates the EE versus the minimum harvested energy required by each UE (i.e., $e_{\min}$). Like Fig.~\ref{Fig_6}, the EE keeps constant when $e_{\min}$ is relatively small but it decreases when $e_{\min} > 15 \text{ mW}$. This is because larger $e_{\min}$ means more power is consumed by UEs themselves, such as internal circuits. Consequently, this will decrease the power dedicated to data transmission, thus yielding lower EE. In other words, developing low-power consumption terminals benefits higher EE in future massive access networks. 

\section{Conclusions}
\label{Sec_con}
In this paper, we designed an energy-efficient resource allocation for non-orthogonal multicast and unicast transmission in cell-free massive MIMO systems with SWIPT. To suit massive access applications, first-order algorithms were designed for obtaining both the initial point and the final solution, which are Hessian-free and involve only the first-order gradient information and, thus, have very low computational complexity. Moreover, to improve the convergence speed of the proposed first-order algorithm, an accelerated algorithm was developed. Simulation results demonstrate that the proposed algorithms achieve almost the same energy efficiency as the traditional second-order algorithms. In addition, the impacts of some key system parameters on the energy efficiency were disclosed. Thanks to the fast convergence speed and lower computational complexity, the proposed first-order algorithms are promising in massive access applications, such as intensive IoT networks. For further research, in light of the limited capacity of backhaul links in practical cell-free MIMO systems, it is essential to jointly optimize the backhaul compression and transmit beamforming for maximal energy efficiency. Also, the joint AP clustering and UE scheduling as well as beamforming design is promising to satisfy future massive connectivity in real-world applications.

\numberwithin{equation}{section}
\appendices
\section{Channel Estimation}
\label{Appendix-A}
In the phase of channel estimation, each AP autonomously acquires the CSI between itself and all UEs through uplink training. The APs do not cooperate on the channel estimation and no channel estimates are interchanged among the APs. To reduce the overhead of pilot sequences, a common pilot is assigned to all UEs in the same multicast group whereas the pilots assigned to the UEs within different groups are mutually orthogonal. Therefore, only $G$ pilots are required to acquire the CSI, with $G \le K$. Applying the minimum mean squared error (MMSE) channel estimation criterion \cite{CE_11}, the $n^{\rm th}$ AP can estimate $\bm{g}_{n,k}$ as
\begin{equation}
	\hat{\bm{g}}_{n,k}
	=\bar{\beta}_{n,k}\Big(\sum_{j \in \mathcal{K}_{g_{k}}}\sqrt{\tau_{\rm p}\rho_{\rm p}}\bm{g}_{n,j}+\bm{n}_{n,k}\Big),\label{CE_1}
\end{equation}
where $\rho_{\rm p}$ and $\tau_{\rm p}$ are the transmit power and length of each pilot sequence, respectively, $\bm{n}_{n,k}\sim \mathcal{CN}\left(\bm{0},\bm{I}_{M}\right)$ is the normalized additive Gaussian noise, $\bar{\beta}_{n,k}=\sqrt{\tau_{\rm p}\rho_{\rm p}}\beta_{n,k}/\left(1+\tau_{\rm p}\rho_{\rm p}\varrho_{n,g_{k}}\right)$ 
with $\varrho_{n,g_{k}}=\sum_{j\in \mathcal{K}_{g_{k}}}\beta_{n,j}$. Accordingly,  $\hat{\bm{g}}_{n,k} \sim\mathcal{CN}\left(\bm{0},\hat{\beta}_{n,k}\bm{I}_{M}\right)$
with $\hat{\beta}_{n,k}=\tau_{\rm p}\rho_{\rm p}\beta_{n,k}^{2}/\left(1+\tau_{\rm p}\rho_{\rm p}\varrho_{n,g_{k}}\right)$. Given \eqref{CE_1}, it is clear that $\hat{\bm{g}}_{n,k}$ is contaminated by other UEs within the same group.

By using a similar approach as above, we can estimate the linear combination of the channels of all UEs within the $g^{\rm th}$ group (i.e., $\bm{f}_{n,g}=\sum_{k\in\mathcal{K}_{g}}\bm{g}_{n,k}$):
\begin{equation}
	\hat{\bm f}_{n,g}  =\bar{\alpha}_{n,g}\Big(\sqrt{\tau_{\rm p}\rho_{\rm p}}\sum_{k\in\mathcal{K}_{g}}\bm{g}_{n,k}+\bm{n}_{n,g}\Big),\label{GCE_1}
\end{equation}
where $\bar{\alpha}_{n,g}=\sqrt{\tau_{\rm p}\rho_{\rm p}}\varrho_{n,g}/\left(1+\tau_{\rm p}\rho_{\rm p}\varrho_{n,g}\right)$
and $\hat{\bm{f}}_{n,g}$ is distributed as $\hat{\bm{f}}_{n,g}\sim \mathcal{CN}\left(0,\hat{\alpha}_{n,g}\bm{I}_{M}\right)$
with $\hat{\alpha}_{n,g}=\tau_{\rm p}\rho_{\rm p}\varrho_{n,g}^{2}/\left(1+\tau_{\rm p}\rho_{\rm p}\varrho_{n,g}\right)$.

\section{Proof of Theorem 1}
\label{Appendix-B}
Firstly, we derive $R_{k}(\mathcal{V})$. In light of \eqref{rate_m}, we need to compute ${\rm D}_{{\rm m},g_{k}}$, $\mathbb{E}\left[|{\rm V}_{{\rm m},\tilde{g}_{k}}|^2\right]$, $\mathbb{E}\left[|{\rm I}_{{\rm m},\tilde{g}_{k}}|^2\right]$, and $\mathbb{E}\left[|{\rm I}_{\rm u}|^2\right]$. 
\begin{itemize}
\item Compute ${\rm D}_{{\rm m},g_{k}}$: Substituting \eqref{PM_1} and \eqref{GCE_1} into \eqref{DSg} yields
\begin{align}
	{\rm D}_{{\rm m},g_{k}} 
	& =\sum_{n\in\mathcal{N}}\sqrt{\frac{q_{n,g_{k}}}{M\hat{\alpha}_{n,g_{k}}}}\mathbb{E}\left[\bm{g}_{n,k}^{\rm H}
	\hat{\bm{f}}_{n,g_{k}}\right]\nonumber\\
	& =\sum_{n\in\mathcal{N}}\sqrt{\frac{q_{n,g_{k}}}{M\hat{\alpha}_{n,g_{k}}}}\mathbb{E}\bigg[\bm{g}_{n,k}^{\rm H}
	\bar{\alpha}_{n,g_{k}}\Big(\sum_{j\in\mathcal{K}_{g_{k}}}\sqrt{\tau_{\rm p}\rho_{\rm p}}\bm{g}_{n,j}\nonumber\\
	&\quad{}+\bm{n}_{n,g_{k}}\Big)\bigg]=M^{\frac{1}{2}}\sum_{n\in \mathcal{N}}q_{n,g_{k}}^{\frac{1}{2}}\hat{\beta}_{n,k}^{\frac{1}{2}}.\label{DS_1A}
\end{align}
	
\item Compute $\mathbb{E}[|{\rm V}_{{\rm m},g_{k}}|^2]$: Substituting \eqref{PM_1}, \eqref{GCE_1} and \eqref{DS_1A} into \eqref{Vgk}, we have
\begin{align}
	\lefteqn{\mathbb{E}[|{\rm V}_{{\rm m},g_{k}}|^2]} \nonumber \\
	&= \sum_{n\in\mathcal{N}}\!\!q_{n,g_{k}}\left(\mathbb{E}\!\!\left[|\bm{g}_{n,k}^{\rm H}
	\bm{w}_{n,g_{k}}|^{2}\right]-\mathbb{E}^{2}\!\!\left[\bm{g}_{n,k}^{\rm H}\bm{w}_{n,g_{k}}\right]\right)\nonumber\\
	&= \sum_{n\in\mathcal{N}}\frac{q_{n,g_{k}}\bar{\alpha}_{n,g_{k}}^{2}}{M\hat{\alpha}_{n,g_{k}}}\Bigg(\mathbb{E}
	\Bigg[\bigg|\bm{g}_{n,k}^{\rm H}\Big(\sum_{j\in\mathcal{K}_{g_{k}}}\sqrt{\tau_{\rm p}\rho_{\rm p}}
	\bm{g}_{n,j}\nonumber\\
	&\quad{}+\bm{n}_{n,g_{k}}\Big)\bigg|^{2}\Bigg]-M^{2}\tau_{{\rm p}}\rho_{\rm p}\beta_{n,k}^{2}\Bigg)\nonumber\\
	&= \sum_{n\in\mathcal{N}}\frac{q_{n,g_{k}}\bar{\alpha}_{n,g_{k}}^{2}\beta_{n,k}}{\hat{\alpha}_{n,g_{k}}}
	\Big(\tau_{\rm p}\rho_{\rm p}\sum_{j\in\mathcal{K}_{g_{k}}}\beta_{n,j}+1\Big)\nonumber\\
	&= \sum_{n\in\mathcal{N}}q_{n,g_{k}}\beta_{n,k}.\label{V_1A}
\end{align}
	
\item Compute $\mathbb{E}[|{\rm I}_{{\rm m},\tilde{g}_{k}}|^2]$: Since $\bm{g}_{n,k}$ and $\bm{w}_{g}$ are independent if $g\neq g_{k}$, we have
\begin{align}
	\mathbb{E}[|{\rm I}_{{\rm m},\tilde{g}_{k}}|^2] 
	&=\sum_{g\in\mathcal{G}\setminus g_{k}}\sum_{n\in\mathcal{N}}\frac{q_{n,g}}{M\hat{\alpha}_{n,g}}
	\mathbb{E}\left[|\bm{g}_{n,k}^{\rm H}\hat{\bm{f}}_{n,g}|^{2}\right]\nonumber\\
	&=\sum_{g\in\mathcal{G\setminus}g_{k}}\sum_{n\in\mathcal{N}}\frac{q_{n,g}}{M\hat{\alpha}_{n,g}}
	M\beta_{n,k}\hat{\alpha}_{n,g}\nonumber\\
	&=\sum_{g\in\mathcal{G\setminus}g_{k}}\sum_{n\in\mathcal{N}}q_{n,g}\beta_{n,k}.\label{MUI_1A}
\end{align}
	
\item Compute $\mathbb{E}\left[|{\rm I}_{{\rm u}}|^2\right]$: It can be divided into two parts:
\begin{align}
	\mathbb{E}\left[|{\rm I}_{{\rm u}}|^2\right]
	& =\underbrace{\sum_{j\in\mathcal{K}_{g_{k}}}\mathbb{E}\left[\left|\sum_{n\in\mathcal{N}}\bm{g}_{n,k}^{\rm H}
	\sqrt{p_{n,j}}\bm{v}_{n,j}\right|^{2}\right]}_{\ell_{1}}\nonumber\\
	&\quad{}+\underbrace{\sum_{j\in\mathcal{K}
	\setminus\mathcal{K}_{g_{k}}}\mathbb{E}\left[\left|\sum_{n\in\mathcal{N}}\bm{g}_{n,k}^{\rm H}
	\sqrt{p_{n,j}}\bm{v}_{n,j}\right|^{2}\right]}_{\ell_{2}}.\nonumber
\end{align}
As $\bm{g}_{n,k}$ is not independent of $\hat{\bm{g}}_{n,j}$ if $j\in{\cal K}_{g_{k}}$, we have
\begin{align}
	{\ell}_{1} 
	& =\sum_{j\in\mathcal{K}_{g_{k}}}\sum_{n\in\mathcal{N}}\frac{p_{n,j}}{M\hat{\beta}_{n,j}}\mathbb{E}
	\left[|\bm{g}_{n,k}^{\rm H}\hat{\bm{g}}_{n,j}|^{2}\right]\nonumber\\
	&=\sum_{j\in\mathcal{K}_{g_{k}}}\sum_{n\in\mathcal{N}}\frac{p_{n,j}}{M\hat{\beta}_{n,j}}\bar{\beta}_{n,j}^2
	\mathbb{E}\Bigg[\bigg|\bm{g}_{n,k}^{\rm H}\Big(\sum_{m\in\mathcal{K}_{g_{k}}}
	\sqrt{\tau_{\rm p}\rho_{\rm p}}\bm{g}_{n,j}\nonumber\\
	&\quad{}+\bm{n}_{n,g_{k}}\Big)\bigg|^{2}\Bigg]\nonumber\\
	&=\sum_{j\in\mathcal{K}_{g_{k}}}\sum_{n\in\mathcal{N}}\frac{p_{n,j}\bar{\beta}_{n,j}^{2}}{M\hat{\beta}_{n,j}}
	\Big(\tau_{\rm p}\rho_{\rm p}M\beta_{n,k}\sum_{m\in\mathcal{K}_{g_{k}}}\beta_{n,m}\nonumber\\
	&\quad{}+\tau_{\rm p}\rho_{\rm p}M^{2}\beta_{n,k}^{2}+M\beta_{n,k}\Big)\nonumber\\
	& =\sum_{j\in\mathcal{K}_{g}}\sum_{n\in\mathcal{N}}p_{n,j}\left(\beta_{n,k}+M\hat{\beta}_{n,k}\right).\label{NUI_1A1}
\end{align}
	Also, as $\bm{g}_{n,k}$ and $\bm{v}_{n,j}$ are independent if $j \in \mathcal{K}\setminus\mathcal{K}_{g_{k}}$, we get
\begin{align}
	{\ell}_{2}
	&=\sum_{j\in\mathcal{K}\setminus\mathcal{K}_{g_{k}}}\sum_{n\in\mathcal{N}}
	\frac{p_{n,j}}{M\hat{\beta}_{n,j}}\mathbb{E}\left[|\bm{g}_{n,k}^{\rm H}\hat{\bm{g}}_{n,j}|^{2}\right]\nonumber\\
	&{}=\sum_{j\in\mathcal{K}\setminus\mathcal{K}_{g_{k}}}\sum_{n\in\mathcal{N}}p_{n,j}
	\beta_{n,k}.\label{NUI_1A2}
\end{align}
By virtue of \eqref{NUI_1A1} and \eqref{NUI_1A2}, we obtain
\begin{equation}\label{NUI_1A}
	\mathbb{E}[|{\rm I}_{{\rm u}}|^2]=\sum_{j\in\mathcal{K}}\sum_{n\in\mathcal{N}}p_{n,j}\beta_{n,k}
	+M\sum_{j\in\mathcal{K}_{g_{k}}}\sum_{n\in\mathcal{N}}p_{n,j}\hat{\beta}_{n,k}.
\end{equation}
	
Substituting \eqref{DS_1A}-\eqref{MUI_1A} and \eqref{NUI_1A} into \eqref{rate_m} and after performing some algebraic manipulations, we attain \eqref{RA_M}.
\end{itemize}
On the other hand, using a similar analysis to \eqref{RA_M}, we can obtain the closed-form expression \eqref{RA_U} for $R_{{\rm u},k}(\mathcal{V})$. This completes the proof.

\section{Proof of Proposition 2}
\label{Appendix-C}
The dual function of \eqref{P_ptt22} is defined as
\begin{align}
	\mathcal{D}(\mathcal{L}) 
	& \triangleq \min_{\mathcal{V},\mathcal{R}}\ \!\! \mathcal{M}\left(\mathcal{V},\mathcal{R},\mathcal{L}\right)\nonumber\\
	& =\min_{\mathcal{V},\mathcal{R}}\ \!\! \Bigg\{\Gamma(\mathcal{V})+\sum_{g \in \mathcal{G}}\lambda_{{\rm m},g}
	\bar{r}_{{\rm m},g}+\sum_{k \in \mathcal{K}}\bigg[-\bar{\lambda}_{{\rm m},k} \bar{R}_{k}^{(t)}(\mathcal{V})\nonumber\\
	&\quad{}+\lambda_{{\rm u},k}(\bar{r}_{{\rm u},k}-\bar{R}_{{\rm u},k}^{(t)}(\mathcal{V}))+\lambda_{{\rm e},k}\big(\frac{\mathcal{F}^{-1}(\bar{e}_{k})}{1-\rho_{k}}-\bar{\varphi}_{{\rm e},k}^{(t)}(\mathcal{V})\big)\bigg]\nonumber\\
	&\quad{}+\sum_{n\in\mathcal{N}}\!\bigg[\lambda_{{\rm p},n}\left(p_{{\rm tr},n}(\mathcal{V})-\bar{p}_{n,\max}\right)+\lambda_{{\rm c},n}\big(\bar{C}_{{\rm bh},n}^{(t)}(\mathcal{V})\nonumber\\
	&\quad{}-\bar{c}_{n,\max}\big)\bigg]+\sum_{g\in\mathcal{G}}\bigg[\sum_{k\in\mathcal{K}_{g}}\bar{\lambda}_{{\rm m},k}-\eta^{(c)}
	\big(1-\frac{\tau_{\rm p}}{\tau_{\rm c}}\big)\nonumber\\
	&\quad{}-\lambda_{{\rm m},g}\bigg]\mathcal{R}_{{\rm m},g}\Bigg\},\label{2.1}
\end{align}
where $\mathcal{M}\left(\mathcal{V},\mathcal{R},\mathcal{L}\right)$ is the Lagrangian function of problem \eqref{P_ptt22}. Since $\mathcal{M}\left(\mathcal{V},\mathcal{R},\mathcal{L}\right)$ in \eqref{2.1} is strongly convex over $\mathcal{V}$ and linear over $\mathcal{R}$, the minimum of $\mathcal{M}\left(\mathcal{V},\mathcal{R},\mathcal{L}\right)$ over $\mathcal{V, R}$ is finite if 
$\sum_{k\in\mathcal{K}_{g}}\bar{\lambda}_{{\rm m},k}-\eta^{(c)}\left(1-\tau_{\rm p}/\tau_{\rm c}\right)-\lambda_{{\rm m},g}=0, \forall g \in \mathcal{G}$. 
Otherwise, we have two other cases, i.e., 
$\sum_{k \in \mathcal{K}_{g}}\bar{\lambda}_{{\rm m},k}-\eta^{(c)}\left(1-\tau_{\rm p}/\tau_{\rm c}\right)-\lambda_{{\rm m},g}>0\left(\text{or} <0\right)$. 
Since there are no constraints on $\mathcal{R}_{{\rm m},g}$ in \eqref{2.1},  $\mathcal{D}(\mathcal{L})$ would be infinite, if $\mathcal{R}_{{\rm m},g}$ is infinite and $\sum_{k \in \mathcal{K}_{g}}\bar{\lambda}_{{\rm m},k}-\eta^{(c)}\left(1-\tau_{\rm p}/\tau_{\rm c}\right)-\lambda_{{\rm m},g}\neq0$. 
Therefore, $\mathcal{D}(\mathcal{L})$ is finite if and only if
$\sum_{k \in \mathcal{K}_{g}}\bar{\lambda}_{{\rm m},k}-\eta^{(c)}\left(1-\tau_{\rm p}/\tau_{\rm c}\right)-\lambda_{{\rm m},g}=0,\forall g \in \mathcal{G}$. 
Since the domain of $\mathcal{D}(\mathcal{L})$ is defined as the constraint set over $\mathcal{L}$ such that $\mathcal{D}(\mathcal{L})$ is finite, together with the non-negativity of the dual variables, we obtain the domain of $\mathcal{D}(\mathcal{L})$ as shown in \eqref{dom}.

Next, by substituting $\Gamma(\mathcal{V})$, $\bar{R}^{(t)}_{k}(\mathcal{V})$, $\bar{R}^{(t)}_{{\rm u},k}(\mathcal{V})$, $\bar{\varphi}^{(t)}_{{\rm e},k}(\mathcal{V})$, $\bar{C}^{(t)}_{{\rm bh},n}(\mathcal{V})$ and $p_{{\rm tr},n}(\mathcal{V})$ into \eqref{2.1}, and noticing that $\sum_{k \in \mathcal{K}_{g}}\bar{\lambda}_{{\rm m},k}-\eta^{(c)}\left(1-\tau_{\rm p}/\tau_{\rm c}\right)-\lambda_{{\rm m},g}=0$, dropping the constants independent of $\mathcal{V}$, the problem \eqref{2.1} becomes \eqref{2.2}, shown at the top of the next page.  Due to the  separability of the objective function in \eqref{2.2}, the problem \eqref{2.2} can be decomposed into $2K+G$ parallel subproblems. Specifically, there are $G$ subproblems over $\left\{ \bar{\bm{q}}_{g}\right\} _{g\in \mathcal{G}}$, with each written as \eqref{2.3}, shown in the middle of the next page. As the objective function in \eqref{2.3} is strongly convex over $\bar{\bm q}_{g}$, by setting its gradient to zeros, the minimizer $\bar{\bm q}_{g}^{\circ}$ is uniquely given by \eqref{dp_m1}. Moreover, there are $K$ subproblems over $\left\{ \bar{\bm{p}}_{k}\right\} _{k\in \mathcal{K}}$, with each written as \eqref{2.4}, shown in the middle of of the next page.

\begin{figure*}[tbh]
	\begin{align}
		\min_{\mathcal{V}}\  
		& \sum_{n \in \mathcal{N}}\left[\xi_{n}^{-1}\left(\sum_{j \in \mathcal{K}}\bar{\bm{p}}_{j}^{\rm H}
		\bm{E}_{n}\bar{\bm{p}}_{j}+\sum_{g \in \mathcal{G}}\bar{\bm{q}}_{g}^{\rm H}\bm{E}_{n}\bar{\bm{q}}_{g}\right)+\left(\sum_{j \in \mathcal{K}}
		\bm{\varkappa}_{{\rm u},n,j}^{(t)}\bar{\bm{p}}_{j}+\sum_{g \in \mathcal{G}}\bm{\varkappa}_{{\rm m},n,g}^{(t)}
		\bar{\bm{q}}_{g}\right)\Delta p_{n}\right]\nonumber\\
		&{}+\sum_{n \in \mathcal{N}}\left[p_{{\rm bh},n}\left(\sum_{g \in \mathcal{G}}\bm{\vartheta}_{{\rm 	m},n,g}^{(t)}
		\bar{\bm{q}}_{g}\mathcal{R}_{{\rm m},g}^{(t)}+\sum_{j \in \mathcal{K}}\bm{\vartheta}_{{\rm u},n,j}^{(t)}
		\bar{\bm{p}}_{j}R_{{\rm u},j}(\mathcal{V}^{(t)})\right)\right]-\eta^{(c)}\left(1-\frac{\tau_{\rm p}}{\tau_{\rm c}}\right)\nonumber\\
		&{}\times\sum_{j \in \mathcal{K}}\left[2\bm{\psi}_{{\rm u},j}^{(t)}\bar{\bm{p}}_{j}-\phi_{{\rm u},j}^{(t)}\left(\varphi_{{\rm u},j}(\mathcal{V})
		+M|\hat{\bm{\xi}}_{j}^{\rm H}\bar{\bm{p}}_{j}|^{2}\right)\right]-\sum_{k \in \mathcal{K}}\bar{\lambda}_{{\rm m},k}\big[2\bm{\psi}_{{\rm m},g_{k}}^{(t)}\bar{\bm{q}}_{g_{k}}
		-\phi_{{\rm m},g_{k}}^{(t)}\big(\varphi_{{\rm m},g_{k}}(\mathcal{V})\nonumber\\
		&{}+M|\hat{\bm{\xi}}_{k}^{\rm H}\bar{\bm{q}}_{g_{k}}|^{2}\big)\big]-\sum_{k \in \mathcal{K}}\lambda_{{\rm u},k}\left[\bm{\psi}_{{\rm u},k}^{(t)}\bar{\bm{p}}_{k}-\phi_{{\rm u},k}^{(t)}\left(\varphi_{{\rm u},k}(\mathcal{V})+M|\hat{\bm{\xi}}_{k}^{\rm H}\bar{\bm{p}}_{k}|^{2}\right)
		\right]+\sum_{k \in \mathcal{K}}\lambda_{{\rm e},k}\Big(\frac{\mathcal{F}^{-1}\left(\bar{e}_{k}\right)}{1-\rho_{k}}\nonumber\\
		&{}-2\sum_{g \in \mathcal{G}}\bar{\bm{q}}_{g}^{(t){\rm H}}\bm{\Xi}_{k}^{2}\bar{\bm{q}}_{g}
		-2M\bar{\bm{q}}_{g_{k}}^{(t){\rm H}}\hat{\bm{\Xi}}_{k}^{2}\bar{\bm{q}}_{g_{k}}-2\sum_{j \in \mathcal{K}}
		\bar{\bm{p}}_{j}^{(t){\rm H}}\bm{\Xi}_{k}^{2}\bar{\bm{p}}_{j}\Big)-\sum_{k \in \mathcal{K}}\lambda_{{\rm e},k}\left(2M\sum_{j \in \mathcal{K}_{g_{k}}}\bar{\bm{p}}_{j}^{(t){\rm H}}\hat{\bm{\Xi}}_{k}^{2}\bar{\bm{p}}_{j}\right)\nonumber\\
		&{}+\sum_{n \in \mathcal{N}}\lambda_{{\rm p},n}\left(\sum_{j \in \mathcal{K}}\bar{\bm{p}}_{j}^{\rm H}\bm{E}_{n}\bar{\bm{p}}_{j}
		+\sum_{g \in \mathcal{G}}\bar{\bm{q}}_{g}^{{\rm H}}\bm{E}_{n}\bar{\bm{q}}_{g}\right)+\sum_{n \in \mathcal{N}}\lambda_{{\rm c},n}\left(\sum_{g \in \mathcal{G}}\bm{\vartheta}_{{\rm m},n,g}^{(t)}\bar{\bm{q}}_{g}\mathcal{R}_{{\rm m},g}^{(t)}+\sum_{j \in \mathcal{K}}\bm{\vartheta}_{{\rm u},n,j}^{(t)}
		\bar{\bm{p}}_{j}R_{{\rm u},j}(\mathcal{V}^{(t)})\right).\label{2.2}
	\end{align}
	\rule[0.5ex]{2\columnwidth}{0.5pt}
\end{figure*}

\begin{figure*}[tbh]
\begin{align}
	\min_{\bar{\bm{q}}_{g}}\  
	& \sum_{n \in \mathcal{N}}\left[\left(\xi_{n}^{-1}+\lambda_{{\rm p},n}\right)\bar{\bm{q}}_{g}^{\rm H}
	\bm{E}_{n}\bar{\bm{q}}_{g}+\Delta p_{n}\bm{\varkappa}_{{\rm m},n,g}^{(t)}\bar{\bm{q}}_{g}+\left(p_{{\rm bh},n}+\lambda_{{\rm c},n}\right)
	\mathcal{R}_{{\rm m},g}^{(t)}	\bm{\vartheta}_{{\rm m},n,g}^{(t)}\bar{\bm{q}}_{g}\right]\nonumber\\
	&{}+\sum_{j \in \mathcal{K}}\left[\eta^{(c)}\left(1-\frac{\tau_{\rm p}}{\tau_{\rm c}}\right)+\lambda_{{\rm u},j}\right]
	\phi_{{\rm u},j}^{(t)}\|\bm{\Xi}_{j}\bar{\bm{q}}_{g}\|^{2}+\sum_{j \notin \mathcal{K}_{g}}\bar{\lambda}_{{\rm m},j}\phi_{{\rm m},g_{j}}^{(t)}
	\|\bm{\Xi}_{j}\bar{\bm{q}}_{g}\|^{2}-\sum_{j \in \mathcal{K}_{g}}\bar{\lambda}_{{\rm m},j}\Big[2\bm{\psi}_{{\rm m},g_{j}}^{(t)}\bar{\bm{q}}_{g}\nonumber\\
	&{}-\phi_{{\rm m},g_{j}}^{(t)}\left(M|\hat{\bm{\xi}}_{j}^{\rm H}\bar{\bm{q}}_{g}|^{2}
	+\|\bm{\Xi}_{j}\bar{\bm{q}}_{g}\|^{2}\right)\Big]-2\sum_{j \in \mathcal{K}_{g}}\lambda_{{\rm e},j}\bar{\bm{q}}_{g}^{(t){\rm H}}\left(
	\bm{\Xi}_{j}^{2}+M\hat{\bm{\Xi}}_{j}^{2}\right)\bar{\bm{q}}_{g}-2\sum_{j \notin \mathcal{K}_{g}}
	\lambda_{{\rm e},j}\bar{\bm{q}}_{g}^{(t){\rm H}}\bm{\Xi}_{j}^{2}\bar{\bm{q}}_{g}.\label{2.3}
\end{align}
\rule[0.5ex]{2\columnwidth}{0.5pt}
\end{figure*}

\begin{figure*}[tbh]
\begin{align}
	\min_{\bar{\bm{p}}_{k}} 
	&\sum_{n \in \mathcal{N}}\left[\left(\xi_{n}^{-1}+\lambda_{{\rm p},n}\right)\bar{\bm{p}}_{k}^{\rm H}
	\bm{E}_{n}\bar{\bm{p}}_{k}+\Delta p_{n}\bm{\varkappa}_{{\rm u},n,k}^{(t)}\bar{\bm{p}}_{k}
	+\left(\lambda_{{\rm c},n}+p_{{\rm bh},n}\right)R_{{\rm u},k}(\mathcal{V}^{(t)})
	\bm{\vartheta}_{{\rm u},n,k}^{(t)}\bar{\bm{p}}_{k}\right]\nonumber\\
	&{}+\sum_{j \in \mathcal{K}_{g_{k}} \setminus k}\left[\eta^{(c)}\left(1-\frac{\tau_{\rm p}}{\tau_{\rm c}}\right)
	+\lambda_{{\rm u},j}\right]\phi_{{\rm u},j}^{(t)}\left(\|\bm{\Xi}_{j}\bar{\bm{p}}_{k}\|^{2}
	+M\|\hat{\bm{\Xi}}_{j}\bar{\bm{p}}_{k}\|^{2}\right)-\left[\eta^{(c)}\left(1-\frac{\tau_{\rm p}}{\tau_{\rm c}}\right)+\lambda_{{\rm u},k}\right]\nonumber\\
	&{}\times\left[2\bm{\psi}_{{\rm u},k}^{(t)}\bar{\bm{p}}_{k}-\phi_{{\rm u},k}^{(t)}
	\left(\|\bm{\Xi}_{k}\bar{\bm{p}}_{k}\|^{2}+M|\hat{\bm{\xi}}_{k}^{\rm H}\bar{\bm{p}}_{k}|^{2}\right)\right]+\sum_{j \notin \mathcal{K}_{g_{k}}}\left[\eta^{(c)}\left(1-\frac{\tau_{\rm p}}{\tau_{\rm c}}\right)+\lambda_{{\rm u},j}\right]\phi_{{\rm u},j}^{(t)}\|\bm{\Xi}_{j}\bar{\bm{p}}_{k}\|^{2}\nonumber\\
	&{}+M\sum_{j \in \mathcal{K}_{g_{k}}}\bar{\lambda}_{{\rm m},j}\phi_{{\rm m},g_{j}}^{(t)}\|\hat{\bm{\Xi}}_{j}\bar{\bm{p}}_{k}\|^{2}+\sum_{j \in \mathcal{K}}\bar{\lambda}_{{\rm m},j}\phi_{{\rm m},g_{j}}^{(t)}\|\bm{\Xi}_{j}\bar{\bm{p}}_{k}\|^{2}-2M\sum_{j \in \mathcal{K}_{g_{k}}}\lambda_{{\rm e},j}\bar{\bm{p}}_{k}^{(t){\rm H}} \hat{\bm{\Xi}}_{j}^{2}\bar{\bm{p}}_{k}-2\sum_{j \in \mathcal{K}}\lambda_{{\rm e},j}\bar{\bm{p}}_{k}^{(t){\rm H}}\bm{\Xi}_{j}^{2}\bar{\bm{p}}_{k}.\label{2.4}
\end{align}
\rule[0.5ex]{2\columnwidth}{0.5pt}
\end{figure*}

Since the objective function in \eqref{2.4} is strongly convex over $\bar{\bm p}_{k}$, by setting its gradient to zeros, the minimizer $\bar{\bm p}_{k}^{\circ}$ is uniquely given by \eqref{dp_m2}. Moreover, there are $K$ subproblems over $\left\{ \rho_{k}\right\} _{k\in\mathcal{K}}$,
with each written as
\begin{align}\label{2.5}
	\min_{\rho_{k}}\ &\frac{\eta^{(c)}\left(1-\frac{\tau_{\rm p}}{\tau_{\rm c}}\right)\phi_{{\rm u},k}^{(t)}
		\sigma_{k}^{2}}{\rho_{k}}+\frac{\bar{\lambda}_{{\rm m},k}\phi_{{\rm m},g_{k}}^{(t)}\sigma_{k}^{2}}{\rho_{k}}\nonumber\\
	&\quad{}+\frac{\lambda_{{\rm u},k}\phi_{{\rm u},k}^{(t)}\sigma_{k}^{2}}{\rho_{k}}+\frac{\lambda_{{\rm e},k}
		\mathcal{F}^{-1}\left(\bar{e}_{k}\right)}{1-\rho_{k}}.
\end{align}
By setting its derivative with respect to $\rho_{k}$ to zeros, the minimizer $\rho_{k}^{\circ}$ is uniquely given by \eqref{dp_m3}. By substituting the optimal solution $\mathcal{V}^{\circ}$ into \eqref{2.1}, the dual function $\mathcal{D}(\mathcal{L})$ is expressed in closed-form as shown in \eqref{dual}.

\section{Closed-form solution to the problem \eqref{pro_1}}
\label{Appendix-D}
Since the constraints $\lambda_{{\rm u},k}\geq0$, $\lambda_{{\rm e},k}\geq0$, $\lambda_{{\rm p},n}\geq0$ and $\lambda_{{\rm c},n}\geq0$ in \eqref{dom} are independent of the other constraints, the optimal solutions of $\lambda_{{\rm u},k}\geq0$, $\lambda_{{\rm e},k}\geq0$, $\lambda_{{\rm p},n}\geq0$ and $\lambda_{{\rm c},n}\geq0$ can be obtained as $\mu_{{\rm u},k}^{+}$, $\mu_{{\rm e},k}^{+}$, $\mu_{{\rm p},n}^{+}$ and $\mu_{{\rm c},n}^{+}$, given by \eqref{pro_2}. Consequently, the problem \eqref{pro_1} reduces to $G$ subproblems, with each expressed as
\begin{subequations}\label{C2.1}
	\begin{align}
		\min_{\lambda_{{\rm m},g},\left\{\bar{\lambda}_{{\rm m},k}\right\} _{k \in \mathcal{K}_{g}}}\!\! 
		& \left(\lambda_{{\rm m},g}-\mu_{{\rm m},g}\right)^{2}+\sum_{k \in \mathcal{K}_{g}}\left(\bar{\lambda}_{{\rm m},k}
		-\bar{\mu}_{{\rm m},k}\right)^{2},\label{C1.1}\\
		{\rm s.\ t.\ } & \ \lambda_{{\rm m},g}\geq0,\ \bar{\lambda}_{{\rm m},k}\geq0,\ \forall k \in \mathcal{K}_{g},\label{C1.2}\\
		&\ \sum_{k\in\mathcal{K}_{g}}\!\!\bar{\lambda}_{{\rm m},k}-\lambda_{{\rm m},g}=\eta^{(c)}\big(1-\frac{\tau_{\rm p}}{\tau_{\rm c}}\big). \label{C1.33}
	\end{align}
\end{subequations}
From \eqref{C2.1}, it is not hard to know that $\lambda_{{\rm m},g}$ and $\bar{\lambda}_{{\rm m},k}$ can be either zero or positive. If they are positive, they must satisfy the following KKT conditions:
\begin{align}\label{C1.3}
	\begin{cases}
	\begin{array}{l}
	\lambda_{{\rm m},g}-\mu_{{\rm m},g}-\frac{\varpi_{g}}{2}=0,\\
	\bar{\lambda}_{{\rm m},k}-\bar{\mu}_{{\rm m},k}+\frac{\varpi_{g}}{2}=0,
	\end{array}
	\end{cases}
\end{align}
where $\varpi_{g}$ is the dual variable corresponding to \eqref{C1.33}. Together with the case of $\lambda_{{\rm m},g}=0$ and $\bar{\lambda}_{{\rm m},k}=0$, we can obtain the optimal solutions of $\lambda_{{\rm m},g}$ and $\bar{\lambda}_{{\rm m},k}$, expressed as \eqref{pro_22}. Finally, substituting \eqref{pro_22} into \eqref{C1.33} yields $\sum_{k \in \mathcal{K}_{g}}\left(\bar{\mu}_{{\rm m},k}-\frac{\varpi_{g}}{2}\right)^{+}-\left(\mu_{{\rm m},g}+\frac{\varpi_{g}}{2}\right)^{+}=\eta^{(c)}\left(1-\frac{\tau_{\rm p}}{\tau_{\rm c}}\right)$, from which the value of $\varpi_{g}$ can be determined by using the bisection method.

\section{Derivations of subgradient}
\label{Appendix-E}
According to \eqref{fea_2}, the subgradients of $\bar{h}^{(t)}(\mathcal{V})$ with respect to $\bar{\bm q}_{g},\forall g\in{\cal G}$, $\bar{\bm p}_{k}$ and $\rho_{k},k\in{\cal K}$ are given by
\begin{subequations} \label{Eq-AppendixE}
	\begin{align}
		\frac{\partial \bar{h}^{(t)}(\mathcal{V})}{\partial\bar{\bm{q}}_{g}} 
		&= \sum_{j \in \mathcal{K}}\!\! \frac{\partial\hbar_{{\rm m},j}+\partial\hbar_{{\rm u},j}+\partial\hbar_{{\rm e},j}}{\partial\bar{\bm{q}}_{g}}
		+ \!\sum_{n\in\mathcal{N}}\!\!\frac{\partial\hbar_{{\rm c},n}}{\partial\bar{\bm{q}}_{g}}, \label{D.4a}\\
		\frac{\partial \bar{h}^{(t)}(\mathcal{V})}{\partial\bar{\bm{p}}_{k}} 
		&= \sum_{j \in \mathcal{K}}\!\!\frac{\partial\hbar_{{\rm m},j}+\partial\hbar_{{\rm u},j}+\partial\hbar_{{\rm e},j}} {\partial\bar{\bm{p}}_{k}}+ \!\sum_{n\in\mathcal{N}}\!\!\frac{\partial\hbar_{{\rm c},n}}{\partial\bar{\bm{p}}_{k}}, \label{D.4b} \\
		\frac{\partial \bar{h}^{(t)}(\mathcal{V})}{\partial\rho_{k}} 
		&= \sum_{j \in \mathcal{K}}\!\!\frac{\partial\hbar_{{\rm m},j}+\partial\hbar_{{\rm u},j}+\partial\hbar_{{\rm e},j}}{\partial\rho_{k}}. \label{D.4c}
	\end{align}
\end{subequations}
Since $\left(x\right)^{+}$ in \eqref{fea_2} is the pointwise maximum between $x$ and 0, the subgradient of $\left\{ \hbar_{{\rm m},j}, \hbar_{{\rm u},j},\hbar_{{\rm e}, j}\right\}_{j\in\mathcal{K}}$ and $\left\{ \hbar_{{\rm c},n}\right\} _{n\in\mathcal{N}}$ with respect to $\bar{\bm q}_{g}$  can be computed as
\begin{subequations}\label{E_q}
	\begin{align}
		\frac{\partial\hbar_{{\rm m},j}}{\partial\bar{\bm{q}}_{g}} &=
		\left\{
		\begin{array}{ll}
		 	2\bm{\kappa}_{{\rm m},j}, & \text{if } \bar{R}_{j}^{(t)}(\mathcal{V})<\bar{r}_{{\rm m}, g_{j}}, g_{j} = g, \\
		 	2\phi_{{\rm m},g_{j}}^{(t)}\bm{\Xi}_{j}^{2}\bar{\bm{q}}_{g}, &\text{if } \bar{R}_{j}^{(t)}(\mathcal{V})<\bar{r}_{{\rm m}, g_{j}}, g_{j} \neq g,\\
			\bm{0}, & \text{otherwise},
		\end{array}
		\right. \\
		\frac{\partial\hbar_{{\rm u},j}}{\partial\bar{\bm{q}}_{g}} &=
		\left\{
		\begin{array}{ll}
			-2\phi_{{\rm u},j}^{(t)}\bm{\Xi}_{j}^{2}\bar{\bm{q}}_{g}, & \text{if } \bar{R}_{{\rm u},j}^{(t)}(\mathcal{V})<\bar{r}_{{\rm u},j}, \\
			\bm{0}, & \text{otherwise},
		\end{array}
		\right. \\
		\frac{\partial\hbar_{{\rm e},j}}{\partial\bar{\bm{q}}_{g}} &=
		\left\{
		\begin{array}{ll}
			-2\bm{\nabla}_{j}\bar{\bm{q}}_{g}^{(t)}, & \text{if } \mathcal{E}_{j}(\mathcal{V})<\bar{e}_{j},g_{j}=g, \\
			-2\bm{\Xi}_{j}^{2}\bar{\bm{q}}_{g}^{(t)}, & \text{if } \mathcal{E}_{j}(\mathcal{V})<\bar{e}_{j},g_{j}\neq g, \\
			\bm{0}, & \text{otherwise},
		\end{array}
		\right. \\
		\frac{\partial\hbar_{{\rm c},n}}{\partial\bar{\bm{q}}_{g}} &=
		\left\{
		\begin{array}{ll}
			\mathcal{R}_{{\rm m},g}^{(t)}\bm{\vartheta}_{{\rm m},n,g}^{(t){\rm H}}, & \text{if } \bar{C}_{{\rm bh},n}^{(t)}(\mathcal{V})>\bar{c}_{n,\max}, \\
			\bm{0}, & \text{otherwise},
		\end{array}
		\right.
	\end{align}
\end{subequations}
where $\bm{\kappa}_{{\rm m},j}=-\bm{\psi}_{{\rm m},g_{j}}^{(t){\rm H}}+\phi_{{\rm m},g_{j}}^{(t)} \left(M\hat{\bm{\xi}}_{j}\hat{\bm{\xi}}_{j}^{\rm H}+\bm{\Xi}_{j}^{2}\right)\bar{\bm{q}}_{g}$, $\bm{\nabla}_{j}=\bm{\Xi}_{j}^{2}+M\hat{\bm{\Xi}}_{j}^{2}$.

Then, following similar derivations, we obtain the subgradients of $\hbar_{{\rm m},j}$, $\hbar_{{\rm u},j}$, $\hbar_{{\rm e},j}$ and $\hbar_{{\rm c},n}$ with respect to $\bar{\bm p}_{k}$ and $\rho_{k}$, given by
\begin{subequations}\label{E_p}
	\begin{align}
		\frac{\partial\hbar_{{\rm m},j}}{\partial\bar{\bm{p}}_{k}} &=
		\left\{
		\begin{array}{ll}
			2\phi_{{\rm m},g_{j}}^{(t)} \bm{\nabla}_{j}\bar{\bm{p}}_{k}, & \text{if } \bar{R}_{j}^{(t)}(\mathcal{V})<\bar{r}_{{\rm m}, g_{j}}, g_{j}=g, \\
			2\phi_{{\rm m},g_{j}}^{(t)}\bm{\Xi}_{j}^{2}\bar{\bm{p}}_{k}, & \text{if } \bar{R}_{j}^{(t)}(\mathcal{V})<\bar{r}_{{\rm m}, g_{j}}, g_{j} \neq g,\\
			\bm{0},  & {\rm otherwise},
		\end{array}
		\right. \\
		\frac{\partial\hbar_{{\rm u},j}}{\partial\bar{\bm{p}}_{k}} &=
		\left\{
		\begin{array}{ll}
			2\bm{\kappa}_{{\rm u},k}, & \text{if } \bar{R}_{j}^{(t)}(\mathcal{V})<\bar{r}_{{\rm m}, g_{j}}, j=k, \\
			2\phi_{{\rm u},j}^{(t)}\bm{\nabla}_{j}\bar{\bm{p}}_{k}, & \text{if } \bar{R}_{j}^{(t)}(\mathcal{V})<\bar{r}_{{\rm m}, g_{j}}, j \in \mathcal{K}_{g_{k}} \setminus k,\\
		2\phi_{{\rm u},j}^{(t)}\bm{\Xi}_{j}^{2}\bar{\bm{p}}_{k}, & \text{if } \bar{R}_{j}^{(t)}(\mathcal{V})<\bar{r}_{{\rm m},g_{j}}, j \notin \mathcal{K}_{g_{k}}, \\
		\bm{0}, & \text{otherwise},
		\end{array}
		\right. \\
		\frac{\partial\hbar_{{\rm e},j}}{\partial\bar{\bm{p}}_{k}} &=
		\left\{
		\begin{array}{ll}
			-2\bm{\nabla}_{j}\bar{\bm{p}}_{k}^{(t)}, &\text{if } \mathcal{E}_{j}(\mathcal{V})<\bar{e}_{j}, j \in \mathcal{K}_{g_{k}}, \\
			-2\bm{\Xi}_{j}^{2}\bar{\bm{p}}_{k}^{(t)}, &\text{if } \mathcal{E}_{j}(\mathcal{V})<\bar{e}_{j}, j \notin \mathcal{K}_{g_{k}}, \\
			\bm{0}, & \text{otherwise},
		\end{array}
		\right. \\
		\frac{\partial\hbar_{{\rm c},n}}{\partial\bar{\bm{p}}_{k}} &=
		\left\{
		\begin{array}{ll}
			R_{{\rm u,}k}^{(t)}\bm{\vartheta}_{{\rm u},n,k}^{(t)}, & \text{if } \bar{C}_{{\rm bh},n}^{(t)}(\mathcal{V})>\bar{c}_{n,\max}, \\
			\bm{0}, & \text{otherwise},
		\end{array}
		\right. \\
	\end{align}
\end{subequations}
where $\bm{\kappa}_{{\rm u},k}=-\bm{\psi}_{{\rm u},k}^{(t){\rm H}}+\phi_{{\rm u},k}^{(t)} \left(M\hat{\bm{\xi}}_{k}\hat{\bm{\xi}}_{k}^{\rm H}+\bm{\Xi}_{k}^{2}\right)\bar{\bm{p}}_{k}$, and
\begin{subequations}\label{E_r}
	\begin{align}
		\frac{\partial\hbar_{{\rm m},j}}{\partial\rho_{k}} &\!=\!
		\left\{
		\begin{array}{ll}
			\frac{-\phi_{{\rm m},g_{k}}^{(t)}\sigma_{k}^{2}}{\rho_{k}^{2}}, &\text{if } \bar{R}_{j}^{(t)}(\mathcal{V}) < \bar{r}_{{\rm m}, g_{j}}, j=k, \\
			0, &\text{otherwise},
		\end{array} 
		\right. \\
		\frac{\partial\hbar_{{\rm u},j}}{\partial\rho_{k}} &=
		\left\{
		\begin{array}{ll}
			-\frac{\phi_{{\rm u},k}^{(t)}\sigma_{k}^{2}}{\rho_{k}^{2}}, &\text{if } \bar{R}_{{\rm u},j}^{(t)}(\mathcal{V})<\bar{r}_{{\rm u}, j}, j=k, \\
			0, &\text{otherwise},
		\end{array} 
		\right. \\
		\frac{\partial\hbar_{{\rm e},j}}{\partial\rho_{k}} &=
		\left\{
		\begin{array}{ll}
			\frac{\mathcal{F}^{-1}\left(\bar{e}_{k}\right)}{\left(1-\rho_{k}\right)^{2}}, & \text{if }  \mathcal{E}_{j}(\mathcal{V})<\bar{e}_{j}, j=k, \\
			0, & \text{otherwise},
		\end{array} 
		\right. \\
		\frac{\partial\hbar_{{\rm c},n}}{\partial\rho_{k}} &= 0.
	\end{align}
\end{subequations}
Finally, substituting \eqref{E_q}, \eqref{E_p} and \eqref{E_r} into \eqref{Eq-AppendixE} yields the subgradients of $\bar{h}^{(t)}(\mathcal{V})$ with respect to $\bar{\bm q}_{g}$, $\bar{\bm p}_{k}$ and $\rho_{k}$.

\section{Closed-form solution to the Problem \eqref{fea_4}}
\label{Appendix-F}
Since the problem \eqref{fea_4} is strictly feasible and convex, its optimal solution must satisfy the following KKT conditions:
\begin{subequations}
	\begin{align}
		&2\left(\bar{\bm{q}}_{g}-\bm{z}_{{\rm m},g}\right)+2\sum_{n \in \mathcal{N}}\mu_{n}\bm{E}_{n}\bar{\bm{p}}_{g} 
		 =0, \ \forall g \in \mathcal{G},\label{e1}\\
		&2\left(\bar{\bm{p}}_{k}-\bm{z}_{{\rm u},k}\right)+2\sum_{n \in \mathcal{N}}\mu_{n}\bm{E}_{n}\bar{\bm{p}}_{k} 
		=0, \ \forall k \in \mathcal{K},\label{e2}\\
		&\mu_{n}\Big(\sum_{k \in \mathcal{K}}\bar{\bm{p}}_{k}^{{\rm H}}\bm{E}_{n}\bar{\bm{p}}_{k}+\sum_{g \in \mathcal{G}}\bar{\bm{q}}_{g}^{\rm H}\bm{E}_{n}\bar{\bm{q}}_{g}-\bar{p}_{n,\max}\Big) 
	  	=0, \ \forall n \in \mathcal{N},\label{e5}\\
		&2\left(\rho_{k}-\gamma_{k}\right) 
		 =0, \ \forall k \in \mathcal{K},\label{e3}
	\end{align}
\end{subequations}
where $\mu_n\geq0$ is the dual variables. Next, we discuss two cases: $\mu_{n}=0$ and $\mu_n>0$.

Case I: $\mu_{n}=0$. In this case, \eqref{e5} always holds. Putting $\mu_{n}=0$  into \eqref{e1} and \eqref{e2}, we have
\begin{equation}\label{e6}
	\left\{
	\begin{array}{ll}
		\bm{z}_{{\rm m},g} = \bar{\bm{q}}_{g}, & \forall g \in \mathcal{G}, \\
		\bm{z}_{{\rm u},k} = \bar{\bm{p}}_{k}, & \forall k\in\mathcal{K}, \\
		\gamma_{k} = \rho_{k}, & \forall k \in \mathcal{K}.
	\end{array}
	\right.
\end{equation}
Substituting \eqref{e6} into \eqref{e3}, we obtain a simplified condition for checking whether \eqref{e6} is the optimal solution to \eqref{fea_4}:
\begin{align} \label{e7}
	\Lambda_{n} 
		\triangleq \sum_{k \in \mathcal{K}}\bm{z}_{{\rm u},k}^{\rm H}\bm{E}_{n}\bm{z}_{{\rm u},k} + \sum_{g \in \mathcal{G}}\bm{z}_{{\rm m},g}^{\rm H}\bm{E}_{n}\bm{z}_{{\rm m},g} 
		\leq \bar{p}_{n,\max}.
\end{align}
If the condition \eqref{e7} is satisfied, it means that \eqref{e6} is the optimal solution. Otherwise, it cannot be the optimal solution, and we need to consider the other case $\mu_{n}>0$.

Case II: $\mu_{n}>0$. Now, \eqref{e1}-\eqref{e5} reduce to
\begin{align}\label{e8}
	\left\{
		\begin{array}{ll}
			\bar{\bm{q}}_{g} = \left(\bm{I}_{N}+\sum_{n \in \mathcal{N}}\mu_{n}\bm{E}_{n}\right)^{-1}\bm{z}_{{\rm m},g}, & \forall g\in\mathcal{G}, \\
			\bar{\bm{p}}_{k} = \left(\bm{I}_{N}+\sum_{n \in \mathcal{N}}\mu_{n}\bm{E}_{n}\right)^{-1}\bm{z}_{{\rm u},k}, & \forall k\in\mathcal{K}, \\
			\rho_{k}=\gamma_{k}, & \forall k \in \mathcal{K},
		\end{array}
	\right. \\
	\sum_{k \in \mathcal{K}}\bar{\bm{p}}_{k}^{\rm H}\bm{E}_{n}\bar{\bm{p}}_{k}+\sum_{g \in \mathcal{G}}\bar{\bm{q}}_{g}^{\rm H}\bm{E}_{n}\bar{\bm{q}}_{g} = \bar{p}_{n,\max}, \quad \forall n\in\mathcal{N}.\label{e9}
\end{align}
Substituting \eqref{e8} into \eqref{e9}, we obtain $\mu_{n} = \sqrt{{\Lambda_{n}}/{\bar{p}_{n, \max}}}-1$. Putting $\mu_{n}$ into \eqref{e8} gives the optimal solution:
\begin{align}\label{e10}
	\left\{
	\begin{array}{ll}
		\bar{\bm{q}}_{g} = \sum_{n \in \mathcal{N}}\sqrt{\frac{\Lambda_{n}}{\bar{p}_{n,\max}}}\bm{E}_{n}\bm{z}_{{\rm m},g}, & \forall g \in \mathcal{G},\\
		\bar{\bm{p}}_{k} = \sum_{n \in \mathcal{N}}\sqrt{\frac{\Lambda_{n}}{\bar{p}_{n,\max}}}\bm{E}_{n}\bm{z}_{{\rm u},k}, & \forall k \in \mathcal{K},\\
		\rho_{k} = x_{k}, \quad \forall k\in \mathcal{K}.
	\end{array}
	\right.
\end{align}
Finally, combining \eqref{e6} with \eqref{e10} yields the desired \eqref{fea_5}.

\bibliographystyle{IEEEtran}
\bibliography{References}

\begin{IEEEbiography}[{\includegraphics[width=1in,height=1.25in,clip,keepaspectratio]
	{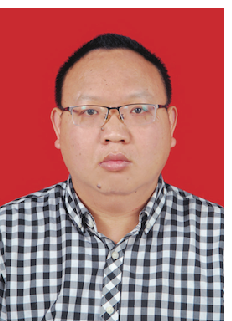}}]{Fangqing Tan}
	received the M.S. degree in communication and information system from Chongqing University of Post and Telecommunications in 2012. He received the Ph.D. degree in Beijing University of Post and Telecommunications in 2017, Beijing, China. From July 2017 to September 2018, he was a Lecturer with the School of Information and Communication, Guilin University of Electronic Technology, Guilin, China. He is now a Postdoctoral Fellow at the School of Electronics and Information Technology, Sun Yat-sen University, Guangzhou, China. His research interests  mainly focus on 5G/6G wireless communications, and Internet of Things.
\end{IEEEbiography}	

\begin{IEEEbiography}[{\includegraphics[width=1in,height=1.25in,clip,keepaspectratio]
		{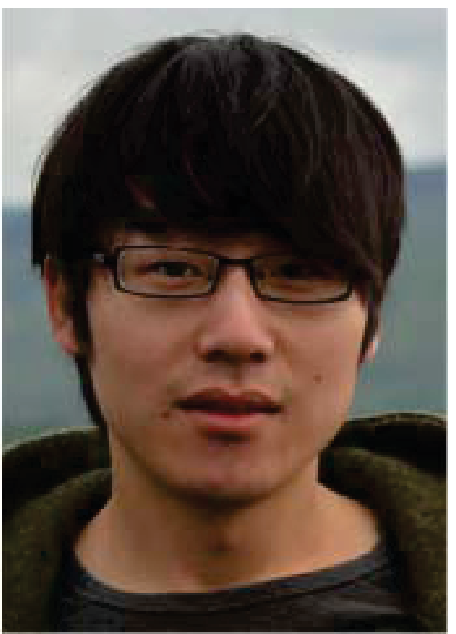}}]{Peiran Wu} (M'16) 
	received the Ph.D. degree in electrical and computer engineering at the University of British Columbia (UBC), Vancouver, Canada, in 2015. From October 2015 to December 2016, he was a Postdoctoral Fellow at the same university. In summer 2014, he was a Visiting Scholar at the Institute for Digital Communications, Friedrich-Alexander-University Erlangen-Nuremberg (FAU), Erlangen, Germany. Since February 2017, he has been with the Sun Yat-sen University, Guangzhou, China, where he is now an Associate Professor. Since 2019, he has been an Adjunct Associate Professor with the Southern Marine Science and Engineering Guangdong Laboratory, Zhuhai, China. His research interests include mobile edge computing, wireless power transfer, and energy-efficient wireless communications. 
	
	He was the recipient of the Fourth-Year Fellowship in 2010, the C. L. Wang Memorial Fellowship in 2011, Graduate Support Initiative (GSI) Award in 2014 from the UBC, German Academic Exchange Service (DAAD) Scholarship in 2014, and the Chinese Government Award for Outstanding Self-Financed Students Abroad in 2014. 
\end{IEEEbiography}

\begin{IEEEbiography}[{\includegraphics[width=1in,height=1.25in,clip,keepaspectratio]
		{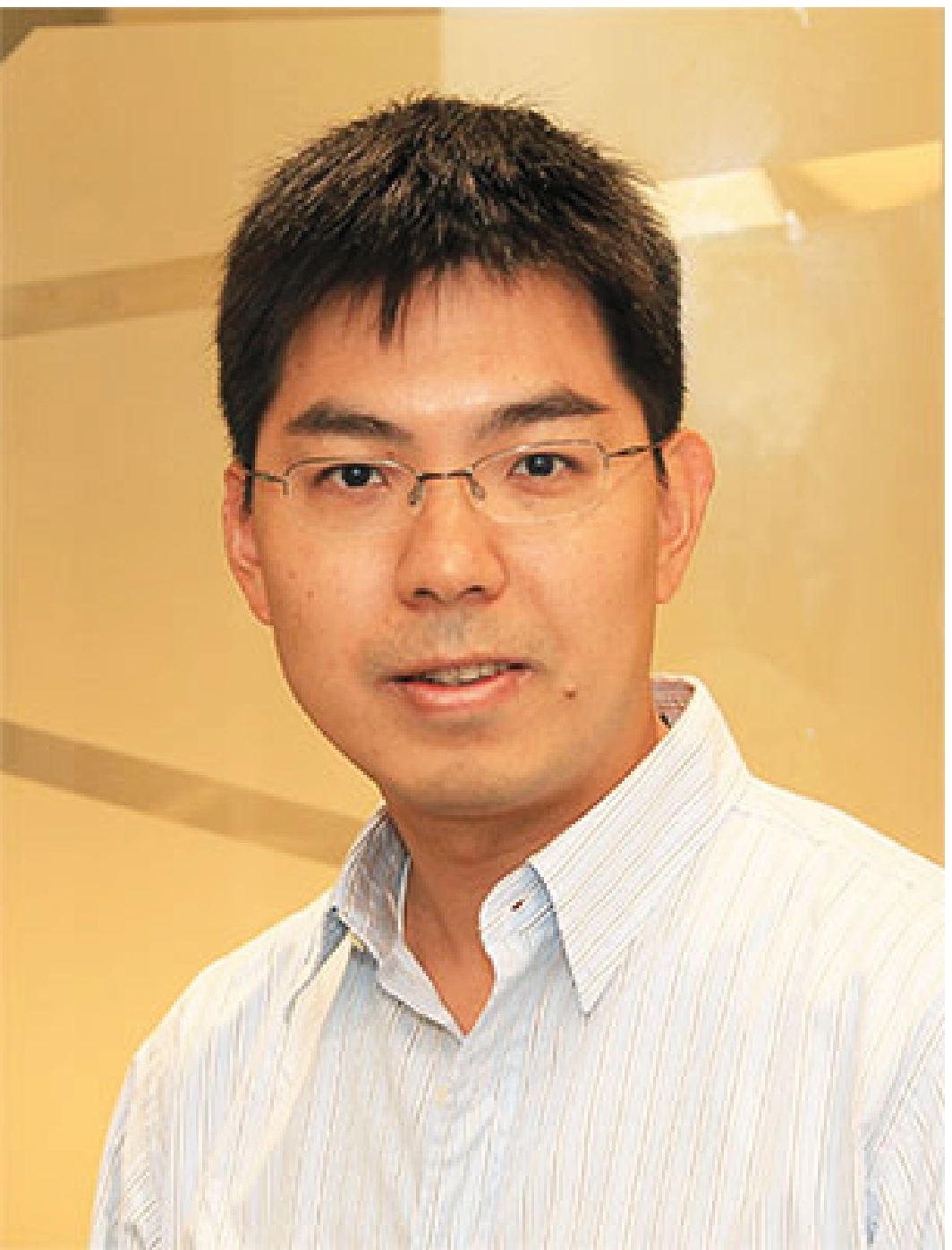}}]{Yik-Chunk Wu} (S'99-M'05-SM'14) 	
	received the B.Eng. (EEE) degree in 1998 and the M.Phil. degree in 2001 from the University of Hong Kong (HKU), and Ph.D. degree from Texas A\&M University in 2005.  From 2005 to 2006, he was with the Thomson Corporate Research, Princeton, NJ, as a Member of Technical Staff.  
	
	Since 2006, he has been with HKU, currently as an Associate Professor.  He was a visiting scholar at Princeton University, in summers of 2015 and 2017.  His research interests are in general areas of signal processing, machine learning and communication systems.  
	
	Dr. Wu served as an editor for {\scshape IEEE Communications Letters} and {\scshape IEEE Transactions on Communications}.  He is currently an editor for {\scshape Journal of Communications and Networks}. 
\end{IEEEbiography}

\begin{IEEEbiography}[{\includegraphics[width=1in,height=1.25in,clip,keepaspectratio]
		{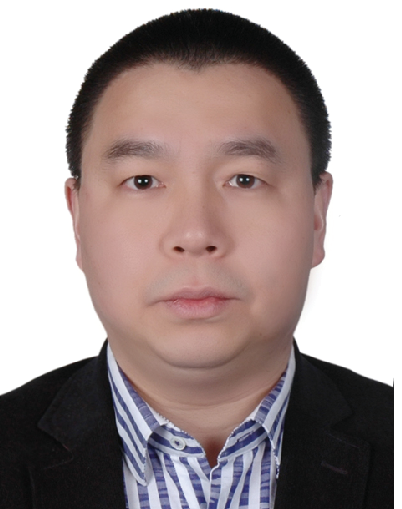}}]{Minghua Xia} (M'12) 
	 received the Ph.D. degree in Telecommunications and Information Systems from Sun Yat-sen University, Guangzhou, China, in 2007. 
	
	From 2007 to 2009, he was with the Electronics and Telecommunications Research Institute (ETRI) of South Korea, Beijing R\&D Center, Beijing, China, where he worked as a member and then as a senior member of engineering staff. From 2010 to 2014, he was in sequence with The University of Hong Kong, Hong Kong, China; King Abdullah University of Science and Technology, Jeddah, Saudi Arabia; and the Institut National de la Recherche Scientifique (INRS), University of Quebec, Montreal, Canada, as a Postdoctoral Fellow. Since 2015, he has been a Professor with Sun Yat-sen University. Since 2019, he has also been an Adjunct Professor with the Southern Marine Science and Engineering Guangdong Laboratory (Zhuhai). His research interests are in the general areas of wireless communications and signal processing. 
	
	Dr. Xia received the Professional Award at the IEEE TENCON, held in Macau, in 2015. He served as a TPC Symposium Chair of IEEE ICC'2019. He currently serves as a TPC Symposium Chair of IEEE ICC'2021, and an Associate Editor for the {\scshape IEEE Transactions on Cognitive Communications and Networking} and the {\it IET Smart Cities}. He was recognized as an Exemplary Reviewer by {\scshape IEEE Transactions on Communications} in 2014, {\scshape IEEE Communications Letters} in 2014, and {\scshape IEEE Wireless Communications Letters} in 2014 and 2015. 
	
\end{IEEEbiography}
\vfill

\end{document}